\documentclass[twocolumn,superscriptaddress,longbibliography,
	aps,prl,preprintnumbers]{revtex4-1}

\usepackage[normalem]{ulem}
\usepackage{graphicx}
\usepackage{bm}
\usepackage{color}
\usepackage{epstopdf}
\usepackage{amsmath}
\usepackage{amssymb}
\usepackage{epstopdf}
\usepackage{lipsum}

\usepackage{multirow}

\usepackage[urlcolor=blue,colorlinks=true,citecolor=blue,linkcolor=blue,pdfstartview={FitH},bookmarks=false]{hyperref}

\usepackage{xcolor}

\graphicspath{{fig/}{./fig/}{.}}

\begin{document}

\title{
Electronic and dynamical properties of CeRh$_{2}$As$_{2}$:\\
Role of Rh$_{2}$As$_{2}$ layers and expected hidden orbital order
}

\author{Andrzej Ptok}
\email[e-mail: ]{aptok@mmj.pl}
\affiliation{\mbox{Institute of Nuclear Physics, Polish Academy of Sciences,
	W. E. Radzikowskiego 152, PL-31342 Krak\'{o}w, Poland}}

\author{Konrad J. Kapcia}
\affiliation{\mbox{Faculty of Physics, Adam Mickiewicz University in Pozna\'{n},
	Uniwersytetu Pozna\'{n}skiego 2, PL-61614 Pozna\'{n}, Poland}}

\author{Pawe\l{} T. Jochym}
\affiliation{\mbox{Institute of Nuclear Physics, Polish Academy of Sciences,
	W. E. Radzikowskiego 152, PL-31342 Krak\'{o}w, Poland}}

\author{\mbox{Jan \L{}a\.{z}ewski}}
\affiliation{\mbox{Institute of Nuclear Physics, Polish Academy of Sciences,
	W. E. Radzikowskiego 152, PL-31342 Krak\'{o}w, Poland}}

\author{Andrzej M. Ole\'{s}}
\affiliation{\mbox{Institute of Theoretical Physics, Jagiellonian University,
Profesora Stanis\l{}awa \L{}ojasiewicza 11, PL-30348  Krak\'{o}w, Poland}}
\affiliation{Max Planck Institute for Solid State Research,
Heisenbergstrasse 1, D-70569 Stuttgart, Germany}

\author{Przemys\l{}aw Piekarz$\,$}
\email[e-mail: ]{piekarz@wolf.ifj.edu.pl}
\affiliation{\mbox{Institute of Nuclear Physics, Polish Academy of Sciences,
	W. E. Radzikowskiego 152, PL-31342 Krak\'{o}w, Poland}}

\date{\today}

\begin{abstract}
Recently discovered heavy fermion CeRh$_{2}$As$_{2}$ compound crystallizes in the 
nonsymmorphic {\it P4/nmm} symmetry, which enables the occurrence of 
topological protection. Experimental results show that this material exhibits 
unusual behavior, which is manifested by the appearance of two superconducting phases.
In this work, we uncover and discuss a role of Rh$_{2}$As$_{2}$ layers and their 
impact on the electronic and dynamical properties of the system. The location of 
Ce atoms between two non-equivalent
layers allows for the realization of hidden orbital order.
We point out that the electronic band structure around the Fermi level is 
associated mostly with Ce $4f$ and Rh $4d$ orbitals and suggest the occurrence 
of the Lifshitz transition induced by the external magnetic field.
We discuss also the role played by the $f$--$d$ orbital hybridization in the electronic band structure.
\end{abstract}

\maketitle

\paragraph*{Introduction.}---
Recently discovered CeRh$_{2}$As$_{2}$ superconductor~\cite{khim.landaeta.21} 
is one of rare examples of a heavy fermion systems crystallizing in {\it P4/nmm} 
space group. In contrast to isostructural SrPt$_{2}$As$_{2}$~\cite{kudo.nishikubo.10} 
or $R$Pt$_{2}$Si$_{2}$ (where $R$=Y, La, Nd, and Lu)~\cite{nagano.araoka.13}, 
it does not exhibit the coexistence of superconductivity and charge density waves.

In the case of CeRh$_{2}$As$_{2}$, two anomalies in the specific heat 
are observed below $1$~K~\cite{khim.landaeta.21}. First anomaly 
(at lower temperature) is associated with the phase transition from the normal
state to the superconducting (SC) phase (connected with a diamagnetic drop of 
the ac-susceptibility and the specific heat jump of the same order of
magnitude as the Bardeen--Cooper--Schrieffer value).
Second anomaly (at higher temperature) is not associated with superconductivity,
but likely signals some other kind of order
(its $T_c$ increases with the in-plane magnetic field).
In the presence of the magnetic field perpendicular to the Rh$_{2}$As$_{2}$ 
layers, the system exhibits a phase transition inside the SC state.
It is suggested that at the transition the parity of superconductivity 
changes from even to odd one.
This leads to $H$-$T$ phase diagram in a characteristic form~\cite{khim.landaeta.21}, which can be treated as a generic one for a realization of spin-singlet 
and spin-triplet SC phases~\cite{schertenleib.fischer.21}. 
Here, we would like to emphasize that this behavior is also observed in 
other systems, where a coexistence of trivial and topological SC phases can occur~\cite{yoshida.sigrist.14,mockli.khodas.18,mockli.khodas.19} due to 
finite spin-orbit coupling (SOC), whereas the ranges, in which both kinds 
of superconductivity exist strongly depend on model parameters, e.g. 
on the ratio of electron hopping and SOC~\cite{mockli.ramires.21}.

\begin{figure}[!b]
\includegraphics[width=\linewidth]{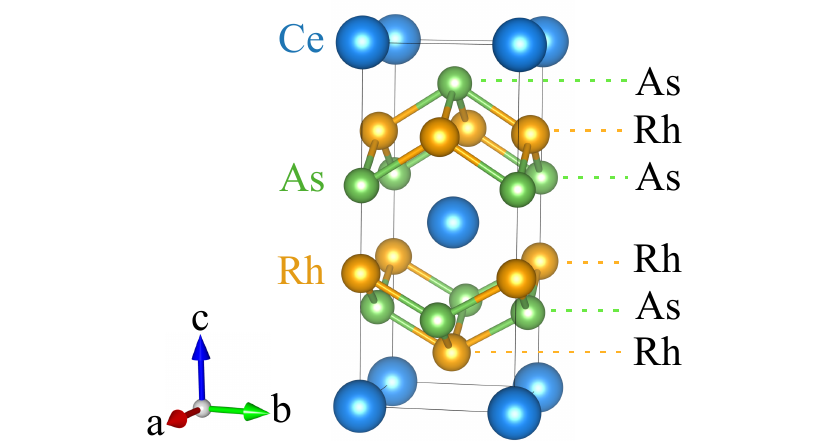}
\caption{
The unit cell of the CeRh$_{2}$As$_{2}$ crystal structure.
The system is composed of two Rh$_{2}$As$_{2}$ layers, which 
separately play a role of the mirrors of the glide symmetry.
\label{fig.crystal}
}
\end{figure}

The symmetry of the system described by {\it P4/nmm} space group is nonsymmorphic 
with multiple symmetries protecting the Dirac points~\cite{young.kane.15}.
The similar situation has been recently reported for Dirac semimetals, 
crystallizing in the same symmetry~\cite{schoop.ali.16,takane.wang.16,hosen.dimitri.17,chen.xu.17,schoop.topp.18,pezzini.vandelft.18,wang.xu.18}.
However, contrary to these materials, where a dominant role is played by the 
square nets~\cite{klemenz.lei.19}, in the CeRh$_{2}$As$_{2}$ compound, 
there exist two different types of Rh$_{2}$As$_{2}$ layers (Fig.~\ref{fig.crystal}).
What is more important, these layers can be treated as planes of the glide symmetry.
The purpose of this Letter is to highlight the important role played by the Rh$_{2}$As$_{2}$ layers in possible realization of the hidden orbital order 
as a result of two distinguishable Ce atom positions with respect to the 
neighboring Rh$_{2}$As$_{2}$ layers.
We point out that an existence of this order leads to essential modification of the phonon dispersion and could be tested by experimental measurements.
We discuss also possibility of the Lifshitz transition induced by the magnetic field as a source of the topological superconducting phase reported experimentally in Ref.~\cite{khim.landaeta.21}.


The density functional theory (DFT) calculations were performed using the 
{\sc Vasp} code~\cite{kresse.hafner.94,kresse.furthmuller.96,kresse.joubert.99}.
Phonon calculations were conducted by {\sc Alamode}~\cite{tadano.gohda.14} 
for the thermal distribution of multi-displacement of atoms at $T = 50$~K, 
generated within the {\sc hecss} procedure ~\cite{jochym.lazewski.21}.
More details of numerical calculation can be found in Refs.~\cite{blochl.94,pardew.burke.96,perdew.ruzsinszky.08,monkhorst.pack.76} and in 
the Supplemental Material (SM)~\footnote{See the Supplemental Material at 
[URL will be inserted by publisher] for the description of numerical methods 
details, additional numerical results discussing: lattice constants, 
band structure, phonons spectra and Lifshitz transition.}.


\paragraph*{Crystal structure.}---
CeRh$_{2}$As$_{2}$ crystallizes in the CaBe$_{2}$Ge$_{2}$~\cite{mu.pan.18} 
tetragonal structure (with symmetry {\it P4/nmm}, space group no.\ $129$).
The stacking sequence along $c$-axis is Ce-Rh$_{2}$As$_{2}$-Ce-Rh$_{2}$As$_{2}$-Ce,
in which Rh$_{2}$As$_{2}$ layers are arranged in two non-equivalent forms: 
square array of Rh atoms sandwiched between two checkerboard layers of As atoms 
(one below and one above the Rh layer, alternately) and {\it vice versa}
(square As layer decorated by Rh atoms, alternately, from top and bottom).
This corresponds to upper and lower Rh$_{2}$As$_{2}$ layers in Fig.~\ref{fig.crystal}.

The unit cell of the studied system consists of two formula units.
From the {\it ab initio} calculations (with the PBEsol pseudopotentials, see SM~\cite{Note1}), we find the lattice constants as $a = 4.2216$~\AA\ and $c=9.8565$~\AA, which are in good agreement with the experimental results~\cite{khim.landaeta.21}.
In the case of Ce $4f$ electrons treated as valence electrons,
the ground state (GS) of the system is found to be nonmagnetic 
(cf. the SM~\cite{Note1}).

\begin{figure}[t!]
\centering
\includegraphics[width=\linewidth]{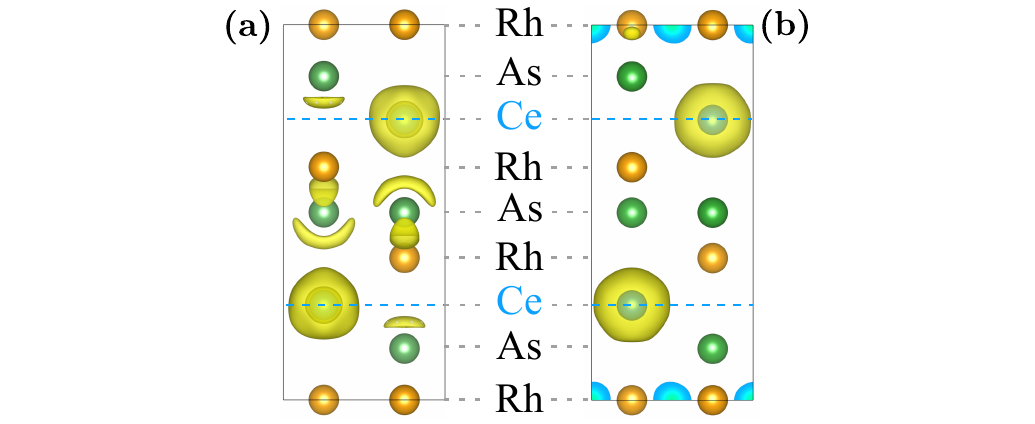}
\caption{
(a)~The electron localization function and (b)~the partial charge density 
(for occupied states $1$~eV below the Fermi level) in CeRh$_2$As$_2$. 
Results were obtained in the presence of SOC and 
with Ce $4f$ electrons treated as valence electrons.
\label{fig.proj}
}
\end{figure}

\paragraph*{Glide symmetry.}---
The nonsymmorphic space group of CeRh$_{2}$As$_{2}$, unusual for heavy fermion systems~\cite{nice.yu.15,nourafkan.tremblay.17},
supports realization of the unconventional SC gap in the electronic structure~\cite{cvetkovic.vafek.13} protected by the space group symmetry~\cite{sumita.yanase.18}.
This space group exhibits the glide symmetry, which is here realized by 
the Rh and As square nets inside Rh$_{2}$As$_{2}$ layers.
Therefore, orbitals of Ce atoms should also exhibit the glide symmetry 
with respect to these planes. Simultaneously, the Ce electron orbitals 
lose the mirror symmetry (along $c$-direction) due to different environments 
from ``top'' and ``bottom'' sides (cf. Fig.~\ref{fig.crystal}).
Indeed, this behavior is clearly manifested directly via the electron localization function~\cite{becke.edgecombe.90,savin.jepsen.92,silvi.savin.94} 
[Fig.~\ref{fig.proj}(a)] or partial charge density [Fig.~\ref{fig.proj}(b)].
As one can see in Fig.~\ref{fig.proj}(a), localization of electrons around Ce atom 
(at the center of the system) does not exhibit symmetry with respect to 
the $ab$ plane (marked by dashed blue line). 
Similar property is observed in the case of partial charge density coming from occupied 
states around the Fermi level 
(we take states $1$~eV below the Fermi level).
In consequence, both Ce atoms of the unit cell are distinguishable due to 
relative position with respect to Rh$_{2}$As$_{2}$ layers and effectively 
realized pseudo-orbitals. This scenario is remarkably similar to URu$_{2}$Si$_{2}$~\cite{chandra.coleman.02}, where the orbitals located at 
U atoms exhibit out-of-plane anisotropy and the ground state is discussed 
as a superposition of different irreducible representations~\cite{sundermann.haverkort.16} resulting in the crystal-field states~\cite{kung.baumbach.15}, and thus we suggest that the orbital order can be called {\it hidden orbital order}.

In our case, 
hidden orbital order can occur due to broken reflection symmetry at Ce atoms.
The onset of hidden order in CeRh$_{2}$As$_{2}$ associated with the symmetry 
breaking at the Ce sites could lower the symmetry from {\it P4/nmm} 
(space group no.\ $129$) to {\it P4mm} (space group no.\ $99$).
The similar effect of symmetry lowering can be achieved by introducing 
antiferromagnetic order on the sublattice of Ce atoms.
This can be crucial in context of the realization of the non-trivial 
topological phase, due to fact that the information about the 
non-trivial phase realization can be attained from the number of the 
bands crossing the Fermi level~\cite{hasan.kane.10}.
Changed degeneracy of the bands by symmetry modifications can lead to 
a different topological phase realized in the system. However, because the 
band structure is relatively dense and a few bands have the maxima located 
around the Fermi level, the results obtained in this way can be ambiguous.


\begin{figure}[!t]
\centering
\includegraphics[width=\linewidth]{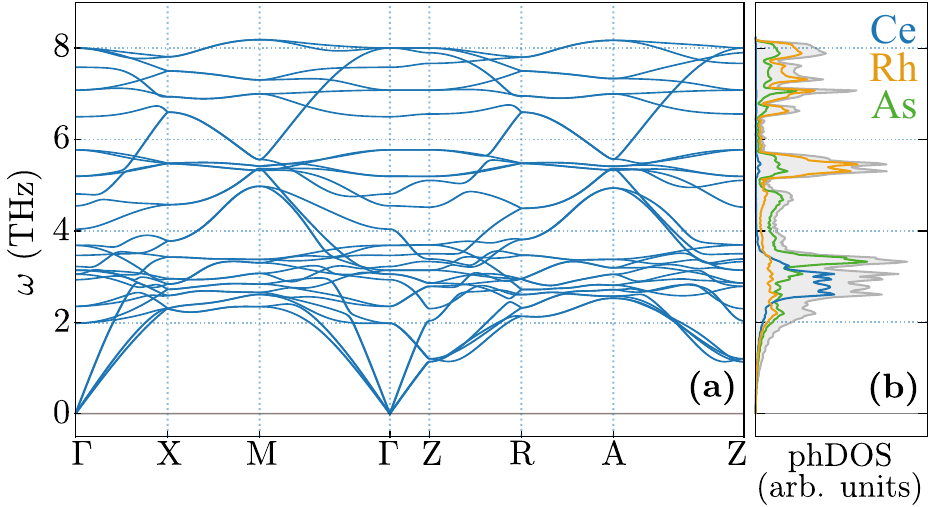}
\caption{
Results of phonon calculations in CeRh$_2$As$_2$:
(a)~phonon dispersion curves along high-symmetry directions;
(b)~total and partial phonon DOSs, shown by gray and color 
(as labeled) solid lines, respectively.
Results were obtained in the presence of SOC 
and with Ce $4f$ electrons treated as valence electrons.
\label{fig.phon}
}
\end{figure}

\paragraph*{Lattice dynamics.}---
The phonon dispersion curves and DOSs are shown in Fig.~\ref{fig.phon}.
As one can see, CeRh$_{2}$As$_{2}$ is stable dynamically, i.e., the soft modes
(imaginary frequencies) are not observed. The irreducible representations at the
the $\Gamma$ point are: $E_{u} + A_{2u}$ for acoustic modes, and
$4E_{u} +4A_{2u} + 5E_{g} + 3A_{1g} + 2B_{1g}$ for optic modes.

Partial DOSs clearly show that the modes associated with Ce atoms are
located mostly at lower frequencies in a range of  $2 \div 3$~THz. This is 
a typical behavior observed in clathrates~\cite{takabatake.suekuni.14}, 
when some heavy atoms are located inside a ``cage''.
In our case, Ce is located inside cage-like dodecahedron constructed
by two Rh$_{2}$As$_{2}$ layers (cf.~Fig.~\ref{fig.crystal}).
Indeed, similar behavior is observed e.g.\ in the case of
KFe$_{2}$As$_{2}$~\cite{ptok.sternik.19} with {\it I4/mmm} symmetry.
There, the modes involving K atom located between Fe$_{2}$Se$_{2}$ layers are 
observed only at lower frequencies [Fig.~\ref{fig.phon}(b)].
Such situation leads to the emergence of nearly flat phonon bands with 
weak contribution to lattice thermal conductivity, due to small group 
velocity and short phonon life-time. Moreover, this physical behavior can 
be examined experimentally by the phonon life-time measurements~\cite{lory.pilhes.17}.
One finds that modes mixing vibrations of Rh and As atoms are of 
particular interest and are visible in the whole range of phonon spectrum.

The effect of hidden orbital order on phonon spectra is noticable by
comparing the results for the {\it P4/nmm} and {\it P4mm} space groups
(without and with orbital order, respectively).
By reducing symmetry, we find small shift of energies and splitting of 
phonon branches resulting from slightly different charge distribution 
on two sites of Ce atoms (Fig.~\ref{fig.oo} in the SM~\cite{Note1}).
This effect could be verified in the future by the inelastic scattering measurements~\cite{delaire.ma.11,ma.delaire.13,li.hong.15}.


\begin{figure}[!t]
\centering
\includegraphics[width=\linewidth]{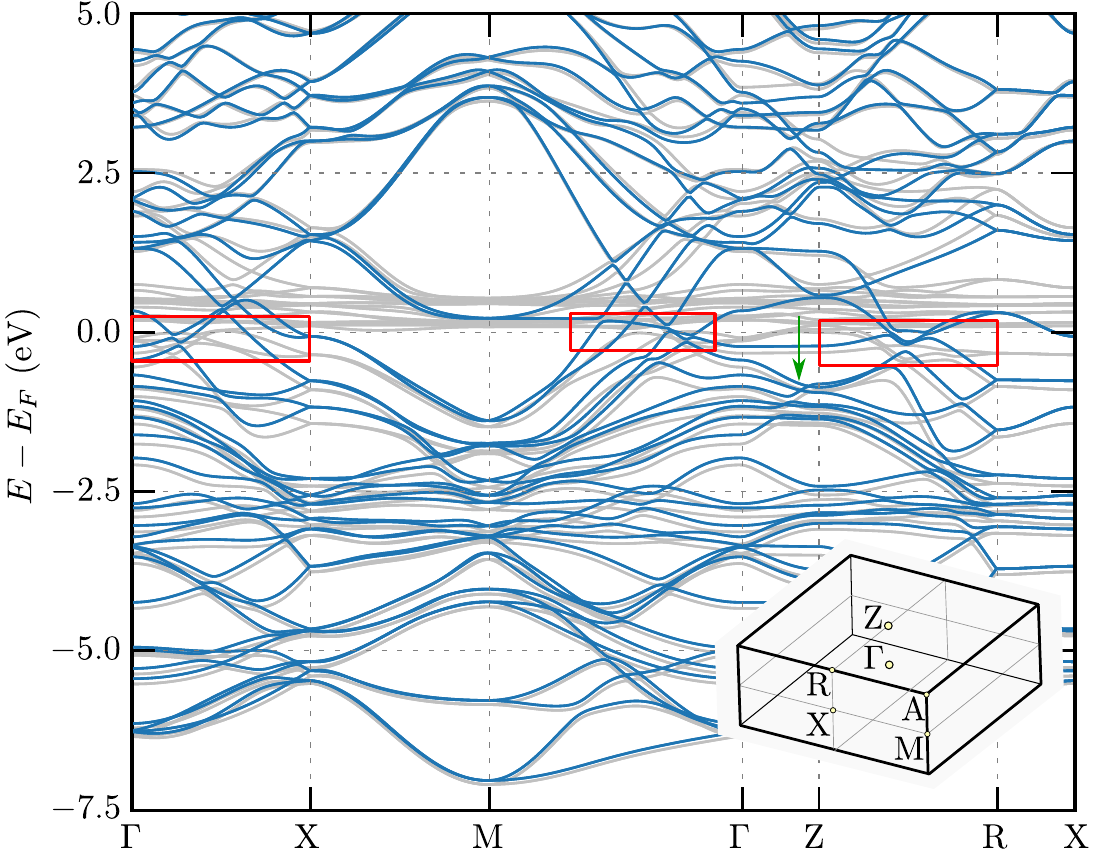}
\caption{
Electronic band structures of CeRh$_{2}$As$_{2}$ in the presence of the SOC.
Solid blue and gray lines correspond to different treatments of the Ce $4f$ 
electrons (as core and as valence electrons, respectively).
\label{fig.bands}
}
\end{figure}

\paragraph*{Electronic band structure.}---
Calculated electronic band structure in the presence of 
SOC is presented in Fig.~\ref{fig.bands}.
The absence of magnetic order in CeRh$_{2}$As$_{2}$ leads to spin 
degeneracy of electronic bands in the {\it P4/nmm} symmetry. For instance, 
at $k_{z}=0$ the degeneracy of the bands is preserved at the X and M points.
Additionally, along the $\Gamma$--Z direction the bands cross below the Fermi 
level (marked by the green arrow in Fig.~\ref{fig.bands}). Similar behavior 
is observed in the nonmagnetic Dirac semimetals (with the same
symmetry), 
like in CeSbTe~\cite{schoop.topp.18}.
However, note that the Dirac point and non-trivial topology can be 
expected even in canonical heavy fermion systems like CeCoIn$_{5}$~\cite{shirer.sun.18} 
with the {\it P4/mmm} symmetry.

CeRh$_{2}$As$_{2}$ can be compared with
SrPt$_{2}$As$_{2}$~\cite{nekrasov.sadovskii.10,kim.kim.15} and
LaPt$_{2}$Si$_{2}$~\cite{kim.kim.15}, both crystallizing in the same
{\it P4/nmm} symmetry. The SOC has weaker impact on the band structure of
CeRh$_{2}$As$_{2}$ than for
SrPt$_{2}$As$_{2}$~\cite{nekrasov.sadovskii.10,kim.kim.15} and is comparable
with the impact on the band structure of LaPt$_{2}$Si$_{2}$~\cite{kim.kim.15}
(cf.~Fig.~\ref{fig.bands_soc} in the SM~\cite{Note1}).
These changes can be a consequence of the modifications of chemical composition,
due to the mass dependence of the SOC~\cite{herman.kuglin.63,shanavas.popovic.14}.
Indeed, the biggest splitting of the bands induced by the SOC is well visible in the unoccupied Ce $4f$ energy levels (around $0.25 \div 0.5$~eV above the Fermi level).

The band structure shows the shift of the Fermi level in CeRh$_{2}$As$_{2}$ to lower
energies w.r.t.\ SrPt$_{2}$As$_{2}$~\cite{nekrasov.sadovskii.10,kim.kim.15} and
LaPt$_{2}$Si$_{2}$~\cite{kim.kim.15}.
In both latter compounds, the~band structures around the Fermi level
are associated with the atoms located in layers, i.e., Pt and As in SrPt$_{2}$As$_{2}$~\cite{nekrasov.sadovskii.10,kim.kim.15}, 
as well as Pt and Si in LaPt$_{2}$Si$_{2}$~\cite{kim.kim.15}.
In our case, the band structure around the Fermi level originates mostly from
Ce $4f$ and Rh $4d$ electrons, what is well visible on the orbital projection 
of bands (see Fig.~\ref{fig.band_proj} in SM~\cite{Note1}).
The Ce $4f$ electrons are located at relatively small range of energies, 
what leads to well visible peak of the density of states 
(see Fig.~\ref{fig.dos} in SM~\cite{Note1}). However, the Rh$_{2}$As$_{2}$ 
layers play important role in the general form of the band structure 
in CeRh$_{2}$As$_{2}$.
This is well visible when we compare the band structure of real CeRh$_{2}$As$_{2}$
with an artificial system without Ce atoms (i.e., containing only two
Rh$_{2}$As$_{2}$ layers), cf. Fig.~\ref{fig.bands_rhas} in the SM~\cite{Note1}.
The main features of the artificial Rh$_{2}$As$_{2}$ system are conserved
and well visible in the band structure of CeRh$_{2}$As$_{2}$. The differences 
arise only from $5d$ electron levels of Ce atoms occupied by one electron.

\begin{figure}[t!]
\includegraphics[width=\linewidth]{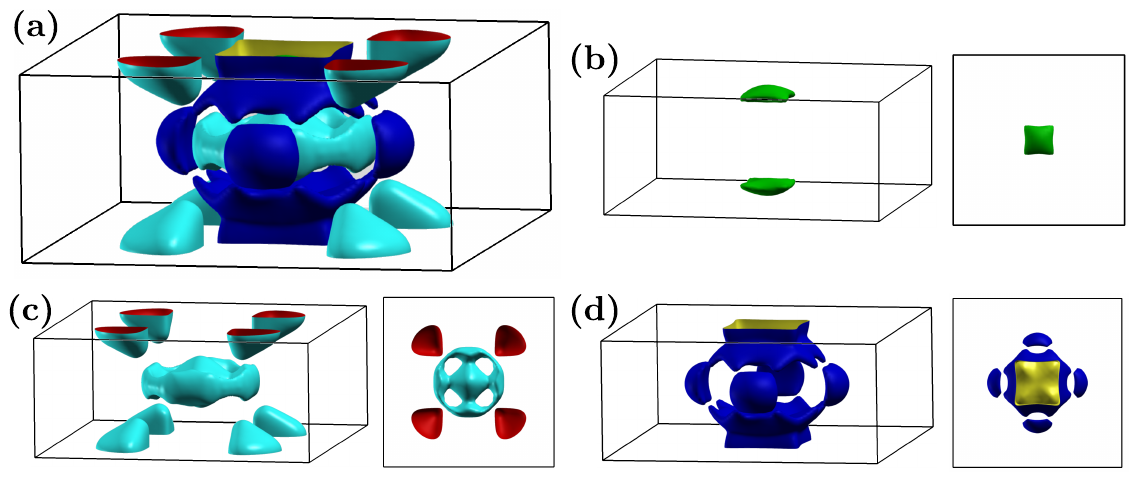}
\caption{
(a) The Fermi surface of CeRh$_{2}$As$_{2}$ and 
(b)--(d) its separate pockets (side and top views).
Results were obtained in the presence of SOC and for Ce $4f$ electrons treated as valence electrons.
\label{fig.fs}
}
\end{figure}

\paragraph*{Fermi surface.}---
The Fermi surface of CeRh$_{2}$As$_{2}$ [Fig.~\ref{fig.fs}(a)] is composed of 
three pockets [see Figs. \mbox{\ref{fig.fs}(b)--\ref{fig.fs}(d)]}, 
which exhibit three dimensional (3D) character.
Contrary to the other heavy fermion systems, e.g. to CeCoIn$_{5}$~\cite{settai.shishido.01,shishido.settai.02,maehira.hotta.03,oppeneer.elgazzar.07,ronning.zhu.12,polyakov.ignatchik.12,ptok.kapcia.17}, there is no pocket with quasi-two dimensional features (i.e., no pockets with weak \mbox{$z$-direction} dependence).
Theoretical calculations of the electronic structure show that the band structure around the Fermi level can be very sensitive to chemical doping [see red boxes in Fig.~\ref{fig.bands}].
This is associated with relatively large number of hybridized bands.
Similarly, the external magnetic field can lead to meaningful modification of the electronic structure, as well as of the character of the Fermi pockets.

\paragraph*{Role of $4f$ electrons.}---
Comparison of the band structures, when Ce $4f$ electrons are treated as core 
or valence electrons in the DFT calculations, are shown in Fig.~\ref{fig.bands} 
(solid blue and gray lines, respectively).
The Ce $4f$ electrons energy levels are located around $0.25$~eV and $0.5$~eV above the Fermi level, due to the relatively strong SOC~\footnote{
We also compared the electronic band structure from {\sc Vasp} with results 
obtained within {\sc Quantum Espresso} software
\cite{giannozzi.baroni.09,giannozzi.andreussi.17} and pseudopotentials 
developed in a frame of {\sc PSlibrary}~\cite{dalcorso.14}.
One needs to notice that the Ce atoms in both cases have different electronic configuration, i.e., [Xe]~$6s^{2}4f^{0.5}5d^{1.5}$ in the case {\sc PSlibrary} and [Xe]~$6s^{2}4f^{1}5d^{1}$ for {\sc Vasp} pseudopotentials.
In consequence, the Ce $4f$ electrons levels can be overestimated (cf. Fig.~\ref{fig.bands} and Fig.~\ref{fig.bands_qe} in SM~\cite{Note1}).}.
he $f$--$d$ orbital hybridization leads to modification of the band structure around the Fermi level.
In practice, all energy levels are slightly shifted to lower energies, while band characters are almost unchanged [cf. grey and blue lines on Fig.~\ref{fig.bands}].

Similarly like in the other heavy fermion systems with {\it I4/mmm} symmetry
(such as, e.g., URu$_{2}$Si$_{2}$~\cite{ikeda.suzuki.14}, CeRh$_{2}$Si$_{2}$~\cite{patil.generalov.16,poelchen.schulz.20}, YbIr$_{2}$Si$_{2}$/YbRh$_{2}$Si$_{2}$~\cite{danzenbacher.kucherenko.07,vyalikh.danzenbacher.10}, or EuRh$_{2}$Si$_{2}$~\cite{hoppner.seiro.13,chikina.hoppner.14,guttler.generalov.19}), the $f$--$d$ orbital hybridization can play important role. Indeed, 
orbital projected band structure show strong impact of the Ce $4f$ and Rh 
$4d$ orbitals onto states around the Fermi level (see Fig.~\ref{fig.band_proj} 
in the SM~\cite{Note1}). This behavior can be important in the context of 
the correct theoretical description of studied system and can be indescribable 
within the simple tight binding model formulation~\cite{schertenleib.fischer.21,khim.landaeta.21,mockli.ramires.21,nogaki.daido.21}.
Nevertheless, the exact form of the electronic band structure around the Fermi 
level should be examined by the future investigations within the high quality
angle-resolved photoemission spectroscopy (ARPES) measurements
\cite{damascelli.hussain.03,dil.09,lv.qian.19}.

The lattice constants found in the numerical calculations depend on the
description of $4f$ electrons of Ce atoms (cf.~Tables~\ref{tab.latt}--\ref{tab.latt2_sol} in the SM~\cite{Note1}). 
Better agreement of numerical results with experimental data is 
obtained when the $4f$ electrons are treated as valence electrons.
This behavior is well known and associated with the formation 
of stronger bonding in systems including $f$ electrons, 
what leads to a decreased volume.

\paragraph*{Magnetic instability.}---
The GS found from the {\it ab initio} calculations is nonmagnetic.
However, Ce atoms have non-zero magnetic moments of magnitude below 
$0.03\;\mu_\text{B}$. This can be a 
symptom of magnetic instability of the system, which could lead to 
magnetic ordering at lower temperatures. Regardless of the GS, 
the numerical calculations with initially enforced different magnetic 
orders uncover structures with lattice constants better matching 
experimental values (cf.~Tab.~\ref{tab.latt_sol} in the SM~\cite{Note1}).
These enforced initial magnetic orders can discriminate between two Ce atomic 
sites and support the realization of previously mentioned hidden orbital order.
However, this case is out of the scope of this work 
and should be studied in the future.

\paragraph*{Lifshitz transition.}---
We emphasize that several bands are located close to the Fermi level.
Here, one should particularly point out the bands with maxima or 
minima around the Fermi level, i.e., bands inside red boxes in 
Fig.~\ref{fig.bands}.
External magnetic field, due to the Zeeman effect and band splitting, 
can lead to emergence of a new, fully polarized Fermi pocket, or to 
disappearance of the existing one. This phenomenon, called {\it magnetic 
Lifshitz transition} (MLT), was described in the context of high-magnetic 
field phase observed in FeSe~\cite{ptok.kapcia.17sr}. During the 
MLT the modification of the Fermi surface topology is observed,
which could be realized here by the external
magnetic field~\cite{ptok.kapcia.17sr,daou.bergemann.06}.

To mimic a tendency of CeRh$_{2}$As$_{2}$ to the realization of MLT,
we performed calculations with non-equal number of spin-up and spin-down 
electrons, i.e., in a spin-polarized state (cf.~Fig.~\ref{fig.bands_mag} 
in the SM~\cite{Note1}). For a comparison, we calculated band structures 
for several values of the spin imbalance.
In such a case, we notice two effects:
(i) the artificially introduced polarization changes the Fermi level, 
and 
(ii) strong modifications of the Fermi pockets even by small change of 
the Fermi level (cf.~Fig.~\ref{fig.mlt} in the SM~\cite{Note1}).
For example, the smallest Fermi pocket [presented on Fig.~\ref{fig.fs}(b)] 
can disappear as a result of a shift of the Fermi level by $0.01$~eV.

Finally, note that the mentioned shift of the Fermi level by $0.01$~eV 
corresponds to relatively large external magnetic field. However, 
one expects that the arising magnetic order may in turn induce sufficient 
internal ``effective'' magnetic field. This resembles the antifferomagnetic-like 
order in other heavy fermion compounds 
(in the presence of the external magnetic field).
As an example one can mention here the $Q$-phase in CeCoIn$_{5}$~\cite{kenzelmann.strassle.08,knezelmann.gerber.10,koutroulakis.steart.10}.
We underline that the magnetic field, at which the MLT occurs, has approximately 
constant value~\cite{ptok.kapcia.17sr,schlottmann.11,daou.bergemann.06}.
Experimentally, the transition observed in CeRh$_{2}$As$_{2}$ occurs
also at approximately constant magnetic field~\cite{khim.landaeta.21}.

The realization of the MLT can change the topological character of  
CeRh$_{2}$As$_{2}$. Similarly to mentioned before case of orbital order, 
the change in degeneracy or the number of bands crossing directly the 
Fermi level, can change the topological character of the system.
Indeed, this behavior can be shown by theoretical calculation based on 
the tight-binding model~\cite{nogaki.daido.21}.
In such situation, the realization of the topological surface edge state 
is expected, what makes this material interesting also in a context of 
topological properties or applications.


\paragraph*{Conclusions.}---
Summarizing, by using the {\it ab initio} techniques we investigated 
the physical properties of the CeRh$_{2}$As$_{2}$ compound. The specific 
crystal structure as well as the previous investigations on heavy fermion 
compounds with the same {\it P4/nmm} symmetry suggest a prominent role of 
the Rh$_{2}$As$_{2}$ layers in the properties of this compound. 
Their impact is visible in the phonon spectra, where Ce modes are 
located only at lower frequencies due to the cage-like structure built 
by Rh$_{2}$As$_{2}$ layers. Additionally, orbital projections of electronic 
band structure suggest the strong hybridization between Ce $4f$ and Rh $4d$ 
electrons. The electronic states located around the Fermi level are mostly 
composed by orbitals of these types, what may be important for construction 
of a realistic tight binding model of this compound.

Due to location of Ce atoms between two nonequivalent Rh$_{2}$As$_{2}$ 
layers the emergence of the hidden orbital ordering is supported. As a result, 
two distinguishable Ce sublattices with different effective ``orbitals'' 
are found.
Similarly as in URu$_{2}$As$_{2}$, such orbitals exhibit out-of-plain 
anisotropy. This orbital order should occur independently of the external 
magnetic field. Moreover, in practice, this orbital order leads to 
lowering of the system symmetry. As a consequence, the measurable 
modifications of the phonon band degeneracy occur.

From the observed small band splitting we conclude that the spin-orbit 
coupling is relatively weak. However, the electronic band structure 
of CeRh$_{2}$As$_{2}$ reveals specific topological properties of this system. 
First, we can observe band crossing (the Dirac points), approximately $1$~eV 
below the Fermi level. Second, the bands near the Fermi level support the 
hypothesis of the magnetic Lifshitz transition. At finite  
magnetic field, the new fully polarized Fermi pocket could emerge, and the 
spin-triplet pairing (topological superconducting phase) would arise.

\begin{acknowledgments}
We warmly thank Dominik Legut for insightful discussions.
Some figures in this work were rendered using {\sc Vesta}~\cite{momma.izumi.11} 
and {\sc XCrySDen}~\cite{kokalj.99} software.
This work was supported by 
National Science Centre (NCN, Poland) under Projects
2016/21/D/ST3/03385 (A.P.),
2017/24/C/ST3/00276 (K.J.K.),
2016/23/B/ST3/00839 (A.M.O.),
and
2017/25/B/ST3/02586 (P.P.).
In addition, A.P. and K.J.K. 
are grateful for the funding from the scholarships of the 
Minister of Science and Higher Education (Poland) for outstanding young
scientists (2019 edition, Nos.~818/STYP/14/2019 and 821/STYP/14/2019, respectively).
\end{acknowledgments}

\bibliography{biblio}

\begin{thebibliography}{81}%
\makeatletter
\providecommand \@ifxundefined [1]{%
 \@ifx{#1\undefined}
}%
\providecommand \@ifnum [1]{%
 \ifnum #1\expandafter \@firstoftwo
 \else \expandafter \@secondoftwo
 \fi
}%
\providecommand \@ifx [1]{%
 \ifx #1\expandafter \@firstoftwo
 \else \expandafter \@secondoftwo
 \fi
}%
\providecommand \natexlab [1]{#1}%
\providecommand \enquote  [1]{``#1''}%
\providecommand \bibnamefont  [1]{#1}%
\providecommand \bibfnamefont [1]{#1}%
\providecommand \citenamefont [1]{#1}%
\providecommand \href@noop [0]{\@secondoftwo}%
\providecommand \href [0]{\begingroup \@sanitize@url \@href}%
\providecommand \@href[1]{\@@startlink{#1}\@@href}%
\providecommand \@@href[1]{\endgroup#1\@@endlink}%
\providecommand \@sanitize@url [0]{\catcode `\\12\catcode `\$12\catcode
  `\&12\catcode `\#12\catcode `\^12\catcode `\_12\catcode `\%12\relax}%
\providecommand \@@startlink[1]{}%
\providecommand \@@endlink[0]{}%
\providecommand \url  [0]{\begingroup\@sanitize@url \@url }%
\providecommand \@url [1]{\endgroup\@href {#1}{\urlprefix }}%
\providecommand \urlprefix  [0]{URL }%
\providecommand \Eprint [0]{\href }%
\providecommand \doibase [0]{http://dx.doi.org/}%
\providecommand \selectlanguage [0]{\@gobble}%
\providecommand \bibinfo  [0]{\@secondoftwo}%
\providecommand \bibfield  [0]{\@secondoftwo}%
\providecommand \translation [1]{[#1]}%
\providecommand \BibitemOpen [0]{}%
\providecommand \bibitemStop [0]{}%
\providecommand \bibitemNoStop [0]{.\EOS\space}%
\providecommand \EOS [0]{\spacefactor3000\relax}%
\providecommand \BibitemShut  [1]{\csname bibitem#1\endcsname}%
\let\auto@bib@innerbib\@empty
\bibitem [{\citenamefont {Khim}\ \emph {et~al.}(2021)\citenamefont {Khim},
  \citenamefont {Landaeta}, \citenamefont {Banda}, \citenamefont {Bannor},
  \citenamefont {Brando}, \citenamefont {Brydon}, \citenamefont {Hafner},
  \citenamefont {K\"{u}chler}, \citenamefont {Cardoso-Gil}, \citenamefont
  {Stockert}, \citenamefont {Mackenzie}, \citenamefont {Agterberg},
  \citenamefont {Geibel},\ and\ \citenamefont {Hassinger}}]{khim.landaeta.21}%
  \BibitemOpen
  \bibfield  {author} {\bibinfo {author} {\bibfnamefont {S.}~\bibnamefont
  {Khim}}, \bibinfo {author} {\bibfnamefont {J.~F.}\ \bibnamefont {Landaeta}},
  \bibinfo {author} {\bibfnamefont {J.}~\bibnamefont {Banda}}, \bibinfo
  {author} {\bibfnamefont {N.}~\bibnamefont {Bannor}}, \bibinfo {author}
  {\bibfnamefont {M.}~\bibnamefont {Brando}}, \bibinfo {author} {\bibfnamefont
  {P.~M.~R.}\ \bibnamefont {Brydon}}, \bibinfo {author} {\bibfnamefont
  {D.}~\bibnamefont {Hafner}}, \bibinfo {author} {\bibfnamefont
  {R.}~\bibnamefont {K\"{u}chler}}, \bibinfo {author} {\bibfnamefont
  {R.}~\bibnamefont {Cardoso-Gil}}, \bibinfo {author} {\bibfnamefont
  {U.}~\bibnamefont {Stockert}}, \bibinfo {author} {\bibfnamefont {A.~P.}\
  \bibnamefont {Mackenzie}}, \bibinfo {author} {\bibfnamefont {D.~F.}\
  \bibnamefont {Agterberg}}, \bibinfo {author} {\bibfnamefont {C.}~\bibnamefont
  {Geibel}}, \ and\ \bibinfo {author} {\bibfnamefont {E.}~\bibnamefont
  {Hassinger}},\ }\href {https://arxiv.org/abs/2101.09522} {\enquote {\bibinfo
  {title} {Field-induced transition from even to odd parity superconductivity
  in {CeRh$_2$As$_2$}},}\ } (\bibinfo {year} {2021}),\ \Eprint
  {http://arxiv.org/abs/arXiv:2101.09522} {arXiv:2101.09522} \BibitemShut
  {NoStop}%
\bibitem [{\citenamefont {Kudo}\ \emph {et~al.}(2010)\citenamefont {Kudo},
  \citenamefont {Nishikubo},\ and\ \citenamefont {Nohara}}]{kudo.nishikubo.10}%
  \BibitemOpen
  \bibfield  {author} {\bibinfo {author} {\bibfnamefont {K.}~\bibnamefont
  {Kudo}}, \bibinfo {author} {\bibfnamefont {Y.}~\bibnamefont {Nishikubo}}, \
  and\ \bibinfo {author} {\bibfnamefont {M.}~\bibnamefont {Nohara}},\
  }\bibfield  {title} {\enquote {\bibinfo {title} {Coexistence of
  superconductivity and charge density wave in {SrPt$_{2}$As$_{2}$}},}\ }\href
  {\doibase 10.1143/JPSJ.79.123710} {\bibfield  {journal} {\bibinfo  {journal}
  {J. Phys. Soc. Jpn.}\ }\textbf {\bibinfo {volume} {79}},\ \bibinfo {pages}
  {123710} (\bibinfo {year} {2010})}\BibitemShut {NoStop}%
\bibitem [{\citenamefont {Nagano}\ \emph {et~al.}(2013)\citenamefont {Nagano},
  \citenamefont {Araoka}, \citenamefont {Mitsuda}, \citenamefont {Yayama},
  \citenamefont {Wada}, \citenamefont {Ichihara}, \citenamefont {Isobe},\ and\
  \citenamefont {Ueda}}]{nagano.araoka.13}%
  \BibitemOpen
  \bibfield  {author} {\bibinfo {author} {\bibfnamefont {Y.}~\bibnamefont
  {Nagano}}, \bibinfo {author} {\bibfnamefont {N.}~\bibnamefont {Araoka}},
  \bibinfo {author} {\bibfnamefont {A.}~\bibnamefont {Mitsuda}}, \bibinfo
  {author} {\bibfnamefont {H.}~\bibnamefont {Yayama}}, \bibinfo {author}
  {\bibfnamefont {H.}~\bibnamefont {Wada}}, \bibinfo {author} {\bibfnamefont
  {M.}~\bibnamefont {Ichihara}}, \bibinfo {author} {\bibfnamefont
  {M.}~\bibnamefont {Isobe}}, \ and\ \bibinfo {author} {\bibfnamefont
  {Y.}~\bibnamefont {Ueda}},\ }\bibfield  {title} {\enquote {\bibinfo {title}
  {Charge density wave and superconductivity of {$R$Pt$_{2}$Si$_{2}$ ($R$ = Y,
  La, Nd, and Lu)}},}\ }\href {\doibase 10.7566/JPSJ.82.064715} {\bibfield
  {journal} {\bibinfo  {journal} {J. Phys. Soc. Jpn.}\ }\textbf {\bibinfo
  {volume} {82}},\ \bibinfo {pages} {064715} (\bibinfo {year}
  {2013})}\BibitemShut {NoStop}%
\bibitem [{\citenamefont {Schertenleib}\ \emph {et~al.}(2021)\citenamefont
  {Schertenleib}, \citenamefont {Fischer},\ and\ \citenamefont
  {Sigrist}}]{schertenleib.fischer.21}%
  \BibitemOpen
  \bibfield  {author} {\bibinfo {author} {\bibfnamefont {E.~G.}\ \bibnamefont
  {Schertenleib}}, \bibinfo {author} {\bibfnamefont {M.~H.}\ \bibnamefont
  {Fischer}}, \ and\ \bibinfo {author} {\bibfnamefont {M.}~\bibnamefont
  {Sigrist}},\ }\href {https://arxiv.org/abs/2101.08821} {\enquote {\bibinfo
  {title} {Unusual {$H$-$T$} phase diagram of {CeRh$_2$As$_2$} -- the role of
  staggered non-centrosymmetriciy},}\ } (\bibinfo {year} {2021}),\ \Eprint
  {http://arxiv.org/abs/arXiv:2101.08821} {arXiv:2101.08821} \BibitemShut
  {NoStop}%
\bibitem [{\citenamefont {Yoshida}\ \emph {et~al.}(2014)\citenamefont
  {Yoshida}, \citenamefont {Sigrist},\ and\ \citenamefont
  {Yanase}}]{yoshida.sigrist.14}%
  \BibitemOpen
  \bibfield  {author} {\bibinfo {author} {\bibfnamefont {T.}~\bibnamefont
  {Yoshida}}, \bibinfo {author} {\bibfnamefont {M.}~\bibnamefont {Sigrist}}, \
  and\ \bibinfo {author} {\bibfnamefont {Y.}~\bibnamefont {Yanase}},\
  }\bibfield  {title} {\enquote {\bibinfo {title} {Parity-mixed
  superconductivity in locally non-centrosymmetric system},}\ }\href {\doibase
  10.7566/JPSJ.83.013703} {\bibfield  {journal} {\bibinfo  {journal} {J. Phys.
  Soc. Jpn.}\ }\textbf {\bibinfo {volume} {83}},\ \bibinfo {pages} {013703}
  (\bibinfo {year} {2014})}\BibitemShut {NoStop}%
\bibitem [{\citenamefont {M\"ockli}\ and\ \citenamefont
  {Khodas}(2018)}]{mockli.khodas.18}%
  \BibitemOpen
  \bibfield  {author} {\bibinfo {author} {\bibfnamefont {D.}~\bibnamefont
  {M\"ockli}}\ and\ \bibinfo {author} {\bibfnamefont {M.}~\bibnamefont
  {Khodas}},\ }\bibfield  {title} {\enquote {\bibinfo {title} {Robust
  parity-mixed superconductivity in disordered monolayer transition metal
  dichalcogenides},}\ }\href {\doibase 10.1103/PhysRevB.98.144518} {\bibfield
  {journal} {\bibinfo  {journal} {Phys. Rev. B}\ }\textbf {\bibinfo {volume}
  {98}},\ \bibinfo {pages} {144518} (\bibinfo {year} {2018})}\BibitemShut
  {NoStop}%
\bibitem [{\citenamefont {M\"ockli}\ and\ \citenamefont
  {Khodas}(2019)}]{mockli.khodas.19}%
  \BibitemOpen
  \bibfield  {author} {\bibinfo {author} {\bibfnamefont {D.}~\bibnamefont
  {M\"ockli}}\ and\ \bibinfo {author} {\bibfnamefont {M.}~\bibnamefont
  {Khodas}},\ }\bibfield  {title} {\enquote {\bibinfo {title} {Magnetic-field
  induced \mbox{$s+if$} pairing in {Ising} superconductors},}\ }\href {\doibase
  10.1103/PhysRevB.99.180505} {\bibfield  {journal} {\bibinfo  {journal} {Phys.
  Rev. B}\ }\textbf {\bibinfo {volume} {99}},\ \bibinfo {pages} {180505(R)}
  (\bibinfo {year} {2019})}\BibitemShut {NoStop}%
\bibitem [{\citenamefont {M\"{o}ckli}\ and\ \citenamefont
  {Ramires}(2021)}]{mockli.ramires.21}%
  \BibitemOpen
  \bibfield  {author} {\bibinfo {author} {\bibfnamefont {D.}~\bibnamefont
  {M\"{o}ckli}}\ and\ \bibinfo {author} {\bibfnamefont {A.}~\bibnamefont
  {Ramires}},\ }\href@noop {} {\enquote {\bibinfo {title} {Two scenarios for
  superconductivity in {CeRh$_2$As$_2$}},}\ } (\bibinfo {year} {2021}),\
  \Eprint {http://arxiv.org/abs/arXiv:2102.09425} {arXiv:2102.09425}
  \BibitemShut {NoStop}%
\bibitem [{\citenamefont {Young}\ and\ \citenamefont
  {Kane}(2015)}]{young.kane.15}%
  \BibitemOpen
  \bibfield  {author} {\bibinfo {author} {\bibfnamefont {S.~M.}\ \bibnamefont
  {Young}}\ and\ \bibinfo {author} {\bibfnamefont {C.~L.}\ \bibnamefont
  {Kane}},\ }\bibfield  {title} {\enquote {\bibinfo {title} {Dirac semimetals
  in two dimensions},}\ }\href {\doibase 10.1103/PhysRevLett.115.126803}
  {\bibfield  {journal} {\bibinfo  {journal} {Phys. Rev. Lett.}\ }\textbf
  {\bibinfo {volume} {115}},\ \bibinfo {pages} {126803} (\bibinfo {year}
  {2015})}\BibitemShut {NoStop}%
\bibitem [{\citenamefont {Schoop}\ \emph {et~al.}(2016)\citenamefont {Schoop},
  \citenamefont {Ali}, \citenamefont {Stra{\ss}er}, \citenamefont {Topp},
  \citenamefont {Varykhalov}, \citenamefont {Marchenko}, \citenamefont
  {Duppel}, \citenamefont {Parkin}, \citenamefont {Lotsch},\ and\ \citenamefont
  {Ast}}]{schoop.ali.16}%
  \BibitemOpen
  \bibfield  {author} {\bibinfo {author} {\bibfnamefont {L.~M.}\ \bibnamefont
  {Schoop}}, \bibinfo {author} {\bibfnamefont {M.~N.}\ \bibnamefont {Ali}},
  \bibinfo {author} {\bibfnamefont {C.}~\bibnamefont {Stra{\ss}er}}, \bibinfo
  {author} {\bibfnamefont {A.}~\bibnamefont {Topp}}, \bibinfo {author}
  {\bibfnamefont {A.}~\bibnamefont {Varykhalov}}, \bibinfo {author}
  {\bibfnamefont {D.}~\bibnamefont {Marchenko}}, \bibinfo {author}
  {\bibfnamefont {V.}~\bibnamefont {Duppel}}, \bibinfo {author} {\bibfnamefont
  {S.~S.~P.}\ \bibnamefont {Parkin}}, \bibinfo {author} {\bibfnamefont {B.~V.}\
  \bibnamefont {Lotsch}}, \ and\ \bibinfo {author} {\bibfnamefont {Ch.~R.}\
  \bibnamefont {Ast}},\ }\bibfield  {title} {\enquote {\bibinfo {title} {Dirac
  cone protected by non-symmorphic symmetry and three-dimensional {Dirac} line
  node in {ZrSiS}},}\ }\href {\doibase 10.1038/ncomms11696} {\bibfield
  {journal} {\bibinfo  {journal} {Nat. Commun.}\ }\textbf {\bibinfo {volume}
  {7}},\ \bibinfo {pages} {11696} (\bibinfo {year} {2016})}\BibitemShut
  {NoStop}%
\bibitem [{\citenamefont {Takane}\ \emph {et~al.}(2016)\citenamefont {Takane},
  \citenamefont {Wang}, \citenamefont {Souma}, \citenamefont {Nakayama},
  \citenamefont {Trang}, \citenamefont {Sato}, \citenamefont {Takahashi},\ and\
  \citenamefont {Ando}}]{takane.wang.16}%
  \BibitemOpen
  \bibfield  {author} {\bibinfo {author} {\bibfnamefont {D.}~\bibnamefont
  {Takane}}, \bibinfo {author} {\bibfnamefont {Zhiwei}\ \bibnamefont {Wang}},
  \bibinfo {author} {\bibfnamefont {S.}~\bibnamefont {Souma}}, \bibinfo
  {author} {\bibfnamefont {K.}~\bibnamefont {Nakayama}}, \bibinfo {author}
  {\bibfnamefont {C.~X.}\ \bibnamefont {Trang}}, \bibinfo {author}
  {\bibfnamefont {T.}~\bibnamefont {Sato}}, \bibinfo {author} {\bibfnamefont
  {T.}~\bibnamefont {Takahashi}}, \ and\ \bibinfo {author} {\bibfnamefont
  {Yoichi}\ \bibnamefont {Ando}},\ }\bibfield  {title} {\enquote {\bibinfo
  {title} {Dirac-node arc in the topological line-node semimetal {HfSiS}},}\
  }\href {\doibase 10.1103/PhysRevB.94.121108} {\bibfield  {journal} {\bibinfo
  {journal} {Phys. Rev. B}\ }\textbf {\bibinfo {volume} {94}},\ \bibinfo
  {pages} {121108(R)} (\bibinfo {year} {2016})}\BibitemShut {NoStop}%
\bibitem [{\citenamefont {Hosen}\ \emph {et~al.}(2017)\citenamefont {Hosen},
  \citenamefont {Dimitri}, \citenamefont {Belopolski}, \citenamefont
  {Maldonado}, \citenamefont {Sankar}, \citenamefont {Dhakal}, \citenamefont
  {Dhakal}, \citenamefont {Cole}, \citenamefont {Oppeneer}, \citenamefont
  {Kaczorowski}, \citenamefont {Chou}, \citenamefont {Hasan}, \citenamefont
  {Durakiewicz},\ and\ \citenamefont {Neupane}}]{hosen.dimitri.17}%
  \BibitemOpen
  \bibfield  {author} {\bibinfo {author} {\bibfnamefont {M.~M.}\ \bibnamefont
  {Hosen}}, \bibinfo {author} {\bibfnamefont {K.}~\bibnamefont {Dimitri}},
  \bibinfo {author} {\bibfnamefont {I.}~\bibnamefont {Belopolski}}, \bibinfo
  {author} {\bibfnamefont {P.}~\bibnamefont {Maldonado}}, \bibinfo {author}
  {\bibfnamefont {R.}~\bibnamefont {Sankar}}, \bibinfo {author} {\bibfnamefont
  {N.}~\bibnamefont {Dhakal}}, \bibinfo {author} {\bibfnamefont
  {G.}~\bibnamefont {Dhakal}}, \bibinfo {author} {\bibfnamefont
  {T.}~\bibnamefont {Cole}}, \bibinfo {author} {\bibfnamefont {P.~M.}\
  \bibnamefont {Oppeneer}}, \bibinfo {author} {\bibfnamefont {D.}~\bibnamefont
  {Kaczorowski}}, \bibinfo {author} {\bibfnamefont {F.}~\bibnamefont {Chou}},
  \bibinfo {author} {\bibfnamefont {M.~Z.}\ \bibnamefont {Hasan}}, \bibinfo
  {author} {\bibfnamefont {T.}~\bibnamefont {Durakiewicz}}, \ and\ \bibinfo
  {author} {\bibfnamefont {M.}~\bibnamefont {Neupane}},\ }\bibfield  {title}
  {\enquote {\bibinfo {title} {Tunability of the topological nodal-line
  semimetal phase in {ZrSi$X$}-type materials ({$X$=S, Se, Te})},}\ }\href
  {\doibase 10.1103/PhysRevB.95.161101} {\bibfield  {journal} {\bibinfo
  {journal} {Phys. Rev. B}\ }\textbf {\bibinfo {volume} {95}},\ \bibinfo
  {pages} {161101(R)} (\bibinfo {year} {2017})}\BibitemShut {NoStop}%
\bibitem [{\citenamefont {Chen}\ \emph {et~al.}(2017)\citenamefont {Chen},
  \citenamefont {Xu}, \citenamefont {Jiang}, \citenamefont {Wu}, \citenamefont
  {Qi}, \citenamefont {Yang}, \citenamefont {Wang}, \citenamefont {Sun},
  \citenamefont {Schr\"oter}, \citenamefont {Yang}, \citenamefont {Schoop},
  \citenamefont {Lv}, \citenamefont {Zhou}, \citenamefont {Chen}, \citenamefont
  {Yao}, \citenamefont {Lu}, \citenamefont {Chen}, \citenamefont {Felser},
  \citenamefont {Yan}, \citenamefont {Liu},\ and\ \citenamefont
  {Chen}}]{chen.xu.17}%
  \BibitemOpen
  \bibfield  {author} {\bibinfo {author} {\bibfnamefont {C.}~\bibnamefont
  {Chen}}, \bibinfo {author} {\bibfnamefont {X.}~\bibnamefont {Xu}}, \bibinfo
  {author} {\bibfnamefont {J.}~\bibnamefont {Jiang}}, \bibinfo {author}
  {\bibfnamefont {S.-C.}\ \bibnamefont {Wu}}, \bibinfo {author} {\bibfnamefont
  {Y.~P.}\ \bibnamefont {Qi}}, \bibinfo {author} {\bibfnamefont {L.~X.}\
  \bibnamefont {Yang}}, \bibinfo {author} {\bibfnamefont {M.~X.}\ \bibnamefont
  {Wang}}, \bibinfo {author} {\bibfnamefont {Y.}~\bibnamefont {Sun}}, \bibinfo
  {author} {\bibfnamefont {N.~B.~M.}\ \bibnamefont {Schr\"oter}}, \bibinfo
  {author} {\bibfnamefont {H.~F.}\ \bibnamefont {Yang}}, \bibinfo {author}
  {\bibfnamefont {L.~M.}\ \bibnamefont {Schoop}}, \bibinfo {author}
  {\bibfnamefont {Y.~Y.}\ \bibnamefont {Lv}}, \bibinfo {author} {\bibfnamefont
  {J.}~\bibnamefont {Zhou}}, \bibinfo {author} {\bibfnamefont {Y.~B.}\
  \bibnamefont {Chen}}, \bibinfo {author} {\bibfnamefont {S.~H.}\ \bibnamefont
  {Yao}}, \bibinfo {author} {\bibfnamefont {M.~H.}\ \bibnamefont {Lu}},
  \bibinfo {author} {\bibfnamefont {Y.~F.}\ \bibnamefont {Chen}}, \bibinfo
  {author} {\bibfnamefont {C.}~\bibnamefont {Felser}}, \bibinfo {author}
  {\bibfnamefont {B.~H.}\ \bibnamefont {Yan}}, \bibinfo {author} {\bibfnamefont
  {Z.~K.}\ \bibnamefont {Liu}}, \ and\ \bibinfo {author} {\bibfnamefont
  {Y.~L.}\ \bibnamefont {Chen}},\ }\bibfield  {title} {\enquote {\bibinfo
  {title} {Dirac line nodes and effect of spin-orbit coupling in the
  nonsymmorphic critical semimetals {$M$SiS} ({$M$=Hf, Zr})},}\ }\href
  {\doibase 10.1103/PhysRevB.95.125126} {\bibfield  {journal} {\bibinfo
  {journal} {Phys. Rev. B}\ }\textbf {\bibinfo {volume} {95}},\ \bibinfo
  {pages} {125126} (\bibinfo {year} {2017})}\BibitemShut {NoStop}%
\bibitem [{\citenamefont {Schoop}\ \emph {et~al.}(2018)\citenamefont {Schoop},
  \citenamefont {Topp}, \citenamefont {Lippmann}, \citenamefont {Orlandi},
  \citenamefont {M{\"u}chler}, \citenamefont {Vergniory}, \citenamefont {Sun},
  \citenamefont {Rost}, \citenamefont {Duppel}, \citenamefont {Krivenkov},
  \citenamefont {Sheoran}, \citenamefont {Manuel}, \citenamefont {Varykhalov},
  \citenamefont {Yan}, \citenamefont {Kremer}, \citenamefont {Ast},\ and\
  \citenamefont {Lotsch}}]{schoop.topp.18}%
  \BibitemOpen
  \bibfield  {author} {\bibinfo {author} {\bibfnamefont {L.~M.}\ \bibnamefont
  {Schoop}}, \bibinfo {author} {\bibfnamefont {A.}~\bibnamefont {Topp}},
  \bibinfo {author} {\bibfnamefont {J.}~\bibnamefont {Lippmann}}, \bibinfo
  {author} {\bibfnamefont {F.}~\bibnamefont {Orlandi}}, \bibinfo {author}
  {\bibfnamefont {L.}~\bibnamefont {M{\"u}chler}}, \bibinfo {author}
  {\bibfnamefont {M.~G.}\ \bibnamefont {Vergniory}}, \bibinfo {author}
  {\bibfnamefont {Y.}~\bibnamefont {Sun}}, \bibinfo {author} {\bibfnamefont
  {A.~W.}\ \bibnamefont {Rost}}, \bibinfo {author} {\bibfnamefont
  {V.}~\bibnamefont {Duppel}}, \bibinfo {author} {\bibfnamefont
  {M.}~\bibnamefont {Krivenkov}}, \bibinfo {author} {\bibfnamefont
  {S.}~\bibnamefont {Sheoran}}, \bibinfo {author} {\bibfnamefont
  {P.}~\bibnamefont {Manuel}}, \bibinfo {author} {\bibfnamefont
  {A.}~\bibnamefont {Varykhalov}}, \bibinfo {author} {\bibfnamefont
  {B.}~\bibnamefont {Yan}}, \bibinfo {author} {\bibfnamefont {R.~K.}\
  \bibnamefont {Kremer}}, \bibinfo {author} {\bibfnamefont {Ch.~R.}\
  \bibnamefont {Ast}}, \ and\ \bibinfo {author} {\bibfnamefont {B.~V.}\
  \bibnamefont {Lotsch}},\ }\bibfield  {title} {\enquote {\bibinfo {title}
  {Tunable {Weyl} and {Dirac} states in the nonsymmorphic compound {CeSbTe}},}\
  }\href {\doibase 10.1126/sciadv.aar2317} {\bibfield  {journal} {\bibinfo
  {journal} {Sci. Adv.}\ }\textbf {\bibinfo {volume} {4}},\ \bibinfo {pages}
  {eaar2317} (\bibinfo {year} {2018})}\BibitemShut {NoStop}%
\bibitem [{\citenamefont {Pezzini}\ \emph {et~al.}(2018)\citenamefont
  {Pezzini}, \citenamefont {van Delft}, \citenamefont {Schoop}, \citenamefont
  {Lotsch}, \citenamefont {Carrington}, \citenamefont {Katsnelson},
  \citenamefont {Hussey},\ and\ \citenamefont
  {Wiedmann}}]{pezzini.vandelft.18}%
  \BibitemOpen
  \bibfield  {author} {\bibinfo {author} {\bibfnamefont {S.}~\bibnamefont
  {Pezzini}}, \bibinfo {author} {\bibfnamefont {M.~R.}\ \bibnamefont {van
  Delft}}, \bibinfo {author} {\bibfnamefont {L.~M.}\ \bibnamefont {Schoop}},
  \bibinfo {author} {\bibfnamefont {B.~V.}\ \bibnamefont {Lotsch}}, \bibinfo
  {author} {\bibfnamefont {A.}~\bibnamefont {Carrington}}, \bibinfo {author}
  {\bibfnamefont {M.~I.}\ \bibnamefont {Katsnelson}}, \bibinfo {author}
  {\bibfnamefont {N.~E.}\ \bibnamefont {Hussey}}, \ and\ \bibinfo {author}
  {\bibfnamefont {S.}~\bibnamefont {Wiedmann}},\ }\bibfield  {title} {\enquote
  {\bibinfo {title} {Unconventional mass enhancement around the {Dirac} nodal
  loop in {ZrSiS}},}\ }\href {\doibase 10.1038/nphys4306} {\bibfield  {journal}
  {\bibinfo  {journal} {Nat. Phys.}\ }\textbf {\bibinfo {volume} {14}},\
  \bibinfo {pages} {178} (\bibinfo {year} {2018})}\BibitemShut {NoStop}%
\bibitem [{\citenamefont {Wang}\ \emph {et~al.}(2018)\citenamefont {Wang},
  \citenamefont {Xu}, \citenamefont {Sun},\ and\ \citenamefont
  {Xia}}]{wang.xu.18}%
  \BibitemOpen
  \bibfield  {author} {\bibinfo {author} {\bibfnamefont {Y.-Y.}\ \bibnamefont
  {Wang}}, \bibinfo {author} {\bibfnamefont {S.}~\bibnamefont {Xu}}, \bibinfo
  {author} {\bibfnamefont {L.-L.}\ \bibnamefont {Sun}}, \ and\ \bibinfo
  {author} {\bibfnamefont {T.-L.}\ \bibnamefont {Xia}},\ }\bibfield  {title}
  {\enquote {\bibinfo {title} {Quantum oscillations and coherent interlayer
  transport in a new topological dirac semimetal candidate {YbMnSb$_{2}$}},}\
  }\href {\doibase 10.1103/PhysRevMaterials.2.021201} {\bibfield  {journal}
  {\bibinfo  {journal} {Phys. Rev. Materials}\ }\textbf {\bibinfo {volume}
  {2}},\ \bibinfo {pages} {021201(R)} (\bibinfo {year} {2018})}\BibitemShut
  {NoStop}%
\bibitem [{\citenamefont {Klemenz}\ \emph {et~al.}(2019)\citenamefont
  {Klemenz}, \citenamefont {Lei},\ and\ \citenamefont
  {Schoop}}]{klemenz.lei.19}%
  \BibitemOpen
  \bibfield  {author} {\bibinfo {author} {\bibfnamefont {S.}~\bibnamefont
  {Klemenz}}, \bibinfo {author} {\bibfnamefont {S.}~\bibnamefont {Lei}}, \ and\
  \bibinfo {author} {\bibfnamefont {L.~M.}\ \bibnamefont {Schoop}},\ }\bibfield
   {title} {\enquote {\bibinfo {title} {Topological semimetals in square-net
  materials},}\ }\href {\doibase 10.1146/annurev-matsci-070218-010114}
  {\bibfield  {journal} {\bibinfo  {journal} {Annu. Rev. Mater. Res.}\ }\textbf
  {\bibinfo {volume} {49}},\ \bibinfo {pages} {185} (\bibinfo {year}
  {2019})}\BibitemShut {NoStop}%
\bibitem [{\citenamefont {Kresse}\ and\ \citenamefont
  {Hafner}(1994)}]{kresse.hafner.94}%
  \BibitemOpen
  \bibfield  {author} {\bibinfo {author} {\bibfnamefont {G.}~\bibnamefont
  {Kresse}}\ and\ \bibinfo {author} {\bibfnamefont {J.}~\bibnamefont
  {Hafner}},\ }\bibfield  {title} {\enquote {\bibinfo {title} {Ab initio
  molecular-dynamics simulation of the liquid-metal--amorphous-semiconductor
  transition in germanium},}\ }\href {\doibase 10.1103/PhysRevB.49.14251}
  {\bibfield  {journal} {\bibinfo  {journal} {Phys. Rev. B}\ }\textbf {\bibinfo
  {volume} {49}},\ \bibinfo {pages} {14251} (\bibinfo {year}
  {1994})}\BibitemShut {NoStop}%
\bibitem [{\citenamefont {Kresse}\ and\ \citenamefont
  {Furthm\"uller}(1996)}]{kresse.furthmuller.96}%
  \BibitemOpen
  \bibfield  {author} {\bibinfo {author} {\bibfnamefont {G.}~\bibnamefont
  {Kresse}}\ and\ \bibinfo {author} {\bibfnamefont {J.}~\bibnamefont
  {Furthm\"uller}},\ }\bibfield  {title} {\enquote {\bibinfo {title} {Efficient
  iterative schemes for ab initio total-energy calculations using a plane-wave
  basis set},}\ }\href {\doibase 10.1103/PhysRevB.54.11169} {\bibfield
  {journal} {\bibinfo  {journal} {Phys. Rev. B}\ }\textbf {\bibinfo {volume}
  {54}},\ \bibinfo {pages} {11169} (\bibinfo {year} {1996})}\BibitemShut
  {NoStop}%
\bibitem [{\citenamefont {Kresse}\ and\ \citenamefont
  {Joubert}(1999)}]{kresse.joubert.99}%
  \BibitemOpen
  \bibfield  {author} {\bibinfo {author} {\bibfnamefont {G.}~\bibnamefont
  {Kresse}}\ and\ \bibinfo {author} {\bibfnamefont {D.}~\bibnamefont
  {Joubert}},\ }\bibfield  {title} {\enquote {\bibinfo {title} {From ultrasoft
  pseudopotentials to the projector augmented-wave method},}\ }\href {\doibase
  10.1103/PhysRevB.59.1758} {\bibfield  {journal} {\bibinfo  {journal} {Phys.
  Rev. B}\ }\textbf {\bibinfo {volume} {59}},\ \bibinfo {pages} {1758}
  (\bibinfo {year} {1999})}\BibitemShut {NoStop}%
\bibitem [{\citenamefont {Tadano}\ \emph {et~al.}(2014)\citenamefont {Tadano},
  \citenamefont {Gohda},\ and\ \citenamefont {Tsuneyuki}}]{tadano.gohda.14}%
  \BibitemOpen
  \bibfield  {author} {\bibinfo {author} {\bibfnamefont {T.}~\bibnamefont
  {Tadano}}, \bibinfo {author} {\bibfnamefont {Y.}~\bibnamefont {Gohda}}, \
  and\ \bibinfo {author} {\bibfnamefont {S.}~\bibnamefont {Tsuneyuki}},\
  }\bibfield  {title} {\enquote {\bibinfo {title} {Anharmonic force constants
  extracted from first-principles molecular dynamics: applications to heat
  transfer simulations},}\ }\href {\doibase 10.1088/0953-8984/26/22/225402}
  {\bibfield  {journal} {\bibinfo  {journal} {J. Phys.: Condens. Matter}\
  }\textbf {\bibinfo {volume} {26}},\ \bibinfo {pages} {225402} (\bibinfo
  {year} {2014})}\BibitemShut {NoStop}%
\bibitem [{\citenamefont {Jochym}\ and\ \citenamefont
  {\L{}a\.{z}ewski}(2021)}]{jochym.lazewski.21}%
  \BibitemOpen
  \bibfield  {author} {\bibinfo {author} {\bibfnamefont {P.~T.}\ \bibnamefont
  {Jochym}}\ and\ \bibinfo {author} {\bibfnamefont {J.}~\bibnamefont
  {\L{}a\.{z}ewski}},\ }\href
  {https://scipost.org/preprints/scipost_202101_00011v1/} {\enquote {\bibinfo
  {title} {{High Efficiency Configuration Space Sampling} -- probing the
  distribution of available states},}\ } (\bibinfo {year} {2021}),\ \bibinfo
  {note} {{SciPost Physics: scipost\_202101\_00011v1 (submission)}}\BibitemShut
  {NoStop}%
\bibitem [{\citenamefont {Bl\"ochl}(1994)}]{blochl.94}%
  \BibitemOpen
  \bibfield  {author} {\bibinfo {author} {\bibfnamefont {P.~E.}\ \bibnamefont
  {Bl\"ochl}},\ }\bibfield  {title} {\enquote {\bibinfo {title} {Projector
  augmented-wave method},}\ }\href {\doibase 10.1103/PhysRevB.50.17953}
  {\bibfield  {journal} {\bibinfo  {journal} {Phys. Rev. B}\ }\textbf {\bibinfo
  {volume} {50}},\ \bibinfo {pages} {17953} (\bibinfo {year}
  {1994})}\BibitemShut {NoStop}%
\bibitem [{\citenamefont {Perdew}\ \emph {et~al.}(1996)\citenamefont {Perdew},
  \citenamefont {Burke},\ and\ \citenamefont {Ernzerhof}}]{pardew.burke.96}%
  \BibitemOpen
  \bibfield  {author} {\bibinfo {author} {\bibfnamefont {J.~P.}\ \bibnamefont
  {Perdew}}, \bibinfo {author} {\bibfnamefont {K.}~\bibnamefont {Burke}}, \
  and\ \bibinfo {author} {\bibfnamefont {M.}~\bibnamefont {Ernzerhof}},\
  }\bibfield  {title} {\enquote {\bibinfo {title} {Generalized gradient
  approximation made simple},}\ }\href {\doibase 10.1103/PhysRevLett.77.3865}
  {\bibfield  {journal} {\bibinfo  {journal} {Phys. Rev. Lett.}\ }\textbf
  {\bibinfo {volume} {77}},\ \bibinfo {pages} {3865} (\bibinfo {year}
  {1996})}\BibitemShut {NoStop}%
\bibitem [{\citenamefont {Perdew}\ \emph {et~al.}(2008)\citenamefont {Perdew},
  \citenamefont {Ruzsinszky}, \citenamefont {Csonka}, \citenamefont {Vydrov},
  \citenamefont {Scuseria}, \citenamefont {Constantin}, \citenamefont {Zhou},\
  and\ \citenamefont {Burke}}]{perdew.ruzsinszky.08}%
  \BibitemOpen
  \bibfield  {author} {\bibinfo {author} {\bibfnamefont {J.~P.}\ \bibnamefont
  {Perdew}}, \bibinfo {author} {\bibfnamefont {A.}~\bibnamefont {Ruzsinszky}},
  \bibinfo {author} {\bibfnamefont {G.~I.}\ \bibnamefont {Csonka}}, \bibinfo
  {author} {\bibfnamefont {O.~A.}\ \bibnamefont {Vydrov}}, \bibinfo {author}
  {\bibfnamefont {G.~E.}\ \bibnamefont {Scuseria}}, \bibinfo {author}
  {\bibfnamefont {L.~A.}\ \bibnamefont {Constantin}}, \bibinfo {author}
  {\bibfnamefont {X.}~\bibnamefont {Zhou}}, \ and\ \bibinfo {author}
  {\bibfnamefont {K.}~\bibnamefont {Burke}},\ }\bibfield  {title} {\enquote
  {\bibinfo {title} {Restoring the density-gradient expansion for exchange in
  solids and surfaces},}\ }\href {\doibase 10.1103/PhysRevLett.100.136406}
  {\bibfield  {journal} {\bibinfo  {journal} {Phys. Rev. Lett.}\ }\textbf
  {\bibinfo {volume} {100}},\ \bibinfo {pages} {136406} (\bibinfo {year}
  {2008})}\BibitemShut {NoStop}%
\bibitem [{\citenamefont {Monkhorst}\ and\ \citenamefont
  {Pack}(1976)}]{monkhorst.pack.76}%
  \BibitemOpen
  \bibfield  {author} {\bibinfo {author} {\bibfnamefont {H.~J.}\ \bibnamefont
  {Monkhorst}}\ and\ \bibinfo {author} {\bibfnamefont {J.~D.}\ \bibnamefont
  {Pack}},\ }\bibfield  {title} {\enquote {\bibinfo {title} {Special points for
  {Brillouin}-zone integrations},}\ }\href {\doibase 10.1103/PhysRevB.13.5188}
  {\bibfield  {journal} {\bibinfo  {journal} {Phys. Rev. B}\ }\textbf {\bibinfo
  {volume} {13}},\ \bibinfo {pages} {5188} (\bibinfo {year}
  {1976})}\BibitemShut {NoStop}%
\bibitem [{Note1()}]{Note1}%
  \BibitemOpen
  \bibinfo {note} {See the Supplemental Material at [URL will be inserted by
  publisher] for the description of numerical methods details, additional
  numerical results discussing: lattice constants, band structure, phonons
  spectra and Lifshitz transition.}\BibitemShut {Stop}%
\bibitem [{\citenamefont {Mu}\ \emph {et~al.}(2018)\citenamefont {Mu},
  \citenamefont {Pan}, \citenamefont {Ruan}, \citenamefont {Liu}, \citenamefont
  {Zhao}, \citenamefont {Shan}, \citenamefont {Chen},\ and\ \citenamefont
  {Ren}}]{mu.pan.18}%
  \BibitemOpen
  \bibfield  {author} {\bibinfo {author} {\bibfnamefont {Q.~G.}\ \bibnamefont
  {Mu}}, \bibinfo {author} {\bibfnamefont {B.~J.}\ \bibnamefont {Pan}},
  \bibinfo {author} {\bibfnamefont {B.~B.}\ \bibnamefont {Ruan}}, \bibinfo
  {author} {\bibfnamefont {T.}~\bibnamefont {Liu}}, \bibinfo {author}
  {\bibfnamefont {K.}~\bibnamefont {Zhao}}, \bibinfo {author} {\bibfnamefont
  {L.}~\bibnamefont {Shan}}, \bibinfo {author} {\bibfnamefont {G.~F.}\
  \bibnamefont {Chen}}, \ and\ \bibinfo {author} {\bibfnamefont {Z.~A.}\
  \bibnamefont {Ren}},\ }\bibfield  {title} {\enquote {\bibinfo {title}
  {Superconductivity in {LaPd$_{2}$Bi$_{2}$} with {CaBe$_{2}$Ge$_{2}$}-type
  structure},}\ }\href {\doibase 10.1007/s11433-018-9285-5} {\bibfield
  {journal} {\bibinfo  {journal} {Sci. China Phys. Mech. Astron.}\ }\textbf
  {\bibinfo {volume} {61}},\ \bibinfo {pages} {127409} (\bibinfo {year}
  {2018})}\BibitemShut {NoStop}%
\bibitem [{\citenamefont {Nica}\ \emph {et~al.}(2015)\citenamefont {Nica},
  \citenamefont {Yu},\ and\ \citenamefont {Si}}]{nice.yu.15}%
  \BibitemOpen
  \bibfield  {author} {\bibinfo {author} {\bibfnamefont {E.~M.}\ \bibnamefont
  {Nica}}, \bibinfo {author} {\bibfnamefont {R.}~\bibnamefont {Yu}}, \ and\
  \bibinfo {author} {\bibfnamefont {Q.}~\bibnamefont {Si}},\ }\bibfield
  {title} {\enquote {\bibinfo {title} {Glide reflection symmetry, {Brillouin}
  zone folding, and superconducting pairing for the {{\it P4/nmm}} space
  group},}\ }\href {\doibase 10.1103/PhysRevB.92.174520} {\bibfield  {journal}
  {\bibinfo  {journal} {Phys. Rev. B}\ }\textbf {\bibinfo {volume} {92}},\
  \bibinfo {pages} {174520} (\bibinfo {year} {2015})}\BibitemShut {NoStop}%
\bibitem [{\citenamefont {Nourafkan}\ and\ \citenamefont
  {Tremblay}(2017)}]{nourafkan.tremblay.17}%
  \BibitemOpen
  \bibfield  {author} {\bibinfo {author} {\bibfnamefont {R.}~\bibnamefont
  {Nourafkan}}\ and\ \bibinfo {author} {\bibfnamefont {A.-M.~S.}\ \bibnamefont
  {Tremblay}},\ }\bibfield  {title} {\enquote {\bibinfo {title} {Effect of
  nonsymmorphic space groups on correlation functions in iron-based
  superconductors},}\ }\href {\doibase 10.1103/PhysRevB.96.125140} {\bibfield
  {journal} {\bibinfo  {journal} {Phys. Rev. B}\ }\textbf {\bibinfo {volume}
  {96}},\ \bibinfo {pages} {125140} (\bibinfo {year} {2017})}\BibitemShut
  {NoStop}%
\bibitem [{\citenamefont {Cvetkovic}\ and\ \citenamefont
  {Vafek}(2013)}]{cvetkovic.vafek.13}%
  \BibitemOpen
  \bibfield  {author} {\bibinfo {author} {\bibfnamefont {V.}~\bibnamefont
  {Cvetkovic}}\ and\ \bibinfo {author} {\bibfnamefont {O.}~\bibnamefont
  {Vafek}},\ }\bibfield  {title} {\enquote {\bibinfo {title} {Space group
  symmetry, spin-orbit coupling, and the low-energy effective {Hamiltonian} for
  iron-based superconductors},}\ }\href {\doibase 10.1103/PhysRevB.88.134510}
  {\bibfield  {journal} {\bibinfo  {journal} {Phys. Rev. B}\ }\textbf {\bibinfo
  {volume} {88}},\ \bibinfo {pages} {134510} (\bibinfo {year}
  {2013})}\BibitemShut {NoStop}%
\bibitem [{\citenamefont {Sumita}\ and\ \citenamefont
  {Yanase}(2018)}]{sumita.yanase.18}%
  \BibitemOpen
  \bibfield  {author} {\bibinfo {author} {\bibfnamefont {S.}~\bibnamefont
  {Sumita}}\ and\ \bibinfo {author} {\bibfnamefont {Y.}~\bibnamefont
  {Yanase}},\ }\bibfield  {title} {\enquote {\bibinfo {title} {Unconventional
  superconducting gap structure protected by space group symmetry},}\ }\href
  {\doibase 10.1103/PhysRevB.97.134512} {\bibfield  {journal} {\bibinfo
  {journal} {Phys. Rev. B}\ }\textbf {\bibinfo {volume} {97}},\ \bibinfo
  {pages} {134512} (\bibinfo {year} {2018})}\BibitemShut {NoStop}%
\bibitem [{\citenamefont {Becke}\ and\ \citenamefont
  {Edgecombe}(1990)}]{becke.edgecombe.90}%
  \BibitemOpen
  \bibfield  {author} {\bibinfo {author} {\bibfnamefont {A.~D.}\ \bibnamefont
  {Becke}}\ and\ \bibinfo {author} {\bibfnamefont {K.~E.}\ \bibnamefont
  {Edgecombe}},\ }\bibfield  {title} {\enquote {\bibinfo {title} {A simple
  measure of electron localization in atomic and molecular systems},}\ }\href
  {\doibase 10.1063/1.458517} {\bibfield  {journal} {\bibinfo  {journal} {J.
  Chem. Phys.}\ }\textbf {\bibinfo {volume} {92}},\ \bibinfo {pages} {5397}
  (\bibinfo {year} {1990})}\BibitemShut {NoStop}%
\bibitem [{\citenamefont {Savin}\ \emph {et~al.}(1992)\citenamefont {Savin},
  \citenamefont {Jepsen}, \citenamefont {Flad}, \citenamefont {Andersen},
  \citenamefont {Preuss},\ and\ \citenamefont {von
  Schnering}}]{savin.jepsen.92}%
  \BibitemOpen
  \bibfield  {author} {\bibinfo {author} {\bibfnamefont {A.}~\bibnamefont
  {Savin}}, \bibinfo {author} {\bibfnamefont {O.}~\bibnamefont {Jepsen}},
  \bibinfo {author} {\bibfnamefont {J.}~\bibnamefont {Flad}}, \bibinfo {author}
  {\bibfnamefont {O.~K.}\ \bibnamefont {Andersen}}, \bibinfo {author}
  {\bibfnamefont {H.}~\bibnamefont {Preuss}}, \ and\ \bibinfo {author}
  {\bibfnamefont {H.~G.}\ \bibnamefont {von Schnering}},\ }\bibfield  {title}
  {\enquote {\bibinfo {title} {Electron localization in solid-state structures
  of the elements: the diamond structure},}\ }\href {\doibase
  10.1002/anie.199201871} {\bibfield  {journal} {\bibinfo  {journal} {Angew.
  Chem., Int. Ed. Engl.}\ }\textbf {\bibinfo {volume} {31}},\ \bibinfo {pages}
  {187} (\bibinfo {year} {1992})}\BibitemShut {NoStop}%
\bibitem [{\citenamefont {Silvi}\ and\ \citenamefont
  {Savin}(1994)}]{silvi.savin.94}%
  \BibitemOpen
  \bibfield  {author} {\bibinfo {author} {\bibfnamefont {B.}~\bibnamefont
  {Silvi}}\ and\ \bibinfo {author} {\bibfnamefont {A.}~\bibnamefont {Savin}},\
  }\bibfield  {title} {\enquote {\bibinfo {title} {Classification of chemical
  bonds based on topological analysis of electron localization functions},}\
  }\href {\doibase 10.1038/371683a0} {\bibfield  {journal} {\bibinfo  {journal}
  {Nature}\ }\textbf {\bibinfo {volume} {371}},\ \bibinfo {pages} {683}
  (\bibinfo {year} {1994})}\BibitemShut {NoStop}%
\bibitem [{\citenamefont {Chandra}\ \emph {et~al.}(2002)\citenamefont
  {Chandra}, \citenamefont {Coleman}, \citenamefont {Mydosh},\ and\
  \citenamefont {Tripathi}}]{chandra.coleman.02}%
  \BibitemOpen
  \bibfield  {author} {\bibinfo {author} {\bibfnamefont {P.}~\bibnamefont
  {Chandra}}, \bibinfo {author} {\bibfnamefont {P.}~\bibnamefont {Coleman}},
  \bibinfo {author} {\bibfnamefont {J.~A.}\ \bibnamefont {Mydosh}}, \ and\
  \bibinfo {author} {\bibfnamefont {V.}~\bibnamefont {Tripathi}},\ }\bibfield
  {title} {\enquote {\bibinfo {title} {Hidden orbital order in the heavy
  fermion metal {URu$_{2}$Si$_{2}$}},}\ }\href {\doibase 10.1038/nature00795}
  {\bibfield  {journal} {\bibinfo  {journal} {Nature}\ }\textbf {\bibinfo
  {volume} {417}},\ \bibinfo {pages} {831} (\bibinfo {year}
  {2002})}\BibitemShut {NoStop}%
\bibitem [{\citenamefont {Sundermann}\ \emph {et~al.}(2016)\citenamefont
  {Sundermann}, \citenamefont {Haverkort}, \citenamefont {Agrestini},
  \citenamefont {Al-Zein}, \citenamefont {Moretti~Sala}, \citenamefont {Huang},
  \citenamefont {Golden}, \citenamefont {de~Visser}, \citenamefont {Thalmeier},
  \citenamefont {Tjeng},\ and\ \citenamefont
  {Severing}}]{sundermann.haverkort.16}%
  \BibitemOpen
  \bibfield  {author} {\bibinfo {author} {\bibfnamefont {M.}~\bibnamefont
  {Sundermann}}, \bibinfo {author} {\bibfnamefont {M.~W.}\ \bibnamefont
  {Haverkort}}, \bibinfo {author} {\bibfnamefont {S.}~\bibnamefont
  {Agrestini}}, \bibinfo {author} {\bibfnamefont {A.}~\bibnamefont {Al-Zein}},
  \bibinfo {author} {\bibfnamefont {M.}~\bibnamefont {Moretti~Sala}}, \bibinfo
  {author} {\bibfnamefont {Y.}~\bibnamefont {Huang}}, \bibinfo {author}
  {\bibfnamefont {M.}~\bibnamefont {Golden}}, \bibinfo {author} {\bibfnamefont
  {A.}~\bibnamefont {de~Visser}}, \bibinfo {author} {\bibfnamefont
  {P.}~\bibnamefont {Thalmeier}}, \bibinfo {author} {\bibfnamefont {L.~H.}\
  \bibnamefont {Tjeng}}, \ and\ \bibinfo {author} {\bibfnamefont
  {A.}~\bibnamefont {Severing}},\ }\bibfield  {title} {\enquote {\bibinfo
  {title} {Direct bulk-sensitive probe of 5f symmetry in
  {URu$_{2}$Si$_{2}$}},}\ }\href {\doibase 10.1073/pnas.1612791113} {\bibfield
  {journal} {\bibinfo  {journal} {Proc. Natl. Acad. Sci. U.S.A.}\ }\textbf
  {\bibinfo {volume} {113}},\ \bibinfo {pages} {13989--13994} (\bibinfo {year}
  {2016})}\BibitemShut {NoStop}%
\bibitem [{\citenamefont {Kung}\ \emph {et~al.}(2015)\citenamefont {Kung},
  \citenamefont {Baumbach}, \citenamefont {Bauer}, \citenamefont
  {Thorsm{\o}lle}, \citenamefont {Zhang}, \citenamefont {Haule}, \citenamefont
  {Mydosh},\ and\ \citenamefont {Blumberg}}]{kung.baumbach.15}%
  \BibitemOpen
  \bibfield  {author} {\bibinfo {author} {\bibfnamefont {H.-H.}\ \bibnamefont
  {Kung}}, \bibinfo {author} {\bibfnamefont {R.~E.}\ \bibnamefont {Baumbach}},
  \bibinfo {author} {\bibfnamefont {E.~D.}\ \bibnamefont {Bauer}}, \bibinfo
  {author} {\bibfnamefont {V.~K.}\ \bibnamefont {Thorsm{\o}lle}}, \bibinfo
  {author} {\bibfnamefont {W.-L.}\ \bibnamefont {Zhang}}, \bibinfo {author}
  {\bibfnamefont {K.}~\bibnamefont {Haule}}, \bibinfo {author} {\bibfnamefont
  {J.~A.}\ \bibnamefont {Mydosh}}, \ and\ \bibinfo {author} {\bibfnamefont
  {G.}~\bibnamefont {Blumberg}},\ }\bibfield  {title} {\enquote {\bibinfo
  {title} {Chirality density wave of the ``hidden order'' phase in
  {URu$_2$Si$_2$}},}\ }\href {\doibase 10.1126/science.1259729} {\bibfield
  {journal} {\bibinfo  {journal} {Science}\ }\textbf {\bibinfo {volume}
  {347}},\ \bibinfo {pages} {1339} (\bibinfo {year} {2015})}\BibitemShut
  {NoStop}%
\bibitem [{\citenamefont {Hasan}\ and\ \citenamefont
  {Kane}(2010)}]{hasan.kane.10}%
  \BibitemOpen
  \bibfield  {author} {\bibinfo {author} {\bibfnamefont {M.~Z.}\ \bibnamefont
  {Hasan}}\ and\ \bibinfo {author} {\bibfnamefont {C.~L.}\ \bibnamefont
  {Kane}},\ }\bibfield  {title} {\enquote {\bibinfo {title} {{\it Colloquium:}
  topological insulators},}\ }\href {\doibase 10.1103/RevModPhys.82.3045}
  {\bibfield  {journal} {\bibinfo  {journal} {Rev. Mod. Phys.}\ }\textbf
  {\bibinfo {volume} {82}},\ \bibinfo {pages} {3045--3067} (\bibinfo {year}
  {2010})}\BibitemShut {NoStop}%
\bibitem [{\citenamefont {Takabatake}\ \emph {et~al.}(2014)\citenamefont
  {Takabatake}, \citenamefont {Suekuni}, \citenamefont {Nakayama},\ and\
  \citenamefont {Kaneshita}}]{takabatake.suekuni.14}%
  \BibitemOpen
  \bibfield  {author} {\bibinfo {author} {\bibfnamefont {T.}~\bibnamefont
  {Takabatake}}, \bibinfo {author} {\bibfnamefont {K.}~\bibnamefont {Suekuni}},
  \bibinfo {author} {\bibfnamefont {T.}~\bibnamefont {Nakayama}}, \ and\
  \bibinfo {author} {\bibfnamefont {E.}~\bibnamefont {Kaneshita}},\ }\bibfield
  {title} {\enquote {\bibinfo {title} {Phonon-glass electron-crystal
  thermoelectric clathrates: Experiments and theory},}\ }\href {\doibase
  10.1103/RevModPhys.86.669} {\bibfield  {journal} {\bibinfo  {journal} {Rev.
  Mod. Phys.}\ }\textbf {\bibinfo {volume} {86}},\ \bibinfo {pages} {669}
  (\bibinfo {year} {2014})}\BibitemShut {NoStop}%
\bibitem [{\citenamefont {Ptok}\ \emph {et~al.}(2019)\citenamefont {Ptok},
  \citenamefont {Sternik}, \citenamefont {Kapcia},\ and\ \citenamefont
  {Piekarz}}]{ptok.sternik.19}%
  \BibitemOpen
  \bibfield  {author} {\bibinfo {author} {\bibfnamefont {A.}~\bibnamefont
  {Ptok}}, \bibinfo {author} {\bibfnamefont {M.}~\bibnamefont {Sternik}},
  \bibinfo {author} {\bibfnamefont {K.~J.}\ \bibnamefont {Kapcia}}, \ and\
  \bibinfo {author} {\bibfnamefont {P.}~\bibnamefont {Piekarz}},\ }\bibfield
  {title} {\enquote {\bibinfo {title} {Structural, electronic, and dynamical
  properties of the tetragonal and collapsed tetragonal phases of
  {KFe$_{2}$As$_{2}$}},}\ }\href {\doibase 10.1103/PhysRevB.99.134103}
  {\bibfield  {journal} {\bibinfo  {journal} {Phys. Rev. B}\ }\textbf {\bibinfo
  {volume} {99}},\ \bibinfo {pages} {134103} (\bibinfo {year}
  {2019})}\BibitemShut {NoStop}%
\bibitem [{\citenamefont {Lory}\ \emph {et~al.}(2017)\citenamefont {Lory},
  \citenamefont {Pailh{\`e}s}, \citenamefont {Giordano}, \citenamefont
  {Euchner}, \citenamefont {Nguyen}, \citenamefont {Ramlau}, \citenamefont
  {Borrmann}, \citenamefont {Schmidt}, \citenamefont {Baitinger}, \citenamefont
  {Ikeda}, \citenamefont {Tome{\v{s}}}, \citenamefont {Mihalkovi{\v{c}}},
  \citenamefont {Allio}, \citenamefont {Johnson}, \citenamefont {Schober},
  \citenamefont {Sidis}, \citenamefont {Bourdarot}, \citenamefont {Regnault},
  \citenamefont {Ollivier}, \citenamefont {Paschen}, \citenamefont {Grin},\
  and\ \citenamefont {de~Boissieu}}]{lory.pilhes.17}%
  \BibitemOpen
  \bibfield  {author} {\bibinfo {author} {\bibfnamefont {P.-F.}\ \bibnamefont
  {Lory}}, \bibinfo {author} {\bibfnamefont {S.}~\bibnamefont {Pailh{\`e}s}},
  \bibinfo {author} {\bibfnamefont {V.~M.}\ \bibnamefont {Giordano}}, \bibinfo
  {author} {\bibfnamefont {H.}~\bibnamefont {Euchner}}, \bibinfo {author}
  {\bibfnamefont {H.~D.}\ \bibnamefont {Nguyen}}, \bibinfo {author}
  {\bibfnamefont {R.}~\bibnamefont {Ramlau}}, \bibinfo {author} {\bibfnamefont
  {H.}~\bibnamefont {Borrmann}}, \bibinfo {author} {\bibfnamefont
  {M.}~\bibnamefont {Schmidt}}, \bibinfo {author} {\bibfnamefont
  {M.}~\bibnamefont {Baitinger}}, \bibinfo {author} {\bibfnamefont
  {M.}~\bibnamefont {Ikeda}}, \bibinfo {author} {\bibfnamefont
  {P.}~\bibnamefont {Tome{\v{s}}}}, \bibinfo {author} {\bibfnamefont
  {M.}~\bibnamefont {Mihalkovi{\v{c}}}}, \bibinfo {author} {\bibfnamefont
  {C.}~\bibnamefont {Allio}}, \bibinfo {author} {\bibfnamefont {M.~R.}\
  \bibnamefont {Johnson}}, \bibinfo {author} {\bibfnamefont {H.}~\bibnamefont
  {Schober}}, \bibinfo {author} {\bibfnamefont {Y.}~\bibnamefont {Sidis}},
  \bibinfo {author} {\bibfnamefont {F.}~\bibnamefont {Bourdarot}}, \bibinfo
  {author} {\bibfnamefont {L.~P.}\ \bibnamefont {Regnault}}, \bibinfo {author}
  {\bibfnamefont {J.}~\bibnamefont {Ollivier}}, \bibinfo {author}
  {\bibfnamefont {S.}~\bibnamefont {Paschen}}, \bibinfo {author} {\bibfnamefont
  {Y.}~\bibnamefont {Grin}}, \ and\ \bibinfo {author} {\bibfnamefont
  {M.}~\bibnamefont {de~Boissieu}},\ }\bibfield  {title} {\enquote {\bibinfo
  {title} {Direct measurement of individual phonon lifetimes in the clathrate
  compound {Ba$_{7.81}$Ge$_{40.67}$Au$_{5.33}$}},}\ }\href {\doibase
  10.1038/s41467-017-00584-7} {\bibfield  {journal} {\bibinfo  {journal} {Nat.
  Commun.}\ }\textbf {\bibinfo {volume} {8}},\ \bibinfo {pages} {491} (\bibinfo
  {year} {2017})}\BibitemShut {NoStop}%
\bibitem [{\citenamefont {Delaire}\ \emph {et~al.}(2011)\citenamefont
  {Delaire}, \citenamefont {Ma}, \citenamefont {Marty}, \citenamefont {May},
  \citenamefont {McGuire}, \citenamefont {Du}, \citenamefont {Singh},
  \citenamefont {Podlesnyak}, \citenamefont {Ehlers}, \citenamefont {Lumsden},\
  and\ \citenamefont {Sales}}]{delaire.ma.11}%
  \BibitemOpen
  \bibfield  {author} {\bibinfo {author} {\bibfnamefont {O.}~\bibnamefont
  {Delaire}}, \bibinfo {author} {\bibfnamefont {J.}~\bibnamefont {Ma}},
  \bibinfo {author} {\bibfnamefont {K.}~\bibnamefont {Marty}}, \bibinfo
  {author} {\bibfnamefont {A.~F.}\ \bibnamefont {May}}, \bibinfo {author}
  {\bibfnamefont {M.~A.}\ \bibnamefont {McGuire}}, \bibinfo {author}
  {\bibfnamefont {M.-H.}\ \bibnamefont {Du}}, \bibinfo {author} {\bibfnamefont
  {D.~J.}\ \bibnamefont {Singh}}, \bibinfo {author} {\bibfnamefont
  {A.}~\bibnamefont {Podlesnyak}}, \bibinfo {author} {\bibfnamefont
  {G.}~\bibnamefont {Ehlers}}, \bibinfo {author} {\bibfnamefont {M.~D.}\
  \bibnamefont {Lumsden}}, \ and\ \bibinfo {author} {\bibfnamefont {B.~C.}\
  \bibnamefont {Sales}},\ }\bibfield  {title} {\enquote {\bibinfo {title}
  {Giant anharmonic phonon scattering in {PbTe}},}\ }\href {\doibase
  10.1038/nmat3035} {\bibfield  {journal} {\bibinfo  {journal} {Nat. Mater.}\
  }\textbf {\bibinfo {volume} {10}},\ \bibinfo {pages} {614} (\bibinfo {year}
  {2011})}\BibitemShut {NoStop}%
\bibitem [{\citenamefont {Ma}\ \emph {et~al.}(2013)\citenamefont {Ma},
  \citenamefont {Delaire}, \citenamefont {May}, \citenamefont {Carlton},
  \citenamefont {McGuire}, \citenamefont {VanBebber}, \citenamefont
  {Abernathy}, \citenamefont {Ehlers}, \citenamefont {Hong}, \citenamefont
  {Huq}, \citenamefont {Tian}, \citenamefont {Keppens}, \citenamefont
  {Shao-Horn},\ and\ \citenamefont {Sales}}]{ma.delaire.13}%
  \BibitemOpen
  \bibfield  {author} {\bibinfo {author} {\bibfnamefont {J.}~\bibnamefont
  {Ma}}, \bibinfo {author} {\bibfnamefont {O.}~\bibnamefont {Delaire}},
  \bibinfo {author} {\bibfnamefont {A.~F.}\ \bibnamefont {May}}, \bibinfo
  {author} {\bibfnamefont {C.~E.}\ \bibnamefont {Carlton}}, \bibinfo {author}
  {\bibfnamefont {M.~A.}\ \bibnamefont {McGuire}}, \bibinfo {author}
  {\bibfnamefont {L.~H.}\ \bibnamefont {VanBebber}}, \bibinfo {author}
  {\bibfnamefont {D.~L.}\ \bibnamefont {Abernathy}}, \bibinfo {author}
  {\bibfnamefont {G.}~\bibnamefont {Ehlers}}, \bibinfo {author} {\bibfnamefont
  {Tao}\ \bibnamefont {Hong}}, \bibinfo {author} {\bibfnamefont
  {A.}~\bibnamefont {Huq}}, \bibinfo {author} {\bibfnamefont {Wei}\
  \bibnamefont {Tian}}, \bibinfo {author} {\bibfnamefont {V.~M.}\ \bibnamefont
  {Keppens}}, \bibinfo {author} {\bibfnamefont {Y.}~\bibnamefont {Shao-Horn}},
  \ and\ \bibinfo {author} {\bibfnamefont {B.~C.}\ \bibnamefont {Sales}},\
  }\bibfield  {title} {\enquote {\bibinfo {title} {Glass-like phonon scattering
  from a spontaneous nanostructure in {AgSbTe$_{2}$}},}\ }\href {\doibase
  10.1038/nnano.2013.95} {\bibfield  {journal} {\bibinfo  {journal} {Nat.
  Nanotech.}\ }\textbf {\bibinfo {volume} {8}},\ \bibinfo {pages} {445}
  (\bibinfo {year} {2013})}\BibitemShut {NoStop}%
\bibitem [{\citenamefont {Li}\ \emph {et~al.}(2015)\citenamefont {Li},
  \citenamefont {Hong}, \citenamefont {May}, \citenamefont {Bansal},
  \citenamefont {Chi}, \citenamefont {Hong}, \citenamefont {Ehlers},\ and\
  \citenamefont {Delaire}}]{li.hong.15}%
  \BibitemOpen
  \bibfield  {author} {\bibinfo {author} {\bibfnamefont {C.~W.}\ \bibnamefont
  {Li}}, \bibinfo {author} {\bibfnamefont {J.}~\bibnamefont {Hong}}, \bibinfo
  {author} {\bibfnamefont {A.~F.}\ \bibnamefont {May}}, \bibinfo {author}
  {\bibfnamefont {D.}~\bibnamefont {Bansal}}, \bibinfo {author} {\bibfnamefont
  {S.}~\bibnamefont {Chi}}, \bibinfo {author} {\bibfnamefont {T.}~\bibnamefont
  {Hong}}, \bibinfo {author} {\bibfnamefont {G.}~\bibnamefont {Ehlers}}, \ and\
  \bibinfo {author} {\bibfnamefont {O.}~\bibnamefont {Delaire}},\ }\bibfield
  {title} {\enquote {\bibinfo {title} {Orbitally driven giant phonon
  anharmonicity in {SnSe}},}\ }\href {\doibase 10.1038/nphys3492} {\bibfield
  {journal} {\bibinfo  {journal} {Nat. Phys.}\ }\textbf {\bibinfo {volume}
  {11}},\ \bibinfo {pages} {1063} (\bibinfo {year} {2015})}\BibitemShut
  {NoStop}%
\bibitem [{\citenamefont {Shirer}\ \emph {et~al.}(2018)\citenamefont {Shirer},
  \citenamefont {Sun}, \citenamefont {Bachmann}, \citenamefont {Putzke},
  \citenamefont {Helm}, \citenamefont {Winter}, \citenamefont {Balakirev},
  \citenamefont {McDonald}, \citenamefont {Analytis}, \citenamefont {Nair},
  \citenamefont {Bauer}, \citenamefont {Ronning}, \citenamefont {Felser},
  \citenamefont {Meng}, \citenamefont {Yan},\ and\ \citenamefont
  {Moll}}]{shirer.sun.18}%
  \BibitemOpen
  \bibfield  {author} {\bibinfo {author} {\bibfnamefont {K.~R.}\ \bibnamefont
  {Shirer}}, \bibinfo {author} {\bibfnamefont {Y.}~\bibnamefont {Sun}},
  \bibinfo {author} {\bibfnamefont {M.~D.}\ \bibnamefont {Bachmann}}, \bibinfo
  {author} {\bibfnamefont {C.}~\bibnamefont {Putzke}}, \bibinfo {author}
  {\bibfnamefont {T.}~\bibnamefont {Helm}}, \bibinfo {author} {\bibfnamefont
  {L.~E.}\ \bibnamefont {Winter}}, \bibinfo {author} {\bibfnamefont {F.~F.}\
  \bibnamefont {Balakirev}}, \bibinfo {author} {\bibfnamefont {R.~D.}\
  \bibnamefont {McDonald}}, \bibinfo {author} {\bibfnamefont {J.~G.}\
  \bibnamefont {Analytis}}, \bibinfo {author} {\bibfnamefont {N.~L.}\
  \bibnamefont {Nair}}, \bibinfo {author} {\bibfnamefont {E.~D.}\ \bibnamefont
  {Bauer}}, \bibinfo {author} {\bibfnamefont {F.}~\bibnamefont {Ronning}},
  \bibinfo {author} {\bibfnamefont {C.}~\bibnamefont {Felser}}, \bibinfo
  {author} {\bibfnamefont {T.}~\bibnamefont {Meng}}, \bibinfo {author}
  {\bibfnamefont {B.}~\bibnamefont {Yan}}, \ and\ \bibinfo {author}
  {\bibfnamefont {P.~J.~W.}\ \bibnamefont {Moll}},\ }\href
  {https://arxiv.org/abs/1808.00403} {\enquote {\bibinfo {title} {Dirac
  fermions in the heavy-fermion superconductors {Ce(Co,Rh,Ir)In$_5$}},}\ }
  (\bibinfo {year} {2018}),\ \Eprint {http://arxiv.org/abs/arXiv:1808.00403}
  {arXiv:1808.00403} \BibitemShut {NoStop}%
\bibitem [{\citenamefont {Nekrasov}\ and\ \citenamefont
  {Sadovskii}(2010)}]{nekrasov.sadovskii.10}%
  \BibitemOpen
  \bibfield  {author} {\bibinfo {author} {\bibfnamefont {I.~A.}\ \bibnamefont
  {Nekrasov}}\ and\ \bibinfo {author} {\bibfnamefont {M.~V.}\ \bibnamefont
  {Sadovskii}},\ }\bibfield  {title} {\enquote {\bibinfo {title} {Electronic
  structure of novel multiple-band superconductor {SrPt$_{2}$As$_{2}$}},}\
  }\href {\doibase 10.1134/S0021364010230074} {\bibfield  {journal} {\bibinfo
  {journal} {JETP Letters}\ }\textbf {\bibinfo {volume} {92}},\ \bibinfo
  {pages} {751} (\bibinfo {year} {2010})}\BibitemShut {NoStop}%
\bibitem [{\citenamefont {Kim}\ \emph {et~al.}(2015)\citenamefont {Kim},
  \citenamefont {Kim},\ and\ \citenamefont {Min}}]{kim.kim.15}%
  \BibitemOpen
  \bibfield  {author} {\bibinfo {author} {\bibfnamefont {S.}~\bibnamefont
  {Kim}}, \bibinfo {author} {\bibfnamefont {K.}~\bibnamefont {Kim}}, \ and\
  \bibinfo {author} {\bibfnamefont {B.~I.}\ \bibnamefont {Min}},\ }\bibfield
  {title} {\enquote {\bibinfo {title} {The mechanism of charge density wave in
  {Pt-based} layered superconductors: {SrPt$_{2}$As$_{2}$} and
  {LaPt$_{2}$Si$_{2}$}},}\ }\href {\doibase 10.1038/srep15052} {\bibfield
  {journal} {\bibinfo  {journal} {Sci. Rep.}\ }\textbf {\bibinfo {volume}
  {5}},\ \bibinfo {pages} {15052} (\bibinfo {year} {2015})}\BibitemShut
  {NoStop}%
\bibitem [{\citenamefont {Herman}\ \emph {et~al.}(1963)\citenamefont {Herman},
  \citenamefont {Kuglin}, \citenamefont {Cuff},\ and\ \citenamefont
  {Kortum}}]{herman.kuglin.63}%
  \BibitemOpen
  \bibfield  {author} {\bibinfo {author} {\bibfnamefont {F.}~\bibnamefont
  {Herman}}, \bibinfo {author} {\bibfnamefont {Ch.~D.}\ \bibnamefont {Kuglin}},
  \bibinfo {author} {\bibfnamefont {K.~F.}\ \bibnamefont {Cuff}}, \ and\
  \bibinfo {author} {\bibfnamefont {R.~L.}\ \bibnamefont {Kortum}},\ }\bibfield
   {title} {\enquote {\bibinfo {title} {Relativistic corrections to the band
  structure of tetrahedrally bonded semiconductors},}\ }\href {\doibase
  10.1103/PhysRevLett.11.541} {\bibfield  {journal} {\bibinfo  {journal} {Phys.
  Rev. Lett.}\ }\textbf {\bibinfo {volume} {11}},\ \bibinfo {pages} {541}
  (\bibinfo {year} {1963})}\BibitemShut {NoStop}%
\bibitem [{\citenamefont {Shanavas}\ \emph {et~al.}(2014)\citenamefont
  {Shanavas}, \citenamefont {Popovi'{c}},\ and\ \citenamefont
  {Satpathy}}]{shanavas.popovic.14}%
  \BibitemOpen
  \bibfield  {author} {\bibinfo {author} {\bibfnamefont {K.~V.}\ \bibnamefont
  {Shanavas}}, \bibinfo {author} {\bibfnamefont {Z.~S.}\ \bibnamefont
  {Popovi'{c}}}, \ and\ \bibinfo {author} {\bibfnamefont {S.}~\bibnamefont
  {Satpathy}},\ }\bibfield  {title} {\enquote {\bibinfo {title} {Theoretical
  model for {Rashba} spin-orbit interaction in $d$ electrons},}\ }\href
  {\doibase 10.1103/PhysRevB.90.165108} {\bibfield  {journal} {\bibinfo
  {journal} {Phys. Rev. B}\ }\textbf {\bibinfo {volume} {90}},\ \bibinfo
  {pages} {165108} (\bibinfo {year} {2014})}\BibitemShut {NoStop}%
\bibitem [{\citenamefont {Settai}\ \emph {et~al.}(2001)\citenamefont {Settai},
  \citenamefont {Shishido}, \citenamefont {Ikeda}, \citenamefont {Murakawa},
  \citenamefont {Nakashima}, \citenamefont {Aoki}, \citenamefont {Haga},
  \citenamefont {Harima},\ and\ \citenamefont {Onuki}}]{settai.shishido.01}%
  \BibitemOpen
  \bibfield  {author} {\bibinfo {author} {\bibfnamefont {R.}~\bibnamefont
  {Settai}}, \bibinfo {author} {\bibfnamefont {H.}~\bibnamefont {Shishido}},
  \bibinfo {author} {\bibfnamefont {S.}~\bibnamefont {Ikeda}}, \bibinfo
  {author} {\bibfnamefont {Y.}~\bibnamefont {Murakawa}}, \bibinfo {author}
  {\bibfnamefont {M.}~\bibnamefont {Nakashima}}, \bibinfo {author}
  {\bibfnamefont {D.}~\bibnamefont {Aoki}}, \bibinfo {author} {\bibfnamefont
  {Y.}~\bibnamefont {Haga}}, \bibinfo {author} {\bibfnamefont {H.}~\bibnamefont
  {Harima}}, \ and\ \bibinfo {author} {\bibfnamefont {Y.}~\bibnamefont
  {Onuki}},\ }\bibfield  {title} {\enquote {\bibinfo {title}
  {Quasi-two-dimensional fermi surfaces and the de haas-van alphen oscillation
  in both the normal and superconducting mixed states of {CeCoIn$_{5}$}},}\
  }\href {\doibase 10.1088/0953-8984/13/27/103} {\bibfield  {journal} {\bibinfo
   {journal} {J. Phys.: Condens. Matter}\ }\textbf {\bibinfo {volume} {13}},\
  \bibinfo {pages} {L627} (\bibinfo {year} {2001})}\BibitemShut {NoStop}%
\bibitem [{\citenamefont {Shishido}\ \emph {et~al.}(2002)\citenamefont
  {Shishido}, \citenamefont {Settai}, \citenamefont {Aoki}, \citenamefont
  {Ikeda}, \citenamefont {Nakawaki}, \citenamefont {Nakamura}, \citenamefont
  {Iizuka}, \citenamefont {Inada}, \citenamefont {Sugiyama}, \citenamefont
  {Takeuchi}, \citenamefont {Kindo}, \citenamefont {Kobayashi}, \citenamefont
  {Haga}, \citenamefont {Harima}, \citenamefont {Aoki}, \citenamefont {Sato},\
  and\ \citenamefont {Onuki}}]{shishido.settai.02}%
  \BibitemOpen
  \bibfield  {author} {\bibinfo {author} {\bibfnamefont {H.}~\bibnamefont
  {Shishido}}, \bibinfo {author} {\bibfnamefont {R.}~\bibnamefont {Settai}},
  \bibinfo {author} {\bibfnamefont {D.}~\bibnamefont {Aoki}}, \bibinfo {author}
  {\bibfnamefont {Shugo}\ \bibnamefont {Ikeda}}, \bibinfo {author}
  {\bibfnamefont {H.}~\bibnamefont {Nakawaki}}, \bibinfo {author}
  {\bibfnamefont {N.}~\bibnamefont {Nakamura}}, \bibinfo {author}
  {\bibfnamefont {T.}~\bibnamefont {Iizuka}}, \bibinfo {author} {\bibfnamefont
  {Y.}~\bibnamefont {Inada}}, \bibinfo {author} {\bibfnamefont
  {K.}~\bibnamefont {Sugiyama}}, \bibinfo {author} {\bibfnamefont
  {T.}~\bibnamefont {Takeuchi}}, \bibinfo {author} {\bibfnamefont
  {K.}~\bibnamefont {Kindo}}, \bibinfo {author} {\bibfnamefont {T.~C.}\
  \bibnamefont {Kobayashi}}, \bibinfo {author} {\bibfnamefont {Y.}~\bibnamefont
  {Haga}}, \bibinfo {author} {\bibfnamefont {H.}~\bibnamefont {Harima}},
  \bibinfo {author} {\bibfnamefont {T.}~\bibnamefont {Aoki}, \bibfnamefont
  {Y.and~Namiki}}, \bibinfo {author} {\bibfnamefont {H.}~\bibnamefont {Sato}},
  \ and\ \bibinfo {author} {\bibfnamefont {Y.}~\bibnamefont {Onuki}},\
  }\bibfield  {title} {\enquote {\bibinfo {title} {Fermi surface, magnetic and
  superconducting properties of {LaRhIn$_{5}$} and {Ce$T$In$_{5}$} ({$T$: Co,
  Rh and Ir})},}\ }\href {\doibase 10.1143/JPSJ.71.162} {\bibfield  {journal}
  {\bibinfo  {journal} {J. Phys. Soc. Jpn.}\ }\textbf {\bibinfo {volume}
  {71}},\ \bibinfo {pages} {162} (\bibinfo {year} {2002})}\BibitemShut
  {NoStop}%
\bibitem [{\citenamefont {Maehira}\ \emph {et~al.}(2003)\citenamefont
  {Maehira}, \citenamefont {Hotta}, \citenamefont {Ueda},\ and\ \citenamefont
  {Hasegawa}}]{maehira.hotta.03}%
  \BibitemOpen
  \bibfield  {author} {\bibinfo {author} {\bibfnamefont {T.}~\bibnamefont
  {Maehira}}, \bibinfo {author} {\bibfnamefont {T.}~\bibnamefont {Hotta}},
  \bibinfo {author} {\bibfnamefont {K.}~\bibnamefont {Ueda}}, \ and\ \bibinfo
  {author} {\bibfnamefont {A.}~\bibnamefont {Hasegawa}},\ }\bibfield  {title}
  {\enquote {\bibinfo {title} {Relativistic band-structure calculations for
  {Ce$T$In$_{5}$} ({$T$ = Ir and Co}) and analysis of the energy bands by using
  tight-binding method},}\ }\href {\doibase 10.1143/JPSJ.72.854} {\bibfield
  {journal} {\bibinfo  {journal} {J. Phys. Soc. Jpn.}\ }\textbf {\bibinfo
  {volume} {72}},\ \bibinfo {pages} {854} (\bibinfo {year} {2003})}\BibitemShut
  {NoStop}%
\bibitem [{\citenamefont {Oppeneer}\ \emph {et~al.}(2007)\citenamefont
  {Oppeneer}, \citenamefont {Elgazzar}, \citenamefont {Shick}, \citenamefont
  {Opahle}, \citenamefont {Rusz},\ and\ \citenamefont
  {Hayn}}]{oppeneer.elgazzar.07}%
  \BibitemOpen
  \bibfield  {author} {\bibinfo {author} {\bibfnamefont {P.~M.}\ \bibnamefont
  {Oppeneer}}, \bibinfo {author} {\bibfnamefont {S.}~\bibnamefont {Elgazzar}},
  \bibinfo {author} {\bibfnamefont {A.B.}\ \bibnamefont {Shick}}, \bibinfo
  {author} {\bibfnamefont {I.}~\bibnamefont {Opahle}}, \bibinfo {author}
  {\bibfnamefont {J.}~\bibnamefont {Rusz}}, \ and\ \bibinfo {author}
  {\bibfnamefont {R.}~\bibnamefont {Hayn}},\ }\bibfield  {title} {\enquote
  {\bibinfo {title} {Fermi surface changes due to localized--delocalized
  $f$-state transitions in {Ce-115} and {Pu-115} compounds},}\ }\href {\doibase
  10.1016/j.jmmm.2006.10.763} {\bibfield  {journal} {\bibinfo  {journal} {J.
  Magn. Magn. Mater.}\ }\textbf {\bibinfo {volume} {310}},\ \bibinfo {pages}
  {1684} (\bibinfo {year} {2007})}\BibitemShut {NoStop}%
\bibitem [{\citenamefont {Ronning}\ \emph {et~al.}(2012)\citenamefont
  {Ronning}, \citenamefont {Zhu}, \citenamefont {Das}, \citenamefont {Graf},
  \citenamefont {Albers}, \citenamefont {Rhee},\ and\ \citenamefont
  {Pickett}}]{ronning.zhu.12}%
  \BibitemOpen
  \bibfield  {author} {\bibinfo {author} {\bibfnamefont {F.}~\bibnamefont
  {Ronning}}, \bibinfo {author} {\bibfnamefont {J.-X.}\ \bibnamefont {Zhu}},
  \bibinfo {author} {\bibfnamefont {T.}~\bibnamefont {Das}}, \bibinfo {author}
  {\bibfnamefont {M.~J.}\ \bibnamefont {Graf}}, \bibinfo {author}
  {\bibfnamefont {R.~C.}\ \bibnamefont {Albers}}, \bibinfo {author}
  {\bibfnamefont {H.~B.}\ \bibnamefont {Rhee}}, \ and\ \bibinfo {author}
  {\bibfnamefont {W.~E.}\ \bibnamefont {Pickett}},\ }\bibfield  {title}
  {\enquote {\bibinfo {title} {Superconducting gap structure of the 115s
  revisited},}\ }\href {\doibase 10.1088/0953-8984/24/29/294206} {\bibfield
  {journal} {\bibinfo  {journal} {J. Phys.: Condens. Matter}\ }\textbf
  {\bibinfo {volume} {24}},\ \bibinfo {pages} {294206} (\bibinfo {year}
  {2012})}\BibitemShut {NoStop}%
\bibitem [{\citenamefont {Polyakov}\ \emph {et~al.}(2012)\citenamefont
  {Polyakov}, \citenamefont {Ignatchik}, \citenamefont {Bergk}, \citenamefont
  {G\"otze}, \citenamefont {Bianchi}, \citenamefont {Blackburn}, \citenamefont
  {Pr\'evost}, \citenamefont {Seyfarth}, \citenamefont {C\^ot\'e},
  \citenamefont {Hurt}, \citenamefont {Capan}, \citenamefont {Fisk},
  \citenamefont {Goodrich}, \citenamefont {Sheikin}, \citenamefont {Richter},\
  and\ \citenamefont {Wosnitza}}]{polyakov.ignatchik.12}%
  \BibitemOpen
  \bibfield  {author} {\bibinfo {author} {\bibfnamefont {A.}~\bibnamefont
  {Polyakov}}, \bibinfo {author} {\bibfnamefont {O.}~\bibnamefont {Ignatchik}},
  \bibinfo {author} {\bibfnamefont {B.}~\bibnamefont {Bergk}}, \bibinfo
  {author} {\bibfnamefont {K.}~\bibnamefont {G\"otze}}, \bibinfo {author}
  {\bibfnamefont {A.~D.}\ \bibnamefont {Bianchi}}, \bibinfo {author}
  {\bibfnamefont {S.}~\bibnamefont {Blackburn}}, \bibinfo {author}
  {\bibfnamefont {B.}~\bibnamefont {Pr\'evost}}, \bibinfo {author}
  {\bibfnamefont {G.}~\bibnamefont {Seyfarth}}, \bibinfo {author}
  {\bibfnamefont {M.}~\bibnamefont {C\^ot\'e}}, \bibinfo {author}
  {\bibfnamefont {D.}~\bibnamefont {Hurt}}, \bibinfo {author} {\bibfnamefont
  {C.}~\bibnamefont {Capan}}, \bibinfo {author} {\bibfnamefont
  {Z.}~\bibnamefont {Fisk}}, \bibinfo {author} {\bibfnamefont {R.~G.}\
  \bibnamefont {Goodrich}}, \bibinfo {author} {\bibfnamefont {I.}~\bibnamefont
  {Sheikin}}, \bibinfo {author} {\bibfnamefont {Manuel}\ \bibnamefont
  {Richter}}, \ and\ \bibinfo {author} {\bibfnamefont {J.}~\bibnamefont
  {Wosnitza}},\ }\bibfield  {title} {\enquote {\bibinfo {title} {Fermi-surface
  evolution in {Yb}-substituted {CeCoIn$_{5}$}},}\ }\href {\doibase
  10.1103/PhysRevB.85.245119} {\bibfield  {journal} {\bibinfo  {journal} {Phys.
  Rev. B}\ }\textbf {\bibinfo {volume} {85}},\ \bibinfo {pages} {245119}
  (\bibinfo {year} {2012})}\BibitemShut {NoStop}%
\bibitem [{\citenamefont {Ptok}\ \emph
  {et~al.}(2017{\natexlab{a}})\citenamefont {Ptok}, \citenamefont {Kapcia},
  \citenamefont {Piekarz},\ and\ \citenamefont {Ole{\'{s}}}}]{ptok.kapcia.17}%
  \BibitemOpen
  \bibfield  {author} {\bibinfo {author} {\bibfnamefont {A.}~\bibnamefont
  {Ptok}}, \bibinfo {author} {\bibfnamefont {K.~J.}\ \bibnamefont {Kapcia}},
  \bibinfo {author} {\bibfnamefont {P.}~\bibnamefont {Piekarz}}, \ and\
  \bibinfo {author} {\bibfnamefont {A.~M}\ \bibnamefont {Ole{\'{s}}}},\
  }\bibfield  {title} {\enquote {\bibinfo {title} {The ab initio study of
  unconventional superconductivity in {CeCoIn$_{5}$} and {FeSe}},}\ }\href
  {\doibase 10.1088/1367-2630/aa6d9d} {\bibfield  {journal} {\bibinfo
  {journal} {New J. Phys.}\ }\textbf {\bibinfo {volume} {19}},\ \bibinfo
  {pages} {063039} (\bibinfo {year} {2017}{\natexlab{a}})}\BibitemShut
  {NoStop}%
\bibitem [{Note2()}]{Note2}%
  \BibitemOpen
  \bibinfo {note} {We also compared the electronic band structure from
  {\protect \sc Vasp} with results obtained within {\protect \sc Quantum
  Espresso} software \cite {giannozzi.baroni.09,giannozzi.andreussi.17} and
  pseudopotentials developed in a frame of {\protect \sc PSlibrary}~\cite
  {dalcorso.14}. One needs to notice that the Ce atoms in both cases have
  different electronic configuration, i.e., [Xe]~$6s^{2}4f^{0.5}5d^{1.5}$ in
  the case {\protect \sc PSlibrary} and [Xe]~$6s^{2}4f^{1}5d^{1}$ for {\protect
  \sc Vasp} pseudopotentials. In consequence, the Ce $4f$ electrons levels can
  be overestimated (cf. Fig.~\ref {fig.bands} and Fig.~\ref {fig.bands_qe} in
  SM~\cite {Note1}).}\BibitemShut {Stop}%
\bibitem [{\citenamefont {Ikeda}\ \emph {et~al.}(2014)\citenamefont {Ikeda},
  \citenamefont {Suzuki}, \citenamefont {Arita},\ and\ \citenamefont
  {Takimoto}}]{ikeda.suzuki.14}%
  \BibitemOpen
  \bibfield  {author} {\bibinfo {author} {\bibfnamefont {H.}~\bibnamefont
  {Ikeda}}, \bibinfo {author} {\bibfnamefont {M.-T.}\ \bibnamefont {Suzuki}},
  \bibinfo {author} {\bibfnamefont {R.}~\bibnamefont {Arita}}, \ and\ \bibinfo
  {author} {\bibfnamefont {T.}~\bibnamefont {Takimoto}},\ }\bibfield  {title}
  {\enquote {\bibinfo {title} {Multipole fluctuations of itinerant f electrons
  and triakontadipole order in {URu$_{2}$Si$_{2}$}},}\ }\href {\doibase
  10.1016/j.crhy.2014.07.002} {\bibfield  {journal} {\bibinfo  {journal}
  {Comptes Rendus Physique}\ }\textbf {\bibinfo {volume} {15}},\ \bibinfo
  {pages} {587} (\bibinfo {year} {2014})}\BibitemShut {NoStop}%
\bibitem [{\citenamefont {Patil}\ \emph {et~al.}(2016)\citenamefont {Patil},
  \citenamefont {Generalov}, \citenamefont {G{\"u}ttler}, \citenamefont
  {Kushwaha}, \citenamefont {Chikina}, \citenamefont {Kummer}, \citenamefont
  {R{\"o}del}, \citenamefont {Santander-Syro}, \citenamefont {Caroca-Canales},
  \citenamefont {Geibel}, \citenamefont {Danzenb{\"a}cher}, \citenamefont
  {Kucherenko}, \citenamefont {Laubschat}, \citenamefont {Allen},\ and\
  \citenamefont {Vyalikh}}]{patil.generalov.16}%
  \BibitemOpen
  \bibfield  {author} {\bibinfo {author} {\bibfnamefont {S.}~\bibnamefont
  {Patil}}, \bibinfo {author} {\bibfnamefont {A.}~\bibnamefont {Generalov}},
  \bibinfo {author} {\bibfnamefont {M.}~\bibnamefont {G{\"u}ttler}}, \bibinfo
  {author} {\bibfnamefont {P.}~\bibnamefont {Kushwaha}}, \bibinfo {author}
  {\bibfnamefont {A.}~\bibnamefont {Chikina}}, \bibinfo {author} {\bibfnamefont
  {K.}~\bibnamefont {Kummer}}, \bibinfo {author} {\bibfnamefont {T.~C.}\
  \bibnamefont {R{\"o}del}}, \bibinfo {author} {\bibfnamefont {A.~F.}\
  \bibnamefont {Santander-Syro}}, \bibinfo {author} {\bibfnamefont
  {N.}~\bibnamefont {Caroca-Canales}}, \bibinfo {author} {\bibfnamefont
  {C.}~\bibnamefont {Geibel}}, \bibinfo {author} {\bibfnamefont
  {S.}~\bibnamefont {Danzenb{\"a}cher}}, \bibinfo {author} {\bibfnamefont
  {Yu.}\ \bibnamefont {Kucherenko}}, \bibinfo {author} {\bibfnamefont
  {C.}~\bibnamefont {Laubschat}}, \bibinfo {author} {\bibfnamefont {J.~W.}\
  \bibnamefont {Allen}}, \ and\ \bibinfo {author} {\bibfnamefont {D.~V.}\
  \bibnamefont {Vyalikh}},\ }\bibfield  {title} {\enquote {\bibinfo {title}
  {{ARPES} view on surface and bulk hybridization phenomena in the
  antiferromagnetic kondo lattice {CeRh$_{2}$Si$_{2}$}},}\ }\href {\doibase
  10.1038/ncomms11029} {\bibfield  {journal} {\bibinfo  {journal} {Nat.
  Commun.}\ }\textbf {\bibinfo {volume} {7}},\ \bibinfo {pages} {11029}
  (\bibinfo {year} {2016})}\BibitemShut {NoStop}%
\bibitem [{\citenamefont {Poelchen}\ \emph {et~al.}(2020)\citenamefont
  {Poelchen}, \citenamefont {Schulz}, \citenamefont {Mende}, \citenamefont
  {G{\"u}ttler}, \citenamefont {Generalov}, \citenamefont {Fedorov},
  \citenamefont {Caroca-Canales}, \citenamefont {Geibel}, \citenamefont
  {Kliemt}, \citenamefont {Krellner}, \citenamefont {Danzenb{\"a}cher},
  \citenamefont {Usachov}, \citenamefont {Dudin}, \citenamefont {Antonov},
  \citenamefont {Allen}, \citenamefont {Laubschat}, \citenamefont {Kummer},
  \citenamefont {Kucherenko},\ and\ \citenamefont
  {Vyalikh}}]{poelchen.schulz.20}%
  \BibitemOpen
  \bibfield  {author} {\bibinfo {author} {\bibfnamefont {G.}~\bibnamefont
  {Poelchen}}, \bibinfo {author} {\bibfnamefont {S.}~\bibnamefont {Schulz}},
  \bibinfo {author} {\bibfnamefont {M.}~\bibnamefont {Mende}}, \bibinfo
  {author} {\bibfnamefont {M.}~\bibnamefont {G{\"u}ttler}}, \bibinfo {author}
  {\bibfnamefont {A.}~\bibnamefont {Generalov}}, \bibinfo {author}
  {\bibfnamefont {A.~V.}\ \bibnamefont {Fedorov}}, \bibinfo {author}
  {\bibfnamefont {N.}~\bibnamefont {Caroca-Canales}}, \bibinfo {author}
  {\bibfnamefont {Ch.}\ \bibnamefont {Geibel}}, \bibinfo {author}
  {\bibfnamefont {K.}~\bibnamefont {Kliemt}}, \bibinfo {author} {\bibfnamefont
  {C.}~\bibnamefont {Krellner}}, \bibinfo {author} {\bibfnamefont
  {S.}~\bibnamefont {Danzenb{\"a}cher}}, \bibinfo {author} {\bibfnamefont
  {D.~Y.}\ \bibnamefont {Usachov}}, \bibinfo {author} {\bibfnamefont
  {P.}~\bibnamefont {Dudin}}, \bibinfo {author} {\bibfnamefont {V.~N.}\
  \bibnamefont {Antonov}}, \bibinfo {author} {\bibfnamefont {J.~W.}\
  \bibnamefont {Allen}}, \bibinfo {author} {\bibfnamefont {C.}~\bibnamefont
  {Laubschat}}, \bibinfo {author} {\bibfnamefont {K.}~\bibnamefont {Kummer}},
  \bibinfo {author} {\bibfnamefont {Y.}~\bibnamefont {Kucherenko}}, \ and\
  \bibinfo {author} {\bibfnamefont {Denis~V.}\ \bibnamefont {Vyalikh}},\
  }\bibfield  {title} {\enquote {\bibinfo {title} {Unexpected differences
  between surface and bulk spectroscopic and implied {Kondo} properties of
  heavy fermion {CeRh$_{2}$Si$_{2}$}},}\ }\href {\doibase
  10.1038/s41535-020-00273-7} {\bibfield  {journal} {\bibinfo  {journal} {npj
  Quantum Mater.}\ }\textbf {\bibinfo {volume} {5}},\ \bibinfo {pages} {70}
  (\bibinfo {year} {2020})}\BibitemShut {NoStop}%
\bibitem [{\citenamefont {Danzenb\"acher}\ \emph {et~al.}(2007)\citenamefont
  {Danzenb\"acher}, \citenamefont {Kucherenko}, \citenamefont {Vyalikh},
  \citenamefont {Holder}, \citenamefont {Laubschat}, \citenamefont {Yaresko},
  \citenamefont {Krellner}, \citenamefont {Hossain}, \citenamefont {Geibel},
  \citenamefont {Zhou}, \citenamefont {Yang}, \citenamefont {Mannella},
  \citenamefont {Hussain}, \citenamefont {Shen}, \citenamefont {Shi},
  \citenamefont {Patthey},\ and\ \citenamefont
  {Molodtsov}}]{danzenbacher.kucherenko.07}%
  \BibitemOpen
  \bibfield  {author} {\bibinfo {author} {\bibfnamefont {S.}~\bibnamefont
  {Danzenb\"acher}}, \bibinfo {author} {\bibfnamefont {Yu.}\ \bibnamefont
  {Kucherenko}}, \bibinfo {author} {\bibfnamefont {D.~V.}\ \bibnamefont
  {Vyalikh}}, \bibinfo {author} {\bibfnamefont {M.}~\bibnamefont {Holder}},
  \bibinfo {author} {\bibfnamefont {C.}~\bibnamefont {Laubschat}}, \bibinfo
  {author} {\bibfnamefont {A.~N.}\ \bibnamefont {Yaresko}}, \bibinfo {author}
  {\bibfnamefont {C.}~\bibnamefont {Krellner}}, \bibinfo {author}
  {\bibfnamefont {Z.}~\bibnamefont {Hossain}}, \bibinfo {author} {\bibfnamefont
  {C.}~\bibnamefont {Geibel}}, \bibinfo {author} {\bibfnamefont {X.~J.}\
  \bibnamefont {Zhou}}, \bibinfo {author} {\bibfnamefont {W.~L.}\ \bibnamefont
  {Yang}}, \bibinfo {author} {\bibfnamefont {N.}~\bibnamefont {Mannella}},
  \bibinfo {author} {\bibfnamefont {Z.}~\bibnamefont {Hussain}}, \bibinfo
  {author} {\bibfnamefont {Z.-X.}\ \bibnamefont {Shen}}, \bibinfo {author}
  {\bibfnamefont {M.}~\bibnamefont {Shi}}, \bibinfo {author} {\bibfnamefont
  {L.}~\bibnamefont {Patthey}}, \ and\ \bibinfo {author} {\bibfnamefont
  {S.~L.}\ \bibnamefont {Molodtsov}},\ }\bibfield  {title} {\enquote {\bibinfo
  {title} {Momentum dependence of $4f$ hybridization in heavy-fermion
  compounds: Angle-resolved photoemission study of {YbIr$_{2}$Si$_{2}$} and
  {YbRh$_{2}$Si$_{2}$}},}\ }\href {\doibase 10.1103/PhysRevB.75.045109}
  {\bibfield  {journal} {\bibinfo  {journal} {Phys. Rev. B}\ }\textbf {\bibinfo
  {volume} {75}},\ \bibinfo {pages} {045109} (\bibinfo {year}
  {2007})}\BibitemShut {NoStop}%
\bibitem [{\citenamefont {Vyalikh}\ \emph {et~al.}(2010)\citenamefont
  {Vyalikh}, \citenamefont {Danzenb\"acher}, \citenamefont {Kucherenko},
  \citenamefont {Kummer}, \citenamefont {Krellner}, \citenamefont {Geibel},
  \citenamefont {Holder}, \citenamefont {Kim}, \citenamefont {Laubschat},
  \citenamefont {Shi}, \citenamefont {Patthey}, \citenamefont {Follath},\ and\
  \citenamefont {Molodtsov}}]{vyalikh.danzenbacher.10}%
  \BibitemOpen
  \bibfield  {author} {\bibinfo {author} {\bibfnamefont {D.~V.}\ \bibnamefont
  {Vyalikh}}, \bibinfo {author} {\bibfnamefont {S.}~\bibnamefont
  {Danzenb\"acher}}, \bibinfo {author} {\bibfnamefont {Yu.}\ \bibnamefont
  {Kucherenko}}, \bibinfo {author} {\bibfnamefont {K.}~\bibnamefont {Kummer}},
  \bibinfo {author} {\bibfnamefont {C.}~\bibnamefont {Krellner}}, \bibinfo
  {author} {\bibfnamefont {C.}~\bibnamefont {Geibel}}, \bibinfo {author}
  {\bibfnamefont {M.~G.}\ \bibnamefont {Holder}}, \bibinfo {author}
  {\bibfnamefont {T.~K.}\ \bibnamefont {Kim}}, \bibinfo {author} {\bibfnamefont
  {C.}~\bibnamefont {Laubschat}}, \bibinfo {author} {\bibfnamefont
  {M.}~\bibnamefont {Shi}}, \bibinfo {author} {\bibfnamefont {L.}~\bibnamefont
  {Patthey}}, \bibinfo {author} {\bibfnamefont {R.}~\bibnamefont {Follath}}, \
  and\ \bibinfo {author} {\bibfnamefont {S.~L.}\ \bibnamefont {Molodtsov}},\
  }\bibfield  {title} {\enquote {\bibinfo {title} {$k$ dependence of the
  crystal-field splittings of $4f$ states in rare-earth systems},}\ }\href
  {\doibase 10.1103/PhysRevLett.105.237601} {\bibfield  {journal} {\bibinfo
  {journal} {Phys. Rev. Lett.}\ }\textbf {\bibinfo {volume} {105}},\ \bibinfo
  {pages} {237601} (\bibinfo {year} {2010})}\BibitemShut {NoStop}%
\bibitem [{\citenamefont {H{\"o}ppner}\ \emph {et~al.}(2013)\citenamefont
  {H{\"o}ppner}, \citenamefont {Seiro}, \citenamefont {Chikina}, \citenamefont
  {Fedorov}, \citenamefont {G{\"u}ttler}, \citenamefont {Danzenb{\"a}cher},
  \citenamefont {Generalov}, \citenamefont {Kummer}, \citenamefont {Patil},
  \citenamefont {Molodtsov}, \citenamefont {Kucherenko}, \citenamefont
  {Geibel}, \citenamefont {Strocov}, \citenamefont {Shi}, \citenamefont
  {Radovic}, \citenamefont {Schmitt}, \citenamefont {Laubschat},\ and\
  \citenamefont {Vyalikh}}]{hoppner.seiro.13}%
  \BibitemOpen
  \bibfield  {author} {\bibinfo {author} {\bibfnamefont {M.}~\bibnamefont
  {H{\"o}ppner}}, \bibinfo {author} {\bibfnamefont {S.}~\bibnamefont {Seiro}},
  \bibinfo {author} {\bibfnamefont {A.}~\bibnamefont {Chikina}}, \bibinfo
  {author} {\bibfnamefont {A.}~\bibnamefont {Fedorov}}, \bibinfo {author}
  {\bibfnamefont {M.}~\bibnamefont {G{\"u}ttler}}, \bibinfo {author}
  {\bibfnamefont {S.}~\bibnamefont {Danzenb{\"a}cher}}, \bibinfo {author}
  {\bibfnamefont {A.}~\bibnamefont {Generalov}}, \bibinfo {author}
  {\bibfnamefont {K.}~\bibnamefont {Kummer}}, \bibinfo {author} {\bibfnamefont
  {S.}~\bibnamefont {Patil}}, \bibinfo {author} {\bibfnamefont {S.~L.}\
  \bibnamefont {Molodtsov}}, \bibinfo {author} {\bibfnamefont {Y.}~\bibnamefont
  {Kucherenko}}, \bibinfo {author} {\bibfnamefont {C.}~\bibnamefont {Geibel}},
  \bibinfo {author} {\bibfnamefont {V.~N.}\ \bibnamefont {Strocov}}, \bibinfo
  {author} {\bibfnamefont {M.}~\bibnamefont {Shi}}, \bibinfo {author}
  {\bibfnamefont {M.}~\bibnamefont {Radovic}}, \bibinfo {author} {\bibfnamefont
  {T.}~\bibnamefont {Schmitt}}, \bibinfo {author} {\bibfnamefont
  {C.}~\bibnamefont {Laubschat}}, \ and\ \bibinfo {author} {\bibfnamefont
  {D.~V.}\ \bibnamefont {Vyalikh}},\ }\bibfield  {title} {\enquote {\bibinfo
  {title} {Interplay of {Dirac} fermions and heavy quasiparticles in solids},}\
  }\href {\doibase 10.1038/ncomms2654} {\bibfield  {journal} {\bibinfo
  {journal} {Nat. Commun.}\ }\textbf {\bibinfo {volume} {4}},\ \bibinfo {pages}
  {1646} (\bibinfo {year} {2013})}\BibitemShut {NoStop}%
\bibitem [{\citenamefont {Chikina}\ \emph {et~al.}(2014)\citenamefont
  {Chikina}, \citenamefont {H{\"o}ppner}, \citenamefont {Seiro}, \citenamefont
  {Kummer}, \citenamefont {Danzenb{\"a}cher}, \citenamefont {Patil},
  \citenamefont {Generalov}, \citenamefont {G{\"u}ttler}, \citenamefont
  {Kucherenko}, \citenamefont {Chulkov}, \citenamefont {Koroteev},
  \citenamefont {Koepernik}, \citenamefont {Geibel}, \citenamefont {Shi},
  \citenamefont {Radovic}, \citenamefont {Laubschat},\ and\ \citenamefont
  {Vyalikh}}]{chikina.hoppner.14}%
  \BibitemOpen
  \bibfield  {author} {\bibinfo {author} {\bibfnamefont {A.}~\bibnamefont
  {Chikina}}, \bibinfo {author} {\bibfnamefont {M.}~\bibnamefont
  {H{\"o}ppner}}, \bibinfo {author} {\bibfnamefont {S.}~\bibnamefont {Seiro}},
  \bibinfo {author} {\bibfnamefont {K.}~\bibnamefont {Kummer}}, \bibinfo
  {author} {\bibfnamefont {S.}~\bibnamefont {Danzenb{\"a}cher}}, \bibinfo
  {author} {\bibfnamefont {S.}~\bibnamefont {Patil}}, \bibinfo {author}
  {\bibfnamefont {A.}~\bibnamefont {Generalov}}, \bibinfo {author}
  {\bibfnamefont {M.}~\bibnamefont {G{\"u}ttler}}, \bibinfo {author}
  {\bibfnamefont {Yu.}\ \bibnamefont {Kucherenko}}, \bibinfo {author}
  {\bibfnamefont {E.~V.}\ \bibnamefont {Chulkov}}, \bibinfo {author}
  {\bibfnamefont {Yu.~M.}\ \bibnamefont {Koroteev}}, \bibinfo {author}
  {\bibfnamefont {K.}~\bibnamefont {Koepernik}}, \bibinfo {author}
  {\bibfnamefont {C.}~\bibnamefont {Geibel}}, \bibinfo {author} {\bibfnamefont
  {M.}~\bibnamefont {Shi}}, \bibinfo {author} {\bibfnamefont {M.}~\bibnamefont
  {Radovic}}, \bibinfo {author} {\bibfnamefont {C.}~\bibnamefont {Laubschat}},
  \ and\ \bibinfo {author} {\bibfnamefont {D.~V.}\ \bibnamefont {Vyalikh}},\
  }\bibfield  {title} {\enquote {\bibinfo {title} {Strong ferromagnetism at the
  surface of an antiferromagnet caused by buried magnetic moments},}\ }\href
  {\doibase 10.1038/ncomms4171} {\bibfield  {journal} {\bibinfo  {journal}
  {Nat. Commun.}\ }\textbf {\bibinfo {volume} {5}},\ \bibinfo {pages} {3171}
  (\bibinfo {year} {2014})}\BibitemShut {NoStop}%
\bibitem [{\citenamefont {G{\"u}ttler}\ \emph {et~al.}(2019)\citenamefont
  {G{\"u}ttler}, \citenamefont {Generalov}, \citenamefont {Fujimori},
  \citenamefont {Kummer}, \citenamefont {Chikina}, \citenamefont {Seiro},
  \citenamefont {Danzenb{\"a}cher}, \citenamefont {Koroteev}, \citenamefont
  {Chulkov}, \citenamefont {Radovic}, \citenamefont {Shi}, \citenamefont
  {Plumb}, \citenamefont {Laubschat}, \citenamefont {Allen}, \citenamefont
  {Krellner}, \citenamefont {Geibel},\ and\ \citenamefont
  {Vyalikh}}]{guttler.generalov.19}%
  \BibitemOpen
  \bibfield  {author} {\bibinfo {author} {\bibfnamefont {M.}~\bibnamefont
  {G{\"u}ttler}}, \bibinfo {author} {\bibfnamefont {A.}~\bibnamefont
  {Generalov}}, \bibinfo {author} {\bibfnamefont {S.~I.}\ \bibnamefont
  {Fujimori}}, \bibinfo {author} {\bibfnamefont {K.}~\bibnamefont {Kummer}},
  \bibinfo {author} {\bibfnamefont {A.}~\bibnamefont {Chikina}}, \bibinfo
  {author} {\bibfnamefont {S.}~\bibnamefont {Seiro}}, \bibinfo {author}
  {\bibfnamefont {S.}~\bibnamefont {Danzenb{\"a}cher}}, \bibinfo {author}
  {\bibfnamefont {Yu.~M.}\ \bibnamefont {Koroteev}}, \bibinfo {author}
  {\bibfnamefont {E.~V.}\ \bibnamefont {Chulkov}}, \bibinfo {author}
  {\bibfnamefont {M.}~\bibnamefont {Radovic}}, \bibinfo {author} {\bibfnamefont
  {M.}~\bibnamefont {Shi}}, \bibinfo {author} {\bibfnamefont {N.~C.}\
  \bibnamefont {Plumb}}, \bibinfo {author} {\bibfnamefont {C.}~\bibnamefont
  {Laubschat}}, \bibinfo {author} {\bibfnamefont {J.~W.}\ \bibnamefont
  {Allen}}, \bibinfo {author} {\bibfnamefont {C.}~\bibnamefont {Krellner}},
  \bibinfo {author} {\bibfnamefont {C.}~\bibnamefont {Geibel}}, \ and\ \bibinfo
  {author} {\bibfnamefont {D.~V.}\ \bibnamefont {Vyalikh}},\ }\bibfield
  {title} {\enquote {\bibinfo {title} {Divalent {EuRh$_{2}$Si$_{2}$} as a
  reference for the luttinger theorem and antiferromagnetism in trivalent
  heavy-fermion {YbRh$_{2}$Si$_{2}$}},}\ }\href {\doibase
  10.1038/s41467-019-08688-y} {\bibfield  {journal} {\bibinfo  {journal} {Nat.
  Commun.}\ }\textbf {\bibinfo {volume} {10}},\ \bibinfo {pages} {796}
  (\bibinfo {year} {2019})}\BibitemShut {NoStop}%
\bibitem [{\citenamefont {Nogaki}\ \emph {et~al.}(2021)\citenamefont {Nogaki},
  \citenamefont {Daido}, \citenamefont {Ishizuka},\ and\ \citenamefont
  {Yanase}}]{nogaki.daido.21}%
  \BibitemOpen
  \bibfield  {author} {\bibinfo {author} {\bibfnamefont {K.}~\bibnamefont
  {Nogaki}}, \bibinfo {author} {\bibfnamefont {A.}~\bibnamefont {Daido}},
  \bibinfo {author} {\bibfnamefont {J.}~\bibnamefont {Ishizuka}}, \ and\
  \bibinfo {author} {\bibfnamefont {Y.}~\bibnamefont {Yanase}},\ }\href@noop {}
  {\enquote {\bibinfo {title} {Topological crystalline superconductivity in
  locally noncentrosymmetric {CeRh$_2$As$_2$}},}\ } (\bibinfo {year} {2021}),\
  \Eprint {http://arxiv.org/abs/arXiv:2103.08088} {arXiv:2103.08088}
  \BibitemShut {NoStop}%
\bibitem [{\citenamefont {Damascelli}\ \emph {et~al.}(2003)\citenamefont
  {Damascelli}, \citenamefont {Hussain},\ and\ \citenamefont
  {Shen}}]{damascelli.hussain.03}%
  \BibitemOpen
  \bibfield  {author} {\bibinfo {author} {\bibfnamefont {A.}~\bibnamefont
  {Damascelli}}, \bibinfo {author} {\bibfnamefont {Z.}~\bibnamefont {Hussain}},
  \ and\ \bibinfo {author} {\bibfnamefont {Z.-X.}\ \bibnamefont {Shen}},\
  }\bibfield  {title} {\enquote {\bibinfo {title} {Angle-resolved photoemission
  studies of the cuprate superconductors},}\ }\href {\doibase
  10.1103/RevModPhys.75.473} {\bibfield  {journal} {\bibinfo  {journal} {Rev.
  Mod. Phys.}\ }\textbf {\bibinfo {volume} {75}},\ \bibinfo {pages} {473}
  (\bibinfo {year} {2003})}\BibitemShut {NoStop}%
\bibitem [{\citenamefont {Dil}(2009)}]{dil.09}%
  \BibitemOpen
  \bibfield  {author} {\bibinfo {author} {\bibfnamefont {J~Hugo}\ \bibnamefont
  {Dil}},\ }\bibfield  {title} {\enquote {\bibinfo {title} {Spin and angle
  resolved photoemission on non-magnetic low-dimensional systems},}\ }\href
  {\doibase 10.1088/0953-8984/21/40/403001} {\bibfield  {journal} {\bibinfo
  {journal} {J. Phys.: Condens. Matter}\ }\textbf {\bibinfo {volume} {21}},\
  \bibinfo {pages} {403001} (\bibinfo {year} {2009})}\BibitemShut {NoStop}%
\bibitem [{\citenamefont {Lv}\ \emph {et~al.}(2019)\citenamefont {Lv},
  \citenamefont {Qian},\ and\ \citenamefont {Ding}}]{lv.qian.19}%
  \BibitemOpen
  \bibfield  {author} {\bibinfo {author} {\bibfnamefont {Baiqing}\ \bibnamefont
  {Lv}}, \bibinfo {author} {\bibfnamefont {Tian}\ \bibnamefont {Qian}}, \ and\
  \bibinfo {author} {\bibfnamefont {Hong}\ \bibnamefont {Ding}},\ }\bibfield
  {title} {\enquote {\bibinfo {title} {Angle-resolved photoemission
  spectroscopy and its application to topological materials},}\ }\href
  {\doibase 10.1038/s42254-019-0088-5} {\bibfield  {journal} {\bibinfo
  {journal} {Nat. Rev. Phys.}\ }\textbf {\bibinfo {volume} {1}},\ \bibinfo
  {pages} {609} (\bibinfo {year} {2019})}\BibitemShut {NoStop}%
\bibitem [{\citenamefont {Ptok}\ \emph
  {et~al.}(2017{\natexlab{b}})\citenamefont {Ptok}, \citenamefont {Kapcia},
  \citenamefont {Cichy}, \citenamefont {Ole{\'{s}}},\ and\ \citenamefont
  {Piekarz}}]{ptok.kapcia.17sr}%
  \BibitemOpen
  \bibfield  {author} {\bibinfo {author} {\bibfnamefont {A.}~\bibnamefont
  {Ptok}}, \bibinfo {author} {\bibfnamefont {K.~J.}\ \bibnamefont {Kapcia}},
  \bibinfo {author} {\bibfnamefont {A.}~\bibnamefont {Cichy}}, \bibinfo
  {author} {\bibfnamefont {A.~M.}\ \bibnamefont {Ole{\'{s}}}}, \ and\ \bibinfo
  {author} {\bibfnamefont {P.}~\bibnamefont {Piekarz}},\ }\bibfield  {title}
  {\enquote {\bibinfo {title} {Magnetic {Lifshitz} transition and its
  consequences in multi-band iron-based superconductors},}\ }\href {\doibase
  10.1038/srep41979} {\bibfield  {journal} {\bibinfo  {journal} {Sci. Rep.}\
  }\textbf {\bibinfo {volume} {7}},\ \bibinfo {pages} {41979} (\bibinfo {year}
  {2017}{\natexlab{b}})}\BibitemShut {NoStop}%
\bibitem [{\citenamefont {Daou}\ \emph {et~al.}(2006)\citenamefont {Daou},
  \citenamefont {Bergemann},\ and\ \citenamefont {Julian}}]{daou.bergemann.06}%
  \BibitemOpen
  \bibfield  {author} {\bibinfo {author} {\bibfnamefont {R.}~\bibnamefont
  {Daou}}, \bibinfo {author} {\bibfnamefont {C.}~\bibnamefont {Bergemann}}, \
  and\ \bibinfo {author} {\bibfnamefont {S.~R.}\ \bibnamefont {Julian}},\
  }\bibfield  {title} {\enquote {\bibinfo {title} {Continuous evolution of the
  {Fermi} surface of {CeRu$_{2}$Si$_{2}$} across the metamagnetic
  transition},}\ }\href {\doibase 10.1103/PhysRevLett.96.026401} {\bibfield
  {journal} {\bibinfo  {journal} {Phys. Rev. Lett.}\ }\textbf {\bibinfo
  {volume} {96}},\ \bibinfo {pages} {026401} (\bibinfo {year}
  {2006})}\BibitemShut {NoStop}%
\bibitem [{\citenamefont {Kenzelmann}\ \emph {et~al.}(2008)\citenamefont
  {Kenzelmann}, \citenamefont {Str{\"a}ssle}, \citenamefont {Niedermayer},
  \citenamefont {Sigrist}, \citenamefont {Padmanabhan}, \citenamefont
  {Zolliker}, \citenamefont {Bianchi}, \citenamefont {Movshovich},
  \citenamefont {Bauer}, \citenamefont {Sarrao},\ and\ \citenamefont
  {Thompson}}]{kenzelmann.strassle.08}%
  \BibitemOpen
  \bibfield  {author} {\bibinfo {author} {\bibfnamefont {M.}~\bibnamefont
  {Kenzelmann}}, \bibinfo {author} {\bibfnamefont {Th.}\ \bibnamefont
  {Str{\"a}ssle}}, \bibinfo {author} {\bibfnamefont {C.}~\bibnamefont
  {Niedermayer}}, \bibinfo {author} {\bibfnamefont {M.}~\bibnamefont
  {Sigrist}}, \bibinfo {author} {\bibfnamefont {B.}~\bibnamefont
  {Padmanabhan}}, \bibinfo {author} {\bibfnamefont {M.}~\bibnamefont
  {Zolliker}}, \bibinfo {author} {\bibfnamefont {A.~D.}\ \bibnamefont
  {Bianchi}}, \bibinfo {author} {\bibfnamefont {R.}~\bibnamefont {Movshovich}},
  \bibinfo {author} {\bibfnamefont {E.~D.}\ \bibnamefont {Bauer}}, \bibinfo
  {author} {\bibfnamefont {J.~L.}\ \bibnamefont {Sarrao}}, \ and\ \bibinfo
  {author} {\bibfnamefont {J.~D.}\ \bibnamefont {Thompson}},\ }\bibfield
  {title} {\enquote {\bibinfo {title} {Coupled superconducting and magnetic
  order in {CeCoIn$_{5}$}},}\ }\href {\doibase 10.1126/science.1161818}
  {\bibfield  {journal} {\bibinfo  {journal} {Science}\ }\textbf {\bibinfo
  {volume} {321}},\ \bibinfo {pages} {1652} (\bibinfo {year}
  {2008})}\BibitemShut {NoStop}%
\bibitem [{\citenamefont {Kenzelmann}\ \emph {et~al.}(2010)\citenamefont
  {Kenzelmann}, \citenamefont {Gerber}, \citenamefont {Egetenmeyer},
  \citenamefont {Gavilano}, \citenamefont {Str\"assle}, \citenamefont
  {Bianchi}, \citenamefont {Ressouche}, \citenamefont {Movshovich},
  \citenamefont {Bauer}, \citenamefont {Sarrao},\ and\ \citenamefont
  {Thompson}}]{knezelmann.gerber.10}%
  \BibitemOpen
  \bibfield  {author} {\bibinfo {author} {\bibfnamefont {M.}~\bibnamefont
  {Kenzelmann}}, \bibinfo {author} {\bibfnamefont {S.}~\bibnamefont {Gerber}},
  \bibinfo {author} {\bibfnamefont {N.}~\bibnamefont {Egetenmeyer}}, \bibinfo
  {author} {\bibfnamefont {J.~L.}\ \bibnamefont {Gavilano}}, \bibinfo {author}
  {\bibfnamefont {Th.}\ \bibnamefont {Str\"assle}}, \bibinfo {author}
  {\bibfnamefont {A.~D.}\ \bibnamefont {Bianchi}}, \bibinfo {author}
  {\bibfnamefont {E.}~\bibnamefont {Ressouche}}, \bibinfo {author}
  {\bibfnamefont {R.}~\bibnamefont {Movshovich}}, \bibinfo {author}
  {\bibfnamefont {E.~D.}\ \bibnamefont {Bauer}}, \bibinfo {author}
  {\bibfnamefont {J.~L.}\ \bibnamefont {Sarrao}}, \ and\ \bibinfo {author}
  {\bibfnamefont {J.~D.}\ \bibnamefont {Thompson}},\ }\bibfield  {title}
  {\enquote {\bibinfo {title} {Evidence for a magnetically driven
  superconducting {$Q$} phase of {CeCoIn$_{5}$}},}\ }\href {\doibase
  10.1103/PhysRevLett.104.127001} {\bibfield  {journal} {\bibinfo  {journal}
  {Phys. Rev. Lett.}\ }\textbf {\bibinfo {volume} {104}},\ \bibinfo {pages}
  {127001} (\bibinfo {year} {2010})}\BibitemShut {NoStop}%
\bibitem [{\citenamefont {Koutroulakis}\ \emph {et~al.}(2010)\citenamefont
  {Koutroulakis}, \citenamefont {Stewart}, \citenamefont {Mitrovi\'{c}},
  \citenamefont {Horvati\'{c}}, \citenamefont {Berthier}, \citenamefont
  {Lapertot},\ and\ \citenamefont {Flouquet}}]{koutroulakis.steart.10}%
  \BibitemOpen
  \bibfield  {author} {\bibinfo {author} {\bibfnamefont {G.}~\bibnamefont
  {Koutroulakis}}, \bibinfo {author} {\bibfnamefont {M.~D.}\ \bibnamefont
  {Stewart}}, \bibinfo {author} {\bibfnamefont {V.~F.}\ \bibnamefont
  {Mitrovi\'{c}}}, \bibinfo {author} {\bibfnamefont {M.}~\bibnamefont
  {Horvati\'{c}}}, \bibinfo {author} {\bibfnamefont {C.}~\bibnamefont
  {Berthier}}, \bibinfo {author} {\bibfnamefont {G.}~\bibnamefont {Lapertot}},
  \ and\ \bibinfo {author} {\bibfnamefont {J.}~\bibnamefont {Flouquet}},\
  }\bibfield  {title} {\enquote {\bibinfo {title} {Field evolution of
  coexisting superconducting and magnetic orders in {CeCoIn$_{5}$}},}\ }\href
  {\doibase 10.1103/PhysRevLett.104.087001} {\bibfield  {journal} {\bibinfo
  {journal} {Phys. Rev. Lett.}\ }\textbf {\bibinfo {volume} {104}},\ \bibinfo
  {pages} {087001} (\bibinfo {year} {2010})}\BibitemShut {NoStop}%
\bibitem [{\citenamefont {Schlottmann}(2011)}]{schlottmann.11}%
  \BibitemOpen
  \bibfield  {author} {\bibinfo {author} {\bibfnamefont {P.}~\bibnamefont
  {Schlottmann}},\ }\bibfield  {title} {\enquote {\bibinfo {title} {Lifshitz
  transition with interactions in high magnetic fields},}\ }\href {\doibase
  10.1103/PhysRevB.83.115133} {\bibfield  {journal} {\bibinfo  {journal} {Phys.
  Rev. B}\ }\textbf {\bibinfo {volume} {83}},\ \bibinfo {pages} {115133}
  (\bibinfo {year} {2011})}\BibitemShut {NoStop}%
\bibitem [{\citenamefont {Momma}\ and\ \citenamefont
  {Izumi}(2011)}]{momma.izumi.11}%
  \BibitemOpen
  \bibfield  {author} {\bibinfo {author} {\bibfnamefont {K.}~\bibnamefont
  {Momma}}\ and\ \bibinfo {author} {\bibfnamefont {F.}~\bibnamefont {Izumi}},\
  }\bibfield  {title} {\enquote {\bibinfo {title} {{{\sc vesta3} for
  three-dimensional visualization of crystal, volumetric and morphology
  data}},}\ }\href {\doibase 10.1107/S0021889811038970} {\bibfield  {journal}
  {\bibinfo  {journal} {J. Appl. Crystallogr.}\ }\textbf {\bibinfo {volume}
  {44}},\ \bibinfo {pages} {1272} (\bibinfo {year} {2011})}\BibitemShut
  {NoStop}%
\bibitem [{\citenamefont {Kokalj}(1999)}]{kokalj.99}%
  \BibitemOpen
  \bibfield  {author} {\bibinfo {author} {\bibfnamefont {A.}~\bibnamefont
  {Kokalj}},\ }\bibfield  {title} {\enquote {\bibinfo {title} {{XCrySDen}--a
  new program for displaying crystalline structures and electron densities},}\
  }\href {\doibase 10.1016/S1093-3263(99)00028-5} {\bibfield  {journal}
  {\bibinfo  {journal} {J. Mol. Graph. Model.}\ }\textbf {\bibinfo {volume}
  {17}},\ \bibinfo {pages} {176} (\bibinfo {year} {1999})}\BibitemShut
  {NoStop}%
\bibitem [{\citenamefont {Giannozzi}\ \emph {et~al.}(2009)\citenamefont
  {Giannozzi}, \citenamefont {Baroni}, \citenamefont {Bonini}, \citenamefont
  {Calandra}, \citenamefont {Car}, \citenamefont {Cavazzoni}, \citenamefont
  {Ceresoli}, \citenamefont {Chiarotti}, \citenamefont {Cococcioni},
  \citenamefont {Dabo}, \citenamefont {Corso}, \citenamefont {de~Gironcoli},
  \citenamefont {Fabris}, \citenamefont {Fratesi}, \citenamefont {Gebauer},
  \citenamefont {Gerstmann}, \citenamefont {Gougoussis}, \citenamefont
  {Kokalj}, \citenamefont {Lazzeri}, \citenamefont {Martin-Samos},
  \citenamefont {Marzari}, \citenamefont {Mauri}, \citenamefont {Mazzarello},
  \citenamefont {Paolini}, \citenamefont {Pasquarello}, \citenamefont
  {Paulatto}, \citenamefont {Sbraccia}, \citenamefont {Scandolo}, \citenamefont
  {Sclauzero}, \citenamefont {Seitsonen}, \citenamefont {Smogunov},
  \citenamefont {Umari},\ and\ \citenamefont
  {Wentzcovitch}}]{giannozzi.baroni.09}%
  \BibitemOpen
  \bibfield  {author} {\bibinfo {author} {\bibfnamefont {P.}~\bibnamefont
  {Giannozzi}}, \bibinfo {author} {\bibfnamefont {S.}~\bibnamefont {Baroni}},
  \bibinfo {author} {\bibfnamefont {N.}~\bibnamefont {Bonini}}, \bibinfo
  {author} {\bibfnamefont {M.}~\bibnamefont {Calandra}}, \bibinfo {author}
  {\bibfnamefont {R.}~\bibnamefont {Car}}, \bibinfo {author} {\bibfnamefont
  {C.}~\bibnamefont {Cavazzoni}}, \bibinfo {author} {\bibfnamefont
  {D.}~\bibnamefont {Ceresoli}}, \bibinfo {author} {\bibfnamefont {G.~L.}\
  \bibnamefont {Chiarotti}}, \bibinfo {author} {\bibfnamefont {M.}~\bibnamefont
  {Cococcioni}}, \bibinfo {author} {\bibfnamefont {I.}~\bibnamefont {Dabo}},
  \bibinfo {author} {\bibfnamefont {A.~Dal}\ \bibnamefont {Corso}}, \bibinfo
  {author} {\bibfnamefont {S.}~\bibnamefont {de~Gironcoli}}, \bibinfo {author}
  {\bibfnamefont {S.}~\bibnamefont {Fabris}}, \bibinfo {author} {\bibfnamefont
  {G.}~\bibnamefont {Fratesi}}, \bibinfo {author} {\bibfnamefont
  {R.}~\bibnamefont {Gebauer}}, \bibinfo {author} {\bibfnamefont
  {U.}~\bibnamefont {Gerstmann}}, \bibinfo {author} {\bibfnamefont {Ch.}\
  \bibnamefont {Gougoussis}}, \bibinfo {author} {\bibfnamefont
  {A.}~\bibnamefont {Kokalj}}, \bibinfo {author} {\bibfnamefont
  {M.}~\bibnamefont {Lazzeri}}, \bibinfo {author} {\bibfnamefont
  {L.}~\bibnamefont {Martin-Samos}}, \bibinfo {author} {\bibfnamefont
  {N.}~\bibnamefont {Marzari}}, \bibinfo {author} {\bibfnamefont
  {F.}~\bibnamefont {Mauri}}, \bibinfo {author} {\bibfnamefont
  {R.}~\bibnamefont {Mazzarello}}, \bibinfo {author} {\bibfnamefont
  {S.}~\bibnamefont {Paolini}}, \bibinfo {author} {\bibfnamefont
  {A.}~\bibnamefont {Pasquarello}}, \bibinfo {author} {\bibfnamefont
  {L.}~\bibnamefont {Paulatto}}, \bibinfo {author} {\bibfnamefont
  {C.}~\bibnamefont {Sbraccia}}, \bibinfo {author} {\bibfnamefont
  {S.}~\bibnamefont {Scandolo}}, \bibinfo {author} {\bibfnamefont
  {G.}~\bibnamefont {Sclauzero}}, \bibinfo {author} {\bibfnamefont {A.~P.}\
  \bibnamefont {Seitsonen}}, \bibinfo {author} {\bibfnamefont {A.}~\bibnamefont
  {Smogunov}}, \bibinfo {author} {\bibfnamefont {P.}~\bibnamefont {Umari}}, \
  and\ \bibinfo {author} {\bibfnamefont {R.~M.}\ \bibnamefont {Wentzcovitch}},\
  }\bibfield  {title} {\enquote {\bibinfo {title} {{QUANTUM} {ESPRESSO}: a
  modular and open-source software project for quantum simulations of
  materials},}\ }\href {\doibase 10.1088/0953-8984/21/39/395502} {\bibfield
  {journal} {\bibinfo  {journal} {J. Phys.: Condens. Matter}\ }\textbf
  {\bibinfo {volume} {21}},\ \bibinfo {pages} {395502} (\bibinfo {year}
  {2009})}\BibitemShut {NoStop}%
\bibitem [{\citenamefont {Giannozzi}\ \emph {et~al.}(2017)\citenamefont
  {Giannozzi}, \citenamefont {Andreussi}, \citenamefont {Brumme}, \citenamefont
  {Bunau}, \citenamefont {Nardelli}, \citenamefont {Calandra}, \citenamefont
  {Car}, \citenamefont {Cavazzoni}, \citenamefont {Ceresoli}, \citenamefont
  {Cococcioni}, \citenamefont {Colonna}, \citenamefont {Carnimeo},
  \citenamefont {Corso}, \citenamefont {de~Gironcoli}, \citenamefont {Delugas},
  \citenamefont {DiStasio}, \citenamefont {Ferretti}, \citenamefont {Floris},
  \citenamefont {Fratesi}, \citenamefont {Fugallo}, \citenamefont {Gebauer},
  \citenamefont {Gerstmann}, \citenamefont {Giustino}, \citenamefont {Gorni},
  \citenamefont {Jia}, \citenamefont {Kawamura}, \citenamefont {Ko},
  \citenamefont {Kokalj}, \citenamefont {K\"{u}{\c{c}}\"{u}kbenli},
  \citenamefont {Lazzeri}, \citenamefont {Marsili}, \citenamefont {Marzari},
  \citenamefont {Mauri}, \citenamefont {Nguyen}, \citenamefont {Nguyen},
  \citenamefont {de-la Roza}, \citenamefont {Paulatto}, \citenamefont
  {Ponc{\'{e}}}, \citenamefont {Rocca}, \citenamefont {Sabatini}, \citenamefont
  {Santra}, \citenamefont {Schlipf}, \citenamefont {Seitsonen}, \citenamefont
  {Smogunov}, \citenamefont {Timrov}, \citenamefont {Thonhauser}, \citenamefont
  {Umari}, \citenamefont {Vast}, \citenamefont {Wu},\ and\ \citenamefont
  {Baroni}}]{giannozzi.andreussi.17}%
  \BibitemOpen
  \bibfield  {author} {\bibinfo {author} {\bibfnamefont {P.}~\bibnamefont
  {Giannozzi}}, \bibinfo {author} {\bibfnamefont {O.}~\bibnamefont
  {Andreussi}}, \bibinfo {author} {\bibfnamefont {T.}~\bibnamefont {Brumme}},
  \bibinfo {author} {\bibfnamefont {O.}~\bibnamefont {Bunau}}, \bibinfo
  {author} {\bibfnamefont {M.~Buongiorno}\ \bibnamefont {Nardelli}}, \bibinfo
  {author} {\bibfnamefont {M.}~\bibnamefont {Calandra}}, \bibinfo {author}
  {\bibfnamefont {R.}~\bibnamefont {Car}}, \bibinfo {author} {\bibfnamefont
  {C.}~\bibnamefont {Cavazzoni}}, \bibinfo {author} {\bibfnamefont
  {D.}~\bibnamefont {Ceresoli}}, \bibinfo {author} {\bibfnamefont
  {M.}~\bibnamefont {Cococcioni}}, \bibinfo {author} {\bibfnamefont
  {N.}~\bibnamefont {Colonna}}, \bibinfo {author} {\bibfnamefont
  {I.}~\bibnamefont {Carnimeo}}, \bibinfo {author} {\bibfnamefont {A.~Dal}\
  \bibnamefont {Corso}}, \bibinfo {author} {\bibfnamefont {S.}~\bibnamefont
  {de~Gironcoli}}, \bibinfo {author} {\bibfnamefont {P.}~\bibnamefont
  {Delugas}}, \bibinfo {author} {\bibfnamefont {R.~A.}\ \bibnamefont
  {DiStasio}}, \bibinfo {author} {\bibfnamefont {A.}~\bibnamefont {Ferretti}},
  \bibinfo {author} {\bibfnamefont {A.}~\bibnamefont {Floris}}, \bibinfo
  {author} {\bibfnamefont {G.}~\bibnamefont {Fratesi}}, \bibinfo {author}
  {\bibfnamefont {G.}~\bibnamefont {Fugallo}}, \bibinfo {author} {\bibfnamefont
  {R.}~\bibnamefont {Gebauer}}, \bibinfo {author} {\bibfnamefont
  {U.}~\bibnamefont {Gerstmann}}, \bibinfo {author} {\bibfnamefont
  {F.}~\bibnamefont {Giustino}}, \bibinfo {author} {\bibfnamefont
  {T.}~\bibnamefont {Gorni}}, \bibinfo {author} {\bibfnamefont
  {J.}~\bibnamefont {Jia}}, \bibinfo {author} {\bibfnamefont {M.}~\bibnamefont
  {Kawamura}}, \bibinfo {author} {\bibfnamefont {H.-Y.}\ \bibnamefont {Ko}},
  \bibinfo {author} {\bibfnamefont {A.}~\bibnamefont {Kokalj}}, \bibinfo
  {author} {\bibfnamefont {E.}~\bibnamefont {K\"{u}{\c{c}}\"{u}kbenli}},
  \bibinfo {author} {\bibfnamefont {M.}~\bibnamefont {Lazzeri}}, \bibinfo
  {author} {\bibfnamefont {M.}~\bibnamefont {Marsili}}, \bibinfo {author}
  {\bibfnamefont {N.}~\bibnamefont {Marzari}}, \bibinfo {author} {\bibfnamefont
  {F.}~\bibnamefont {Mauri}}, \bibinfo {author} {\bibfnamefont {N.~L.}\
  \bibnamefont {Nguyen}}, \bibinfo {author} {\bibfnamefont {H.-V.}\
  \bibnamefont {Nguyen}}, \bibinfo {author} {\bibfnamefont {A.~Otero}\
  \bibnamefont {de-la Roza}}, \bibinfo {author} {\bibfnamefont
  {L.}~\bibnamefont {Paulatto}}, \bibinfo {author} {\bibfnamefont
  {S.}~\bibnamefont {Ponc{\'{e}}}}, \bibinfo {author} {\bibfnamefont
  {D.}~\bibnamefont {Rocca}}, \bibinfo {author} {\bibfnamefont
  {R.}~\bibnamefont {Sabatini}}, \bibinfo {author} {\bibfnamefont
  {B.}~\bibnamefont {Santra}}, \bibinfo {author} {\bibfnamefont
  {M.}~\bibnamefont {Schlipf}}, \bibinfo {author} {\bibfnamefont {A.~P.}\
  \bibnamefont {Seitsonen}}, \bibinfo {author} {\bibfnamefont {A.}~\bibnamefont
  {Smogunov}}, \bibinfo {author} {\bibfnamefont {I.}~\bibnamefont {Timrov}},
  \bibinfo {author} {\bibfnamefont {T.}~\bibnamefont {Thonhauser}}, \bibinfo
  {author} {\bibfnamefont {P.}~\bibnamefont {Umari}}, \bibinfo {author}
  {\bibfnamefont {N.}~\bibnamefont {Vast}}, \bibinfo {author} {\bibfnamefont
  {X.}~\bibnamefont {Wu}}, \ and\ \bibinfo {author} {\bibfnamefont
  {S.}~\bibnamefont {Baroni}},\ }\bibfield  {title} {\enquote {\bibinfo {title}
  {Advanced capabilities for materials modelling with {Quantum} {ESPRESSO}},}\
  }\href {\doibase 10.1088/1361-648x/aa8f79} {\bibfield  {journal} {\bibinfo
  {journal} {J. Phys.: Condens. Matter}\ }\textbf {\bibinfo {volume} {29}},\
  \bibinfo {pages} {465901} (\bibinfo {year} {2017})}\BibitemShut {NoStop}%
\bibitem [{\citenamefont {{Dal Corso}}(2014)}]{dalcorso.14}%
  \BibitemOpen
  \bibfield  {author} {\bibinfo {author} {\bibfnamefont {A.}~\bibnamefont {{Dal
  Corso}}},\ }\bibfield  {title} {\enquote {\bibinfo {title} {Pseudopotentials
  periodic table: From {H} to {Pu}},}\ }\href {\doibase
  10.1016/j.commatsci.2014.07.043} {\bibfield  {journal} {\bibinfo  {journal}
  {Comput. Mater. Sci.}\ }\textbf {\bibinfo {volume} {95}},\ \bibinfo {pages}
  {337} (\bibinfo {year} {2014})}\BibitemShut {NoStop}%
\end{thebibliography}%

\clearpage
\newpage

\onecolumngrid

\begin{center}
  \textbf{\Large Supplemental Material}\\[.2cm]
  \textbf{\large Electronic and dynamical properties of CeRh$_{2}$As$_{2}$:\\[.2cm]
  Role of Rh$_{2}$As$_{2}$ layers and expected hidden orbital order}\\[.2cm]
  Andrzej Ptok,$^{1}$ Konrad J. Kapcia,$^{2}$ Pawe\l{} T. Jochym,$^{1}$ \\ [.1cm]
  Jan \L{}a\.{z}ewski,$^{1}$ Andrzej M. Ole\'{s},$^{3,4}$ and Przemys\l{}aw Piekarz$\,^{1}$\\[.2cm]
  {\itshape
  	\mbox{$^{1}$Institute of Nuclear Physics, Polish Academy of Sciences, 
  	W. E. Radzikowskiego 152, PL-31342 Krak\'{o}w, Poland}\\
	\mbox{$^{2}$Faculty of Physics, Adam Mickiewicz University in Pozna\'{n}, 
	Uniwersytetu Pozna\'{n}skiego 2, PL-61614 Pozna\'{n}, Poland}\\
	\mbox{$^{3}$Institute of Theoretical Physics, Jagiellonian University,  
    Prof. S. \L{}ojasiewicza 11, PL-30348  Krak\'{o}w, Poland}\\
	$^{4}$Max Planck Institute for Solid State Research, 
    Heisenbergstrasse 1, D-70569 Stuttgart, Germany\\}
(Dated: \today)
\\[1cm]
\end{center}

\setcounter{equation}{0}
\renewcommand{\theequation}{S\arabic{equation}}
\setcounter{figure}{0}
\renewcommand{\thefigure}{S\arabic{figure}}
\setcounter{section}{0}
\renewcommand{\thesection}{S\arabic{section}}
\setcounter{table}{0}
\renewcommand{\thetable}{S\arabic{table}}
\setcounter{page}{1}

In this Supplemental Material we present additional results, in particular concerning:
\begin{itemize}
\item The details of numerical calculations in the Section below.
\item Lattice constants obtained from presented 
{\it ab initio} calculations concerning Ce $4f$ electrons as valence and core electrons, respectively (in Tables~\ref{tab.latt} and~\ref{tab.latt2} for the PBE pesudopotentials, and in Tables~\ref{tab.latt_sol} and~\ref{tab.latt2_sol} for the PBEsol pesudopotentials).
\item  Effects of the orbital order (i.e., lowering of the system symmetry) on the phonon band structure in Fig.~\ref{fig.oo}.
\item The electronic density of states with orbital projections, in particular, on $4f$ states, in Fig.~\ref{fig.dos}.
\item The electronic band structure with the orbital projections on Rh and Ce atoms in Fig.~\ref{fig.band_proj}.
\item  Comparison of the band structures in the absence and in the presence of the spin-orbit coupling for CeRh$_{2}$As$_{2}$ compound in Fig.~\ref{fig.bands_soc}.
\item  Comparison between the band structures of nonmagnetic CeRh$_{2}$As$_{2}$ compound and an artificial system without Ce atoms
    (i.e, not existing in the nature  Rh$_2$As$_{2}$ system) in Fig.~\ref{fig.bands_rhas}.
\item  Comparison between the band structures of the non-polarized ground state and system with fixed polarization (for CeRh$_{2}$As$_{2}$ compound) in Fig.~\ref{fig.bands_mag}.
\item Modifications of the Fermi surface by the shift of the Fermi leve (which corresponds to the magnetic Lifshitz transition) shown in Fig.~\ref{fig.mlt}.
\end{itemize}
All presented results are obtained using {\sc VASP} with the $4f$ electrons of Ce atoms treated as valence electrons (except for Tables~\ref{tab.latt2} and~\ref{tab.latt2_sol}). 
Due to the fact that the PBEsol pseudopotetnials correctly reproduce the experimental lattice constant, we use this pseudopotential in all calculations concerning electron and phonon properties.
For additional discussion we present also the electronic band structure obtained using the {\sc Quantum Espresso} software in Fig.~\ref{fig.bands_qe}.

\section{Details of numerical calculation}
\label{sec.det}

The first-principles (DFT) calculations are performed using the projector augmented-wave (PAW) potentials~\cite{blochl.94} implemented in the Vienna Ab initio Simulation Package ({\sc Vasp}) code~\cite{kresse.hafner.94,kresse.furthmuller.96,kresse.joubert.99}.
The calculations are made within generalized gradient approximation (GGA) 
in the Perdew, Burke, and Ernzerhof (PBE) parametrization~\cite{pardew.burke.96} as well as a revised PBE GGA for solids (PBEsol)~\cite{perdew.ruzsinszky.08}
The summation in reciprocal space was performed over
$18 \times 18 \times 8$ {\bf k}-point grid generated with the Monkhorst--Pack scheme~\cite{monkhorst.pack.76}.
The energy cutoff for the plane-wave expansion is set to $350$~eV.
The crystal structure as well as atom positions were optimized in 
the conventional unit cell with Ce $4f$ electrons treated as valence electrons.
As a break condition of the optimization loop, we take energy difference of 
$10^{-6}$~eV and $10^{-7}$~eV for ionic and electronic degrees of freedom, respectively.

A comparison of the obtained lattice constant for the PBE and PBEsol pseudopotentials, can be found in Tables~\ref{tab.latt} and~\ref{tab.latt2}, and Tables~\ref{tab.latt_sol} and~\ref{tab.latt2_sol}, respectively.
From the comparison with experimental data we can conclude, that the PBEsol pseudopotentials more correctly reproduce the experimental data~\cite{khim.landaeta.21}.
The optimized lattice constants read \mbox{$a=4.2216$}~\AA\ and $c=9.8565$~\AA and agree well with experimental data of $a=4.2801$~\AA\ and $c=9.8616$~\AA~\cite{khim.landaeta.21}, respectively. 
Five non-equivalent atoms of the structure occupy following positions: Ce $(1/4,1/4,0.2528)$, 
As $(3/4,1/4,0)$ and $(1/4,1/4,0.8629)$, and Rh $(3/4,1/4,1/2)$ and $(1/4,1/4,0.6212)$, 
which correspond to Wyckoff's positions $2c$, $2a$, $2c$, $2b$, and $2c$, respectively.
In this case, experimentally found positions are given as: Ce $(1/4,1/4,0.25469)$, 
As $(3/4,1/4,0)$ and $(1/4,1/4,0.86407)$,  
and Rh $(3/4,1/4,1/2)$ and $(1/4,1/4,0.61742)$~\cite{khim.landaeta.21}.

\begin{figure}[!t]
\centering
\includegraphics[width=\linewidth]{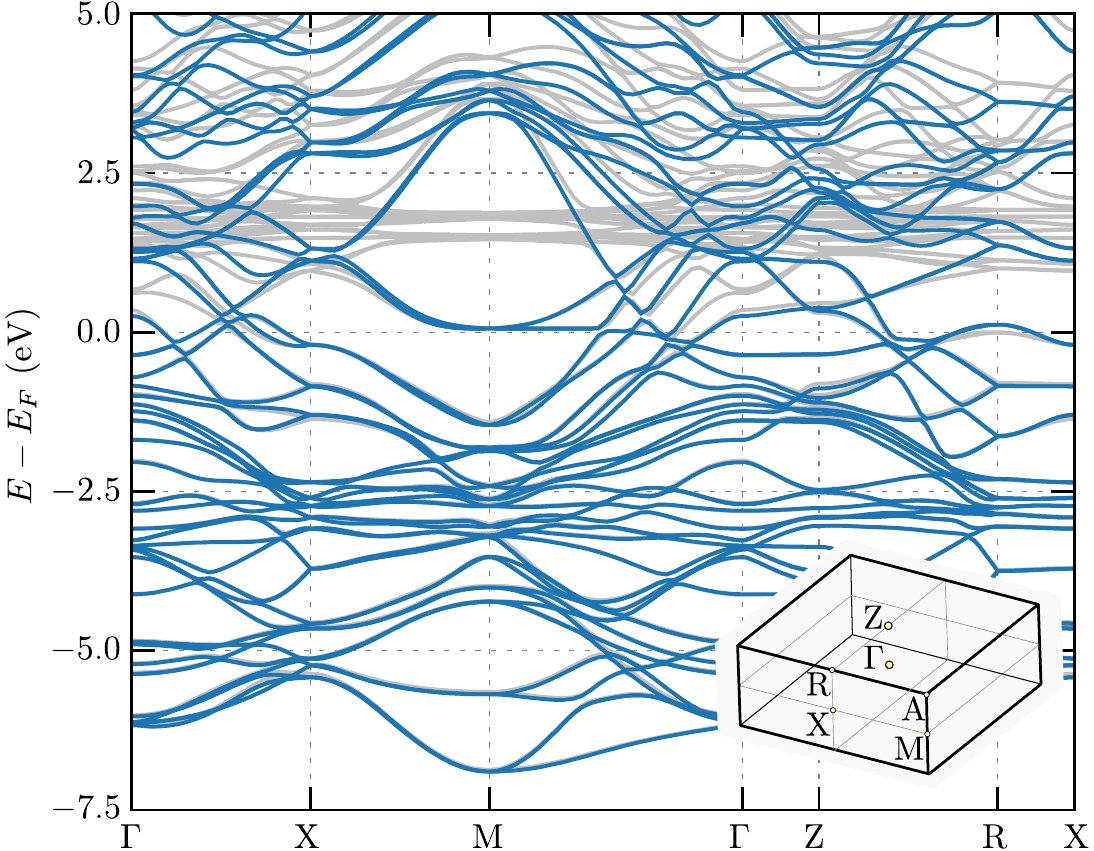}
\caption{
Electronic band structures in the presence of the spin-orbit interaction 
obtained within the {\sc Quantum Espresso} software.
In this calculations, the {\sc PSlibrary} pseudopotentials were used.
Solid blue and gray lines correspond to different treatments
of the Ce $4f$ electrons (as core and valence electrons, respectively).
Due to different electronic configuration of the Ce atoms 
(i.e., [Xe] $6s^{2}4f^{1/2}5d^{3/2}$) than in the {\sc Vasp} pseudopotentials 
(i.e., [Xe] $6s^{2}4f^{1}5d^{1}$) the positions of the $4f$ electron bands 
can be overestimated.
\label{fig.bands_qe}
}
\end{figure}

The electronic band structures are also evaluated within the {\sc Quantum Espresso}~\cite{giannozzi.baroni.09,giannozzi.andreussi.17}.
In this case, during calculations we use pseudopotentials developed in a frame of 
{\sc PSlibrary}~\cite{dalcorso.14}. Additionally, we use the cutoff for charge 
density and wave function with the nominal value increased by $100$ Ry.
However, due to different electronic configuration of the Ce atoms (i.e., [Xe] $6s^{2}4f^{1/2}5d^{3/2}$) than in the case of {\sc Vasp} calculations, the positions of the $4f$ states of Ce atoms can be overestimated -- cf. Fig.~\ref{fig.bands_qe}.
Here, we should notice that the electronic band structure obtained within the full-potential linearized augmented planewave method by  the {\sc Wien2K} software (see Ref.~\cite{nogaki.daido.21}) is similar to this obtained from {\sc Vasp}.

Phonon calculations were performed in the supercell with $40$ atoms 
containing $2\times2\times1$ conventional unit cells. The phonon dispersion 
curves and phonon density of states (DOS) are calculated using {\sc Alamode} software~\cite{tadano.gohda.14}.
Calculations are performed for the thermal distribution of multi-displacements 
of atoms at $T=50$~K, generated within {\sc hecss} procedure ~\cite{jochym.lazewski.21}.
Energy of the one hundred different configurations of the supercells and the Hellmann-Feynman forces acting on all atoms are determined using {\sc Vasp}.
In calculated dynamical properties, we include contributions from harmonic and cubic interatomic force constants to phonon frequencies.

\newpage

\begin{table}[!h]
\caption{
Experimental (taken from Ref.~\cite{khim.landaeta.21}) and theoretical 
({\it ab initio}) lattice constants obtained in the absence of magnetism (NM), 
assuming antiferromagnetic (AFM) order (in two directions), and assuming 
ferromagnetic (FM) order (in two directions) as initial magnetic orders.
During calculations, the final states conserved assumed magnetic orders 
(AFM and FM with final magnetic moments $0.27$~$\mu_\text{B}$ and $0.10$~$\mu_\text{B}$ 
at Ce atoms, respectively). In contrary to this, calculations starting from 
the NM state (all magnetic moments equal zero) lead finally to small magnetic 
moments with magnitude smaller than $0.03$~$\mu_\text{B}$.
For the theoretical results we show also $\delta E$, denoting the 
difference of energy between ground state (GS) and given magnetic order.
Results obtained with PBE pseudopotential treatment of $4f$ electrons 
of Ce atoms as valence electrons
and in the presence of the spin-orbit coupling.
}
\begin{ruledtabular}
\begin{tabular}{ccccccc}
 & exp.~\cite{khim.landaeta.21} & NM & \multicolumn{2}{c}{AFM} & \multicolumn{2}{c}{FM} \\
 & & --- & ${\bm M} || ab$ & ${\bm M} || c$ & ${\bm M} || ab$ & ${\bm M} || c$ \\
 \hline
$\delta E$ (meV) & --- & (GS) 0.000 & 3.135 & 4.082 & 4.447 & 4.606 \\
a (\AA) & 4.2801 & 4.2802 & 4.2804 & 4.2802 & 4.2800 & 4.2798 \\
c (\AA) & 9.8616 & 9.9752 & 9.9788 & 9.9773 & 9.9761 & 9.9766
\end{tabular}
\end{ruledtabular}
\label{tab.latt}
\end{table}

\begin{table}[!h]
\caption{
The same as in Table~\ref{tab.latt}, but from {\it ab initio} calculations with 
PBE pseudopotential treatment of $4f$ electrons of Ce atoms as core electrons.
Independently of the initial magnetic moments configurations, 
the final states do not exhibit any finite 
magnetic moments.
}
\begin{ruledtabular}
\begin{tabular}{ccccccc}
 & exp.~\cite{khim.landaeta.21} & NM & \multicolumn{2}{c}{AFM} & \multicolumn{2}{c}{FM} \\
 & & --- & ${\bm M} || ab$ & ${\bm M} || c$ & ${\bm M} || ab$ & ${\bm M} || c$ \\
 \hline
$\delta E$ (meV) & --- & 0.108 & 0.112 & 0.116 & (GS) 0.000 & 0.047 \\
a (\AA) & 4.2801 & 4.3243 & 4.3243 & 4.3243 & 4.3243 & 4.3243 \\
c (\AA) & 9.8616 & 10.0442 & 10.0442 & 10.0442 & 10.0442 & 10.0442
\end{tabular}
\end{ruledtabular}
\label{tab.latt2}
\end{table}

\begin{table}[!h]
\caption{
The same as in Table~\ref{tab.latt}, but in the case of PBEsol 
pseudopotentials (i.e., $4f$ electrons of Ce atoms treated as valence electrons).
}
\begin{ruledtabular}
\begin{tabular}{ccccccc}
 & exp.~\cite{khim.landaeta.21} & NM & \multicolumn{2}{c}{AFM} & \multicolumn{2}{c}{FM} \\
 & & --- & ${\bm M} || ab$ & ${\bm M} || c$ & ${\bm M} || ab$ & ${\bm M} || c$ \\
 \hline
$\delta E$ (meV) & --- & 0.260 & (GS) 0.000 & 0.078 & 0.148 & 0.160 \\
a (\AA) & 4.2801 & 4.2214 & 4.2216 & 4.2215 & 4.2215 & 4.2214 \\
c (\AA) & 9.8616 & 9.8560 & 9.8565 & 9.8561 & 9.8562 & 9.8562
\end{tabular}
\end{ruledtabular}
\label{tab.latt_sol}
\end{table}

\begin{table}[!h]
\caption{
The same as in Table~\ref{tab.latt2}, but in the case of PBEsol 
pseudopotentials (i.e., $4f$ electrons of Ce atoms treated as core electrons).
}
\begin{ruledtabular}
\begin{tabular}{ccccccc}
 & exp.~\cite{khim.landaeta.21} & NM & \multicolumn{2}{c}{AFM} & \multicolumn{2}{c}{FM} \\
 & & --- & ${\bm M} || ab$ & ${\bm M} || c$ & ${\bm M} || ab$ & ${\bm M} || c$ \\
 \hline
$\delta E$ (meV) & --- & 0.002 & (GS) 0.000 & 6.255 & 0.002 & 0.000 \\
a (\AA) & 4.2801 & 4.2670 & 4.2670 & 4.2666 & 4.2669 & 4.2669 \\
c (\AA) & 9.8616 & 9.9224 & 9.9224 & 9.9225 & 9.9227 & 9.9227
\end{tabular}
\end{ruledtabular}
\label{tab.latt2_sol}
\end{table}

\begin{figure}[!h]
\centering
\includegraphics[width=0.85\linewidth]{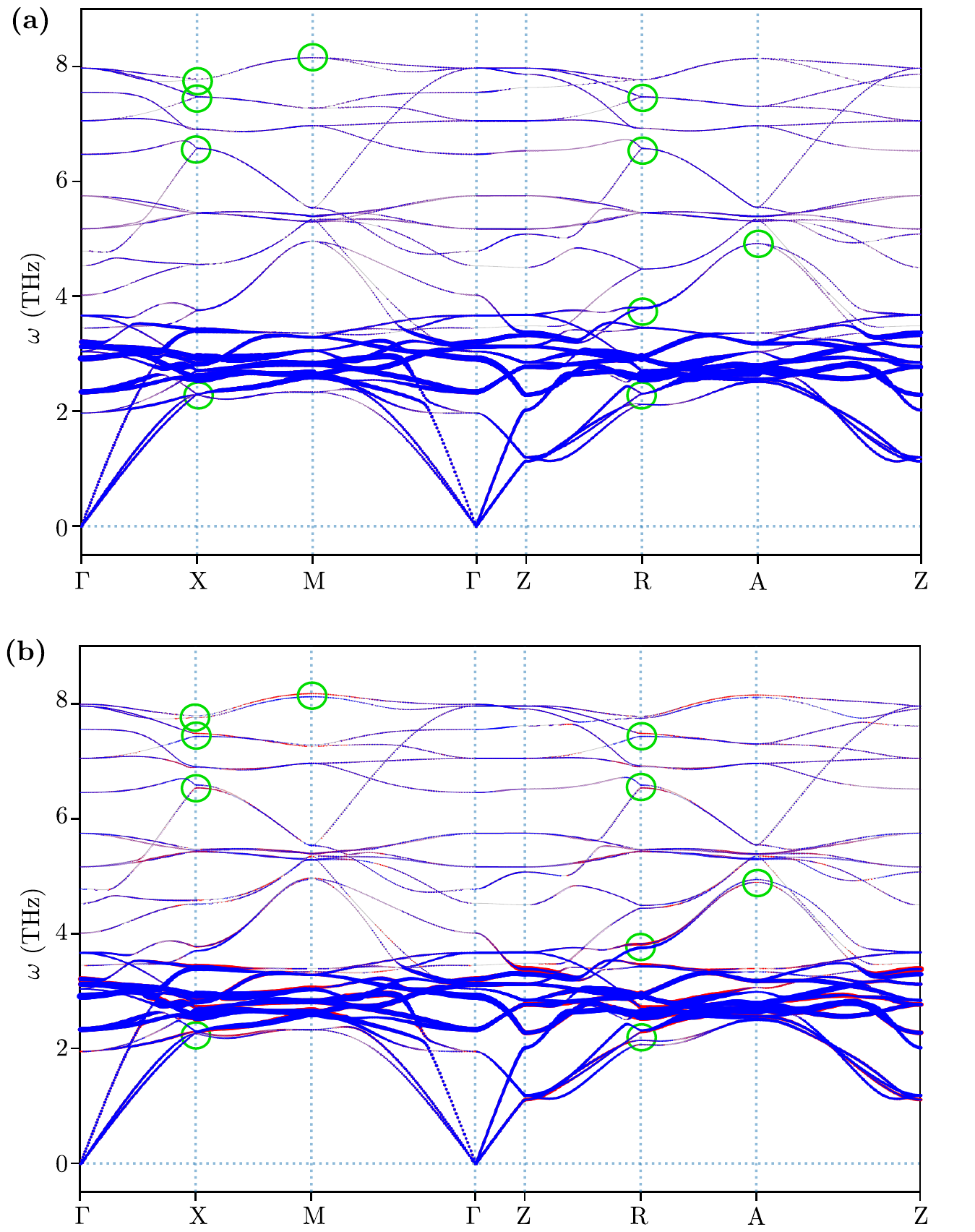}
\caption{
Effects of hidden orbital order on the phonon spectra---shown by width 
and color of line corresponding to the projection of the polarization 
vector on the 1$^{st}$ and 2$^{nd}$ Ce atoms.
In the case of two non-distinguishable Ce atoms the {\it P4/nmm} symmetry exists.
Introduction of the orbital order leads to the realization of two sub-lattices, distinguishable by realized pseudo-orbitals at Ce atoms.
A main role on the realization of this orbital order is played by two 
non-equivalent Rh$_{2}$As$_{2}$ planes surrounding Ce atoms from top and bottom.
Assuming {\it P4/nmm} symmetry (a), the phonon projection on the 1$^{st}$ 
and 2$^{nd}$ Ce atoms (red and blue color, respectively), are non-distinguishable 
(system is described by $64$ symmetry operations).
Introduction of orbital order leads to the lowering of the system 
symmetry  
to {\it P4mm} symmetry (b).
In this case system undergoes only a half of symmetry operations, i.e., 
$32$ operations. This is well visible by projection of polarization vectors 
on the 1$^{st}$ and 2$^{nd}$ Ce atoms in a form of splitting red and blue lines.
From this, degeneracy of the band structure in high symmetry points are lift 
[e.g., cf. green circles at (a) and (b)].
This effect is similar to observed in electronic band structure, e.g. as 
the emergence of the Dirac semi-metals after introducing magnetic order.
Both results were obtained from the same dataset of hundred multidisplacement configurations corresponding to $50$~K.
Results for PBEsol pseudopotentials in the presence of SOC and with 
$4f$ electrons of Ce atoms treated as valence electrons.
\label{fig.oo}
}
\end{figure}

\begin{figure}[t!]
\centering
\includegraphics[width=\linewidth]{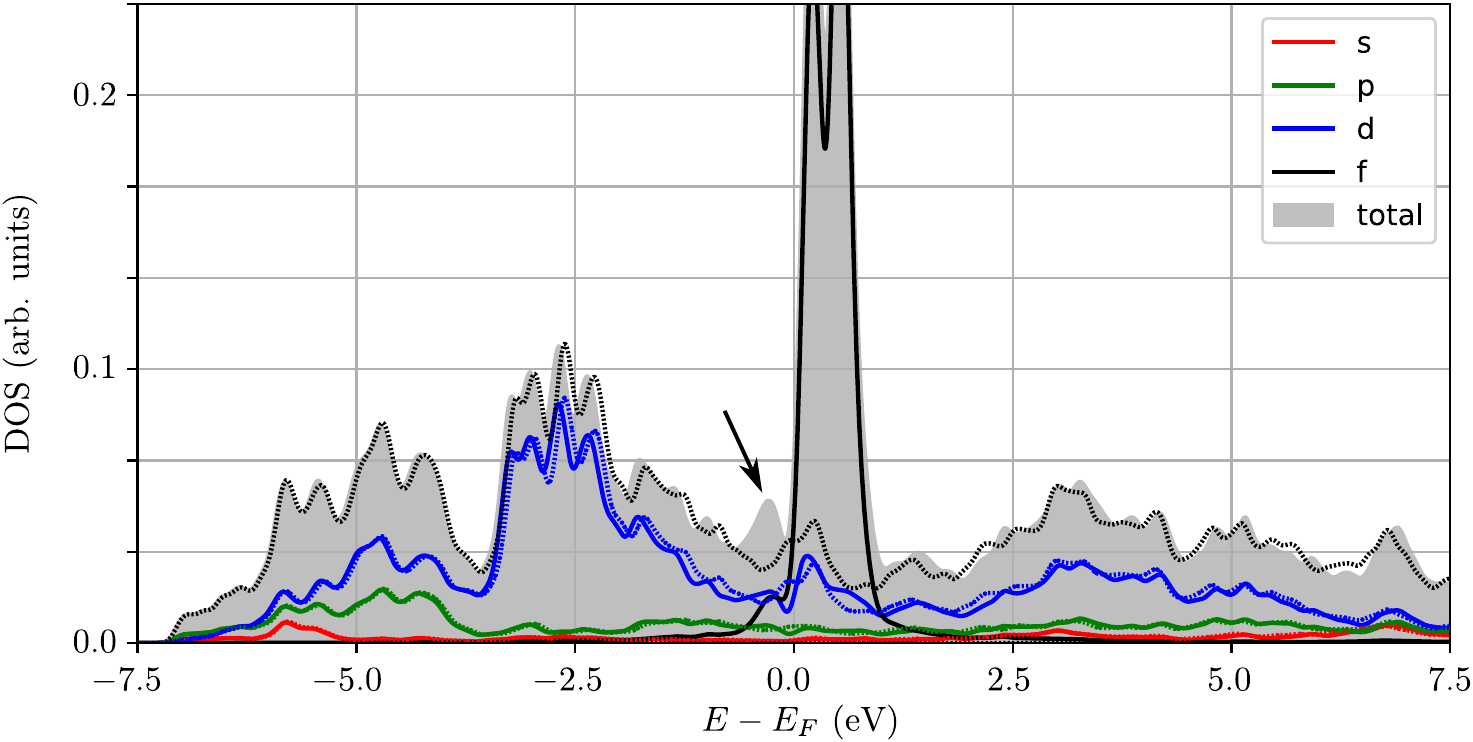}
\caption{
Orbital projected electronic DOS in the presence of SOC. Solid and dashed 
lines (colors as labeled) correspond to different treatments of the $4f$
electrons originating from Ce atoms (as valence and core electrons, respectively).
Dashed black line corresponds to the total DOS in the case of Ce $4f$ electrons treated as core electrons, whereas the grey background denotes the total DOS for 
the case of Ce $4f$ electrons treated as valence electrons.
The modification due to $f$--$d$ orbital hybridization is observed 
around the Fermi level (marked by black arrow).
Results for PBEsol pseudopotentials.
\label{fig.dos}
}
\end{figure}

\begin{figure}[t!]
\centering
\includegraphics[width=\linewidth]{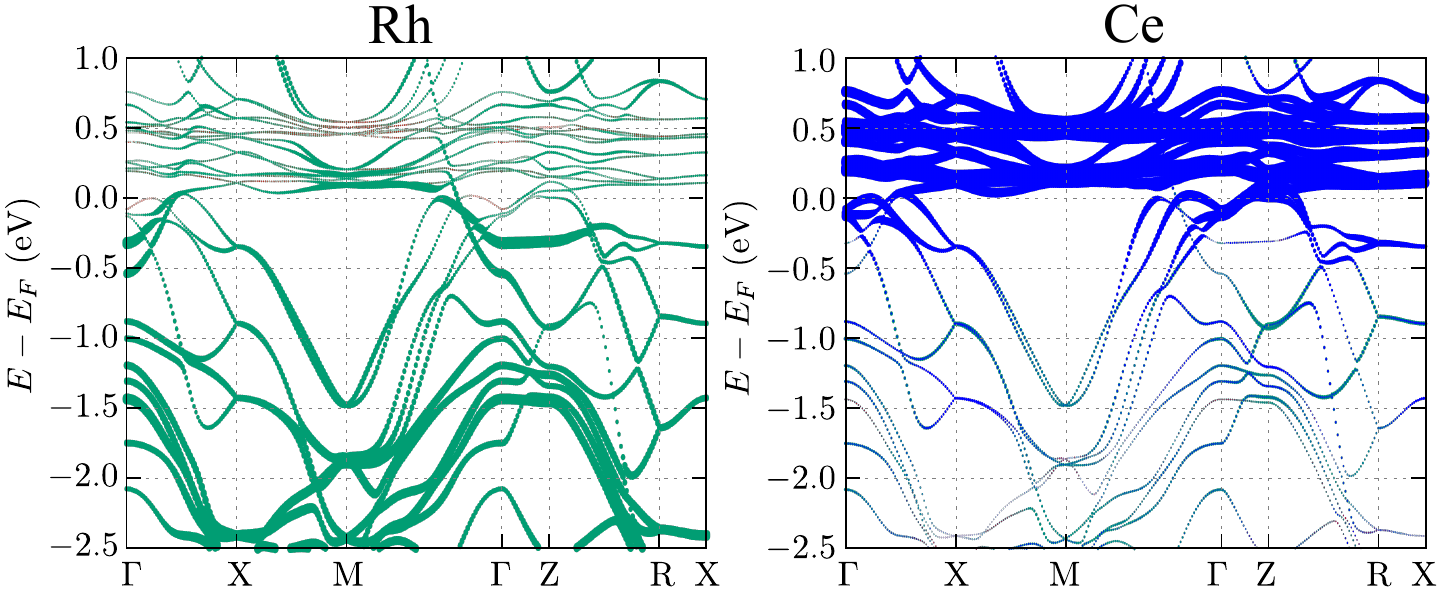}
\caption{
Orbital projected electronic band structure for Rh and Ce atoms (left and right panels, respectively).
Red, green, and blue color correspond to $p$, $d$, and $f$ orbitals, respectively.
Sizes of dots correspond to the value of the contribution of a given orbital.
Results for PBEsol pseudopotentials in the presence of SOC and 
with $4f$ electrons of Ce atoms treated as valence electrons.
\label{fig.band_proj}
}
\end{figure}

\begin{figure}[!t]
\centering
\includegraphics[width=0.89\linewidth]{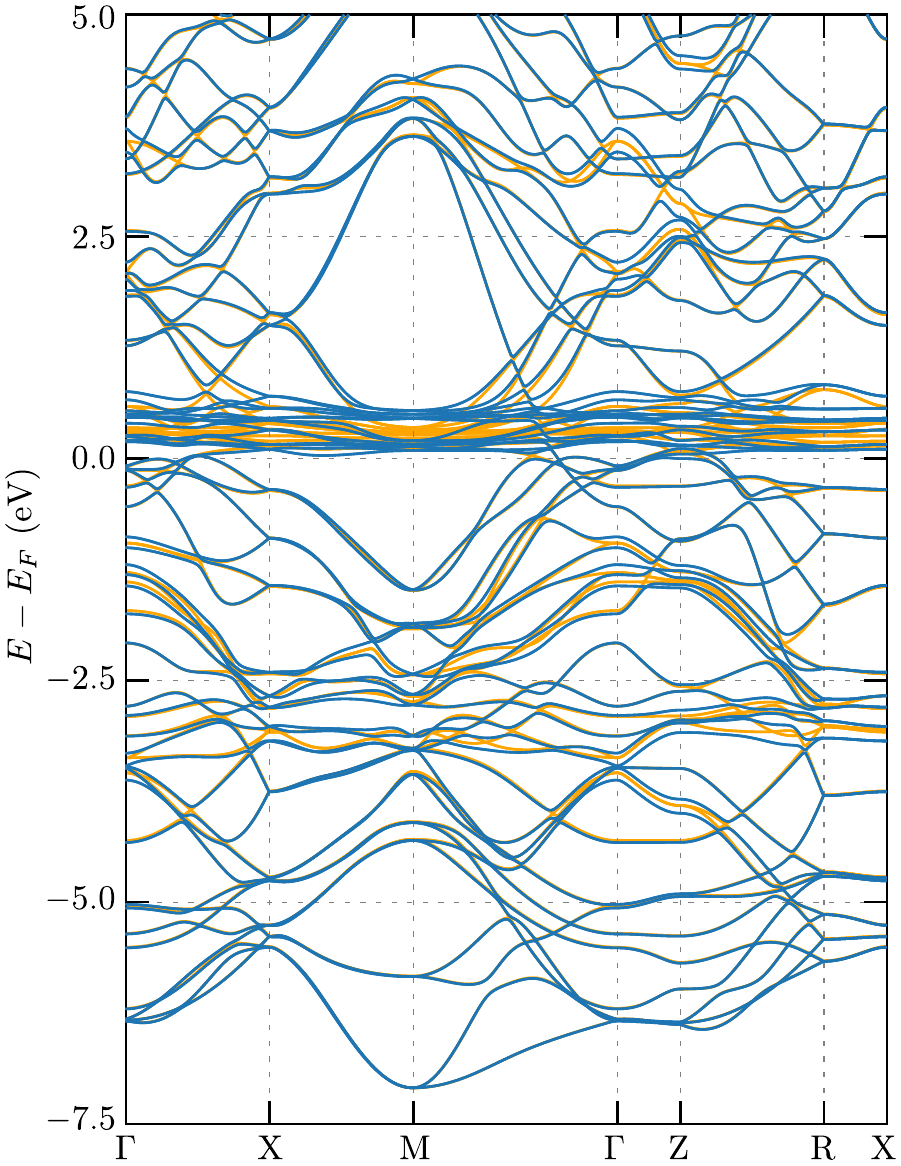}
\caption{
Comparison between the electronic band structures of the non-magnetic 
ground state of CeRh$_{2}$As$_{2}$ in the absence and in the presence of SOC 
(solid orange and blue lines, respectively).
Results for PBEsol pseudopotentials with $4f$ electrons of Ce atoms 
treated as valence electrons.
\label{fig.bands_soc}
}
\end{figure}

\begin{figure}[!t]
\centering
\includegraphics[width=\linewidth]{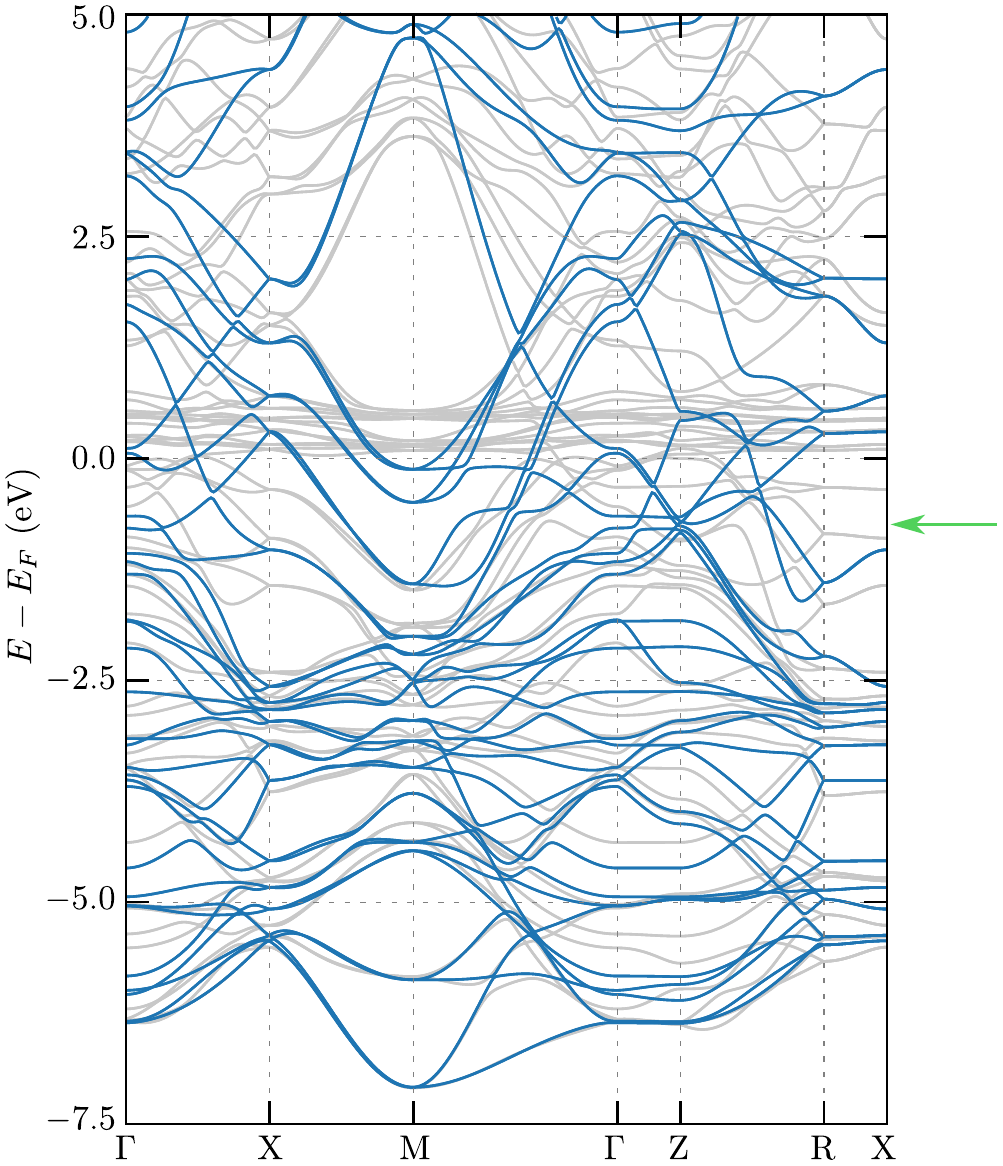}
\caption{
Comparison between the electronic band structures of the non-magnetic ground state of CeRh$_{2}$As$_{2}$ (solid gray lines) and an artificial structure without Ce atoms 
(blue solid lines). Zero-energy denotes the Fermi level of CeRh$_{2}$As$_{2}$.
Green arrow (on the right) show the location of Fermi level for the 
artificial system. Results obtained in the presence of spin-orbit coupling.
Results for PBEsol pseudopotentials in the presence of SOC and with $4f$ 
electrons of Ce atoms treated as valence electrons.
\label{fig.bands_rhas}
}
\end{figure}

\begin{figure}[!t]
\centering
\includegraphics[width=0.875\linewidth]{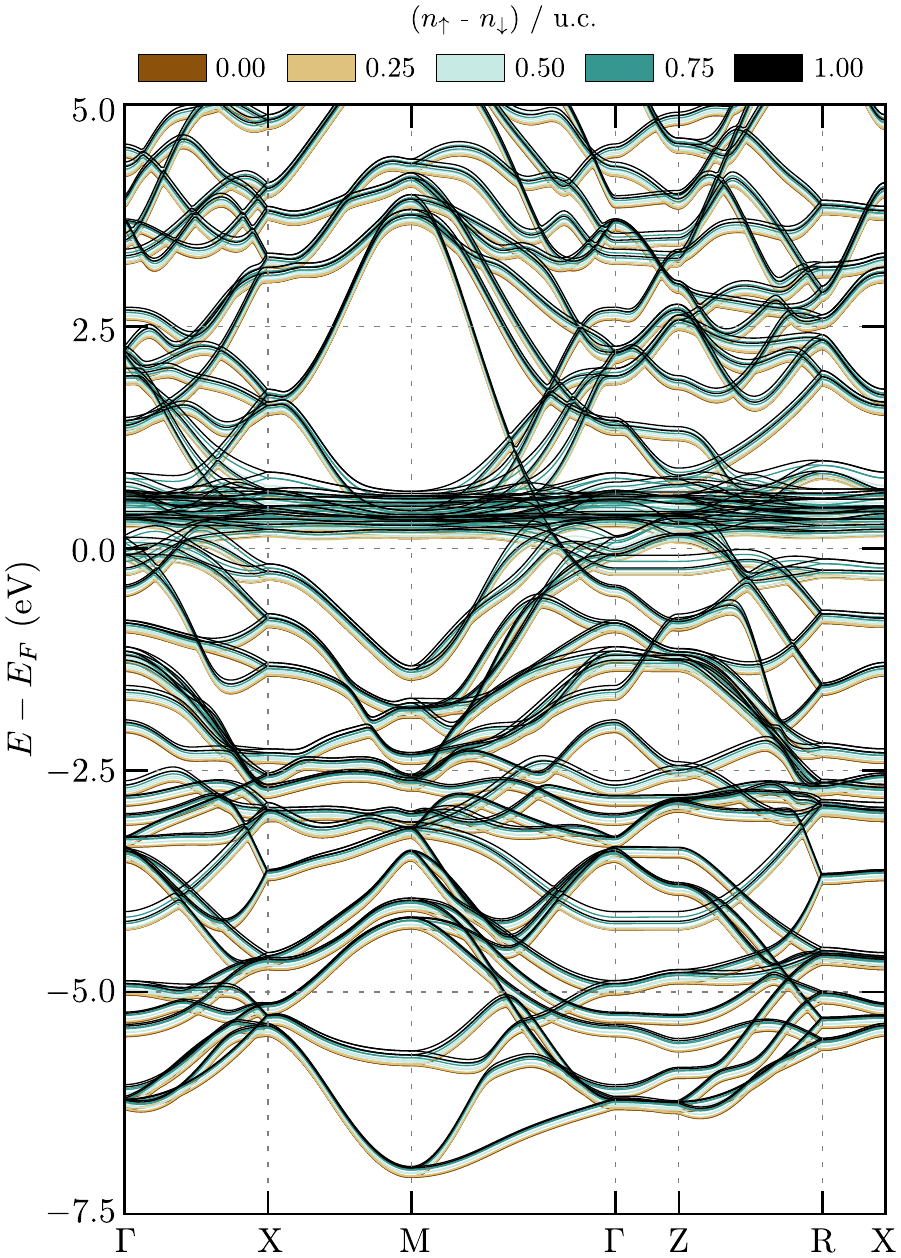}
\caption{
Comparison between the electronic band structures between the non-magnetic ground state (i.e., in the absence of magnetic polarization; solid gray lines) and nonphysically polarized state (difference between electrons with opposite spins per the conventional unit cell is equal as labeled; dispersion curves for spin-up and spin-down electrons are shown by solid red and blue lines respectively).
Results obtained in the absence of the spin-orbit coupling for CeRh$_{2}$As$_{2}$.
Results for PBEsol pseudopotentials in the absence of the SOC and with $4f$ electrons of Ce atoms treated as valence electrons.
\label{fig.bands_mag}
}
\end{figure}

\begin{figure}[!t]
\centering
\includegraphics[width=\linewidth]{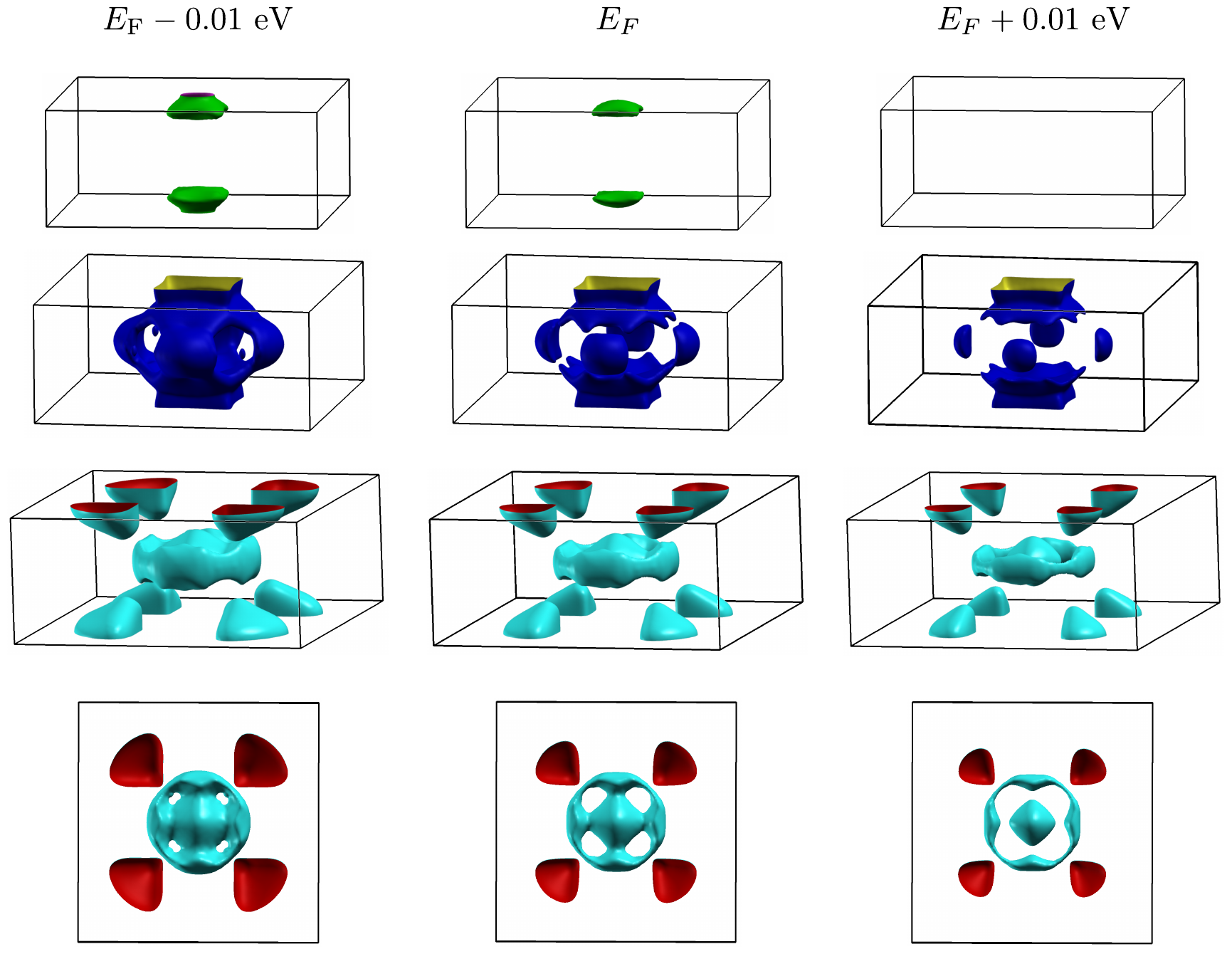}
\caption{
Modification of the Fermi surface by shifting the Fermi level (as labeled),
what corresponds to the realization of the magnetic Lifshitz transition.
Increasing magnetic field leads to lift spin degeneracy and splitting of 
the bands. During this process, the modification of the Fermi surface shape can 
be observed. In the extreme case (top panels) the Fermi pocket can change its 
shape (left top panel) or totally disappear (right top panel).
First three rows correspond to different pockets of the Fermi surface, 
whereas the bottom row shows the view from the top of the pocket shown in third row.
Results for PBEsol pseudopotentials in the presence of SOC and 
with $4f$ electrons of Ce atoms treated as valence electrons.
\label{fig.mlt}
}
\end{figure}

\end{document}